\documentclass[twocolumn]{aastex61}
\usepackage{url}
\usepackage{epsf}

\received{}
\revised{}
\accepted{}
\submitjournal{}
\shorttitle{Optical properties of IR-bright DOGs viewed with Subaru HSC}
\shortauthors{Noboriguchi et al.}

\begin{document}

\title{Optical properties of infrared-bright dust-obscured galaxies viewed with Subaru Hyper Suprime-Cam}

\email{noboriguchi@cosmos.phys.sci.ehime-u.ac.jp}

\author{Akatoki Noboriguchi}
\affiliation{Graduate School of Science and Engineering, Ehime University, Bunkyo-cho, Matsuyama, 790-8577, Japan}
\author{Tohru Nagao}
\affiliation{Research Center for Space and Cosmic Evolution, Ehime University, Bunkyo-cho, Matsuyama, Ehime, 790-8577, Japan}
\author{Yoshiki Toba}
\affiliation{Research Center for Space and Cosmic Evolution, Ehime University, Bunkyo-cho, Matsuyama, Ehime, 790-8577, Japan}
\affiliation{Academia Sinica Institute of Astronomy and Astrophysics, PO Box 23-141, Taipei, 10617, Taiwan}
\affiliation{Department of Astronomy, Kyoto University, Kitashirakawa-Oiwake-cho, Kyoto, 606-8502, Japan}
\author{Mana Niida}
\affiliation{Graduate School of Science and Engineering, Ehime University, Bunkyo-cho, Matsuyama, 790-8577, Japan}
\author{Masaru Kajisawa}
\affiliation{Graduate School of Science and Engineering, Ehime University, Bunkyo-cho, Matsuyama, 790-8577, Japan}
\affiliation{Research Center for Space and Cosmic Evolution, Ehime University, Bunkyo-cho, Matsuyama, Ehime, 790-8577, Japan}
\author{Masafusa Onoue}
\affiliation{National Astronomical Observatory of Japan, 2-21-1 Osawa, Mitaka, Tokyo, 181-8588, Japan}
\affiliation{Department of Astronomy, Graduate University for Advanced Studies (SOKENDAI), 2-21-1 Osawa, Mitaka, Tokyo, 181-8588, Japan}
\affiliation{Max-Planck-Institut f\"ur Astronomie, K\"onigstuhl 17, D-69117, Heidelberg, Germany}
\author{Yoshiki Matsuoka}
\affiliation{Research Center for Space and Cosmic Evolution, Ehime University, Bunkyo-cho, Matsuyama, Ehime, 790-8577, Japan}
\author{Takuji Yamashita}
\affiliation{Research Center for Space and Cosmic Evolution, Ehime University, Bunkyo-cho, Matsuyama, Ehime, 790-8577, Japan}
\affiliation{National Astronomical Observatory of Japan, 2-21-1 Osawa, Mitaka, Tokyo, 181-8588, Japan}
\author{Yu-Yen Chang}
\affiliation{Academia Sinica Institute of Astronomy and Astrophysics, PO Box 23-141, Taipei, 10617, Taiwan}
\author{Toshihiro Kawaguchi}
\affiliation{Department of Economics, Onomichi City University, Onomichi, Hiroshima, 722-8506, Japan}
\author{Yutaka Komiyama}
\affiliation{National Astronomical Observatory of Japan, 2-21-1 Osawa, Mitaka, Tokyo, 181-8588, Japan}
\affiliation{Department of Astronomy, Graduate University for Advanced Studies (SOKENDAI), 2-21-1 Osawa, Mitaka, Tokyo, 181-8588, Japan}
\author{Kodai Nobuhara}
\affiliation{Graduate School of Science and Engineering, Ehime University, Bunkyo-cho, Matsuyama, 790-8577, Japan}
\author{Yuichi Terashima}
\affiliation{Graduate School of Science and Engineering, Ehime University, Bunkyo-cho, Matsuyama, 790-8577, Japan}
\affiliation{Research Center for Space and Cosmic Evolution, Ehime University, Bunkyo-cho, Matsuyama, Ehime, 790-8577, Japan}
\author{Yoshihiro Ueda}
\affiliation{Department of Astronomy, Kyoto University, Kitashirakawa-Oiwake-cho, Kyoto, 606-8502, Japan}

\begin{abstract}

We report on the optical properties of infrared (IR)-bright dust-obscured galaxies (DOGs) that are defined as $(i-{\rm{[22]}})_{\rm AB}\geq7.0$.
Because supermassive black holes (SMBHs) in IR-bright DOGs are expected to be rapidly growing in the major merger scenario, they provide useful clues for understanding the co-evolution of SMBHs and their host galaxies.
However, the optical properties of IR-bright DOGs remain unclear
because the optical emission of a DOG is very faint.
By combining $\sim105\ {\rm{deg^2}}$ images of the optical, near-IR, and mid-IR data obtained from the Subaru Hyper Suprime-Cam (HSC) survey, the VISTA VIKING survey, and the ${\it WISE}$ all-sky survey, respectively, 571 IR-bright DOGs were selected. 
We found that IR-bright DOGs show a redder $(g-z)_{\rm AB}$ color than other populations of dusty galaxies, such as ultra-luminous IR galaxies (ULIRGs) at a similar redshift, with a significantly large dispersion. 
Among the selected DOGs, star-formation (SF) dominated DOGs show a relatively red color, while active galactic nucleus (AGN) dominated DOGs show a rather blue color in optical. 
This result is consistent with the idea that the relative AGN contribution in the optical emission becomes more significant at a later stage in the major merger scenario.
We discovered eight IR-bright DOGs showing a significant blue excess in blue HSC bands (BluDOGs). This blue excess can be interpreted as a leaked AGN emission that is either a directly leaking or a scattered AGN emission, as proposed for some blue-excess Hot DOGs in earlier studies.

\end{abstract}


\section{Introduction} \label{sec:intro}

It is widely recognized that there are some tight scaling relations between the mass of supermassive black holes (SMBHs) and properties of the host galaxy, such as the stellar mass \citep[e.g.,][]{1998AJ....115.2285M,2003ApJ...589L..21M,2009ApJ...698..198G}. Such scaling relationships are now regarded as observational evidence suggesting the so-called co-evolution of galaxies and SMBHs, i.e., galaxies and SMBHs have evolved with a close interplay during the cosmological timescale (see \citealt{2013ARA&A..51..511K} for a review). One important question about this co-evolution is how mass accretion onto SMBHs is triggered, because the angular momentum prevents gas at the nucleus from accreting onto the SMBH.
Therefore, it has been proposed that a major merger of two (or more) gas-rich galaxies is an efficient path to trigger mass accretion onto a SMBH (e.g., {\citealt{1988ApJ...325...74S}; \citealt{2006ApJS..163....1H}}).
In such scenarios, the gas-rich galaxy merger first causes active star formation (SF), and then gas accretion to the nuclear region triggers the activity of SMBHs that will be recognized as an active galactic nucleus (AGN). However, the most active period of this SF phase and AGN phase is generally obscured by heavy dust, which prevents us from investigating these phases observationally. Another difficulty with investigating such active systems is the rareness of galaxies in the very active phase, because the timescale of the most active phase of the co-evolution is expected to be short (e.g., {\citealt{2008ApJ...677..943D}}; \citealt{2008ApJS..175..356H}). 

Recently, dust-obscured galaxies (DOGs, \citealt{2008ApJ...677..943D}) shed light on this issue, because SMBHs in DOGs are expected to be rapidly growing during the co-evolution.
DOGs are originally defined as galaxies that are bright in mid-infrared (MIR), while faint in optical. Specifically, $(R-[24])_{\rm AB} \geq 7.5\ {\rm{mag}}$ (i.e., $F_{\nu} (24{\rm \mu m})/F_{\nu} (R) \gtrsim1000$), \citealt{2008ApJ...677..943D}; \citealt{2008ApJ...672...94F}).
In the context of the gas-rich major merger scenario, it is expected that the SF phase evolves into the AGN phase
because the merging event leads to the active SF, while the gas accretion onto the nucleus caused by such a merger requires some time (see, e.g., \citealt{2007ApJ...671.1388D}; \citealt{2012MNRAS.420L...8H}; \citealt{2017A&A...608A..90M}).
Because such active galaxies are expected to be heavily surrounded by dust, DOGs potentially correspond to galaxies in the SF phase or AGN phase (\citealt{2008ApJ...677..943D}; \citealt{2008ApJS..175..356H}).
DOGs are classified into two sub-classes based on their spectral energy distribution (SED): ``Bump DOGs" and ``Power-Law (PL) DOGs" (\citealt{2008ApJ...677..943D}). 
The bump DOGs show a rest-frame $1.6\ {\rm \mu m}$ stellar bump in their SEDs, while the PL DOGs show a power-law feature on their SEDs. Therefore, it is considered that the bump DOGs correspond to galaxies in SF mode (\citealt{2009ApJ...700.1190D}; \citealt{2011ApJ...733...21B}), while the PL DOGs correspond to galaxies in the AGN phase (\citealt{2008ApJ...672...94F}; \citealt{2009ApJ...705..184B}; \citealt{2012AJ....143..125M}).
The fraction of PL DOGs among all DOGs increases with increasing MIR flux density (e.g., \citealt{2008ApJ...677..943D}; \citealt{2015PASJ...67...86T}), which is similar to the behavior of the luminosity dependence of the AGN fraction in ultraluminous infrared (IR) galaxies (ULIRGs; see {\citealt{1996ARA&A..34..749S}} for a review).
The comoving number density of DOGs shows its peak at $z \sim 1-2$ (e.g., \citealt{2008ApJ...677..943D}; \citealt{2017ApJ...835...36T}) that corresponds to the peak of star formation rate density and the growth rate of SMBHs (e.g., \citealt{2006AJ....131.2766R}; \citealt{2014ARA&A..52..415M}). This strongly suggests that DOGs are related to the most active objects in terms of the co-evolution between galaxies and SMBHs. In this sense, DOGs with a high IR luminosity potentially harbor a rapidly growing SMBH and are therefore important for understanding the co-evolution of galaxies and SMBHs. 

The difficulty in studying statistical properties of DOGs is caused by their low number density and optical faintness, which requires optical imaging surveys with a wide survey area and sufficient sensitivity. Such optical surveys are now feasible, thanks to the Subaru Hyper Suprime-Cam (HSC; \citealt{2012SPIE.8446E..0ZM}; \citealt{2018PASJ...70S...1M}). One of the legacy surveys using the Subaru HSC is the Subaru Strategic Program (SSP; \citealt{2018PASJ...70S...4A}), which consists of three layers (Ultra Deep, Deep, and Wide).
Notably, the HSC-SSP wide-field survey is now going on to observe northern areas of $1400\ {\rm deg^2}$ down to $r_{\rm lim} = 26.1\ {\rm [AB\ mag]}$. This survey is extremely powerful for constructing a statistical sample of IR-bright DOGs\footnote{Here we use the term ``IR-bright DOGs'' without any quantitative definition. See Sections \ref{subsubsec:CSAM} and \ref{subsec:sel_eff} for some descriptions of the IR flux of HSC-selected DOGs.} whose number densities are quite low ($\log{\phi} = -6.59 \pm 0.11\ {\rm{Mpc^{-3}}}$; \citealt{2015PASJ...67...86T}).
\cite{2015PASJ...67...86T} conducted a pilot survey of such IR-bright DOGs for $\sim$9 deg$^2$  by combining optical, near-IR (NIR), and MIR imaging data that were obtained from the HSC-SSP early data release catalog (S14A\footnote{The S14A catalog was released internally within the HSC survey team and is based on data obtained from March 2014 to April 2014.}), the VISTA Kilo-degree Infrared Galaxy survey (VIKING; \citealt{2007Msngr.127...28A}) data release (DR) 1\footnote{https://www.eso.org/sci/observing/phase3/data\_releases/\\viking\_dr1.pdf}, and the Wide-field Infrared Survey Explorer (${\it WISE}$; \citealt{2010AJ....140.1868W}) ALLWISE catalog, respectively. They discovered 48 IR-bright DOGs and investigated their statistical properties in NIR and MIR.
However, they did not investigate the rest-frame ultraviolet (UV) and optical properties of IR-bright DOGs because only two optical bands ($i$ and $y$ band) were available in the HSC-SSP S14A data (see \citealt{2015PASJ...67...86T}).

Recently, \cite{2016ApJ...819..111A} discovered objects with blue excess in the rest-frame UV among the ``hot'' dust-obscured galaxies (Hot DOGs; \citealt{2012ApJ...755..173E}; \citealt{2012ApJ...756...96W}), which are characterized by relatively hotter dust than in normal DOGs (see {\citealt{2012ApJ...756...96W}} for more detail).
\cite{2016ApJ...819..111A} reported that Hot DOGs are not always red in the rest-frame UV/optical; in such Hot DOGs, AGN emission leaked from the nucleus may cause a blue excess (\citealt{2016ApJ...819..111A}).
Thus, it is worth investigating whether such a blue excess is also seen in IR-bright DOGs that are in the most active phase of the co-evolution.
Because the physical nature of IR-bright DOGs is still unexplored, detailed studies of rest-frame UV and optical properties of IR-bright DOGs may provide us with new knowledge on this interesting population of galaxies. 
\cite{2015MNRAS.453.3932R} also discovered such a blue excess in the optical spectra of quasars with a very red SED ($F_{\nu} (24{\rm \mu m})/F_{\nu} (R) \gtrsim1000$). These very red quasars may be in a transition phase from typical DOGs to unobscured quasars.

This research aims at studying, for the first time, the statistical properties of IR-bright DOGs in the rest-frame UV and optical wavelengths based on the latest data release catalog of the HSC-SSP, the VIKING, and the ${\it WISE}$ surveys.
This paper is organized as follows: We describe the sample selection of IR-bright DOGs and the classification into bump DOGs and PL DOGs in Section \ref{sec:data}. In Section \ref{sec:results}, we describe the obtained optical properties of IR-bright DOGs, and we discuss the results in Section \ref{sec:disscussion}.
We give a summary in Section \ref{sec:conclusion}.
Throughout this paper, the adopted cosmology is a flat universe with $H_0 = 70\ {\rm km\ s^{-1}\ Mpc^{-1}}$,\ ${\Omega}_{M}=0.3$,\ and\ ${\Omega}_{\Lambda}=0.7$.
Unless otherwise noted, all magnitudes refer to the AB system.

%%%%%%%%%%%%%%%%%%%%%%%%%
\section{Data analysis} \label{sec:data}

\subsection{Sample selection} \label{sec:ss}
\begin{figure*}[ht!]
	\centering
		\includegraphics[width=15cm]{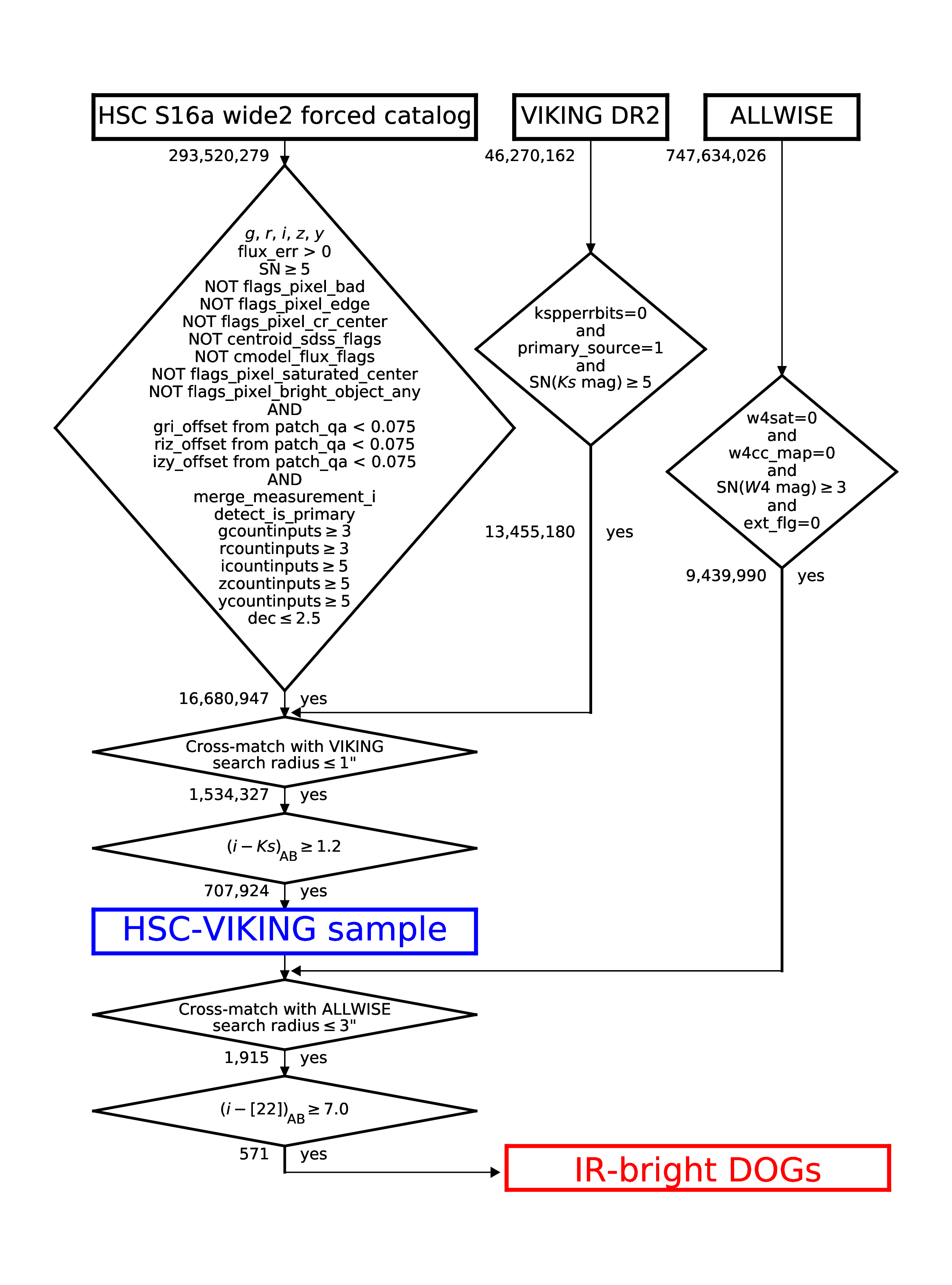} 
		\vspace{0pt}
		\caption{Flow chart of the sample selection process}
		\label{fig:chart}
\end{figure*}

In this study, we selected IR-bright DOGs based on a similar selection procedure as \cite{2015PASJ...67...86T}.
In sum, we selected IR-bright DOGs by combining the ${\it WISE}$ all-sky data and the deep optical imaging data obtained with the Subaru HSC.
However, it is sometimes difficult to identify the optical counterparts of ${\it WISE}$ sources because the spatial resolution of ${\it WISE}$ is far lower than that of HSC.
We therefore utilized NIR imaging data to pre-select relatively red objects among the HSC-detected objects, in the same manner as \cite{2015PASJ...67...86T}.
Figure \ref{fig:chart} shows the flow chart of our sample selection process.
With this algorithm, we found 571 IR-bright DOGs over 105 ${\rm deg^2}$.
The details of the catalogs, matching procedure, and the DOG selection procedure are given in subsequent subsections.

\subsubsection{Catalogs}
In this study, we used the HSC catalog data (optical), VIKING catalog data (NIR), and ${\it WISE}$ catalog data (MIR).
Differences between the catalogs adopted in \cite{2015PASJ...67...86T} and this study are summarized in Table \ref{tab:con}.
Table \ref{tab:limmag} summarizes the details of the photometric bands used in this study.

\begin{deluxetable}{rrr}
\tablecaption{Differences between the catalogs in \cite{2015PASJ...67...86T} and this study}
\tablehead{
	\colhead{}				& \colhead{ \cite{2015PASJ...67...86T} }	& \colhead{This study}
}
\startdata
%\hline
	Optical data			& HSC S14A wide				& HSC S16A wide2				\\
	Photometric bands		& $i, y$ 						& $g, r, i, z, y$ 					\\
	Number of objects		& 16,392,815					& 293,520,279					\\ \hline
	NIR data				& VIKING DR1					& VIKING DR2					\\
	Photometric bands		& $Z, Y, J, H, Ks$ 				& $Z, Y, J, H, Ks$				\\
	Number of objects		& 14,773,385					& 46,270,162					\\ \hline
	MIR data				& ALLWISE					& ALLWISE					\\
	Photometric bands		& $W1, W2, W3, W4$			& $W1, W2, W3, W4$			\\
	Number of objects		& 747,634,026					& 747,634,026					\\ \hline
\enddata
\tablecomments{$W1$, $W2$, $W3$, and $W4$ denote $3.4\ {\rm{\mu m}}$, $4.6\ {\rm{\mu m}}$, $12\ {\rm{\mu m}}$, and $22\ {\rm{\mu m}}$ band, respectively.}
 \label{tab:con}
\end{deluxetable}

\begin{deluxetable}{ccrrrrl}
%\tablenum{4}
\tablecaption{Photometric bands in the catalogs used in this study}
%\tablewidth{0pt}
\tablehead{
\colhead{Catalog}	& \colhead{Band}	& \colhead{$\lambda_{\rm {center}}$}	& \colhead{FWHM} 			& \colhead{$m_{{\rm lim}}$ (5$\sigma$)} 	& \colhead{Ref.}  \\
\colhead{}			& \colhead{}		& \colhead{[${\rm {\mu m}}$]}			& \colhead{[${\rm {\mu m}}$]} 	& \colhead{[AB mag]} 				& \colhead{} 
}
%\decimalcolnumbers
\startdata
%\hline\hline
	HSC S16A 	& $g$ 		& 0.47 					& 0.15	 				& 26.5 						& [1], [2]\\ 
      	wide2		& $r$		& 0.62					& 0.16					& 26.1						& \\ 
				& $i$			& 0.77					& 0.15					& 25.9						& \\ 
				& $z$		& 0.89					& 0.08					& 25.1						& \\ 
				& $y$ 		& 1.00					& 0.14					& 24.4						& \\ \hline
      VIKING DR2 	& $Z$ 		& 0.88 					& 0.10 					& 23.1 						& [3], [4]\\ 
      				& $Y$ 		& 1.02 					& 0.09 					& 22.3 						& \\ 
				& $J$ 		& 1.25 					& 0.17 					& 22.1 						& \\ 
				& $H$ 		& 1.65 					& 0.29 					& 21.5 						& \\ 
				& $Ks$ 		& 2.15 					& 0.31 					& 21.2 						& \\ \hline
      ALLWISE 		& $W1$ 		& 3.47 					& 0.64 					& 19.6						& [5]\\ 
      				& $W2$ 		& 4.64 					& 1.11 					& 19.3						& \\ 
				& $W3$ 		& 13.22 					& 6.28 					& 16.4						& \\ 
				& $W4$ 		& 22.22	 				& 4.74 					& 14.5						& \\ \hline
\enddata
\tablecomments{FWHM and $\lambda_{\rm center}$ are ${\rm FWHM}=\lambda_{\rm red}-\lambda_{\rm blue}$, and $\lambda_{\rm center} = \lambda_{\rm blue} + 0.5\times{\rm FWHM}$, respectively, where $\lambda_{\rm blue}$ is the blue side wavelength with the half maximum transmission, and $\lambda_{\rm red}$ is the red side wavelength with the half maximum transmission. \newline
[1] http://hsc.mtk.nao.ac.jp/ssp/survey/ \newline [2] https://www.subarutelescope.org/Observing/Instruments/{\newline}HSC/sensitivity.html \newline [3] https://www.eso.org/sci/publications/messenger/archive/{\newline}no.154-dec13/messenger-no154-32-34.pdf \newline[4] http://casu.ast.cam.ac.uk/surveys-projects/vista/technical/{\newline}filter-set \newline [5] http://wise2.ipac.caltech.edu/docs/release/allsky/}
\label{tab:limmag}
\end{deluxetable}

HSC is a wide-field optical imaging camera installed at the prime focus of the Subaru Telescope, which has a wide field-of-view with a $1.5\ {\rm deg}$ diameter (\citealt{2018PASJ...70S...1M}; \citealt{2018PASJ...70S...2K}; Kawanomoto et al. in prep.; \citealt{2018PASJ...70S...3F}).
The total survey area of S16A data observed by HSC-SSP is wider than that of S14A data (the total survey area of S14A wide is $24\ {\rm deg^2}$). 
The S16A catalog was released internally within the HSC survey team and is based on data obtained from March 2014 to April 2016.
The total survey area of the S16A wide2 is $456\ {\rm deg^2}$, which is observed with five bands (full color).
Among the observed area, 178 ${\rm deg^2}$ have the planned full depth data for all five bands.
S14A data has only two bands ($i$ and $y$), while S16A data has five bands ($g$, $r$, $i$, $z$, and $y$) (\citealt{2018PASJ...70S...8A}).
Therefore, the optical properties of the IR-bright DOGs in \cite{2015PASJ...67...86T} were not investigated.
In this study, we use a forced photometric catalog of the S16A release (\citealt{2018PASJ...70S...8A}) because the same physical region should be investigated for measuring photometric colors (i.e., multi-band magnitudes).
The data observed by HSC were analyzed through an HSC pipeline (hscPipe version 4.0.2; \citealt{2018PASJ...70S...5B}) developed by the HSC software team using codes from the Large Synoptic Survey Telescope (LSST; \citealt{2009arXiv0912.0201L}) software pipeline \footnote{\url{https://www.lsst.org/files/docs/LSSToverview.pdf}} (\citealt{2008arXiv0805.2366I}; \citealt{2010SPIE.7740E..15A}).
The photometric calibration is based on the Panoramic Survey Telescope and Rapid Response System (Pan-STARRS) 1 imaging survey data (\citealt{2013ApJS..205...20M}; \citealt{2012ApJ...756..158S}; \citealt{2012ApJ...750...99T}).
In this study, we use the Galaxy And Mass Assembly (GAMA: \citealt{2009A&G....50e..12D}; \citealt{2011MNRAS.413..971D}) 9hr field (GAMA09H), 15hr field (GAMA15H), WIDE12H, and the X-ray Multi-Mirror Mission Large Scale Structure Survey (XMM-LSS: \citealt{2016A&A...592A...1P}) (Table \ref{tab:field}), because these fields overlap with VIKING fields.
The limiting magnitudes ($5\sigma,\ 2''$ diameter aperture) of the HSC $g$ band, $r$ band, $i$ band, $z$ band, and $y$ band are 26.5, 26.1, 25.9, 25.1, and 24.4 mag, respectively (Table \ref{tab:limmag}).
As shown in Table \ref{tab:con}, the number of HSC sources in the S16A database is larger than that in the S14A database, which is because of the wider area covered by the S16A database (note that the limiting depth of an HSC image is the same in S14A and S16A). 
Hereafter, we use the cmodel magnitude, which is estimated by a weighted combination of exponential and de Vaucouleurs fits to the light profile of each object (\citealt{2001ASPC..238..269L}; \citealt{2004AJ....128..502A}), for investigating photometric properties of the sample after correcting for Galactic extinction (\citealt{1998ApJ...500..525S}).
\begin{deluxetable}{crr}
\tablecaption{Survey area of this study\label{tab:field}}
\tablehead{
	\colhead{Field}				& \colhead{R.A. (J2000) [deg]}	& \colhead{Decl. (J2000) [deg]}
}
\startdata
%\hline
      XMM-LSS & \ 28.0 ... 42.0\ \ \  & $-$8.0 ... \ \ 0.0  \\ 
      GAMA09H & 125.0 ... 145.0 & $-$3.0 ... $+$2.5  \\
      WIDE12H & 170.0 ... 190.0 & $-$3.0 ... $+$2.5  \\ 
      GAMA15H & 205.0 ... 230.0 & $-$3.0 ... $+$2.5   \\ \hline
\enddata
\end{deluxetable}

The VIKING is a wide area NIR imaging survey with five bands ($Z$, $Y$, $J$, $H$, and $Ks$) observed with the VISTA Infrared Camera on the VISTA telescope (\citealt{2006SPIE.6269E..0XD}).
We use the DR2\footnote{https://www.eso.org/sci/observing/phase3/data\_releases/\\viking\_dr2.pdf} catalog in this study.
The limiting magnitudes ($5\sigma,\ 2''$ diameter aperture) of $Z$ band, $Y$ band, $J$ band, $H$ band, and $Ks$ band are 23.1, 22.3, 22.1, 21.5, and 21.2 mag, respectively (Table \ref{tab:limmag}).
As the survey depths of VIKING DR1 and DR2 are the same, the larger number of objects in DR2 data is caused by the increased survey area in DR2.
We use $2''$-aperture magnitudes in our analysis.
The VIKING NIR magnitudes are also corrected for Galactic extinction (\citealt{1998ApJ...500..525S}).

The ${\it WISE}$ is a satellite that has obtained sensitive all-sky images in the MIR bands (3.4, 4.6, 12, and 22 $\rm{\mu m}$).
We used the ALLWISE catalog (\citealt{2010AJ....140.1868W}; \citealt{2014yCat.2328....0C}) in this study.
Although the sensitivity depends on the sky position, the sensitivity at $3.4$, $4.6$, $12$, and $22$ ${\rm{\mu m}}$ is generally better than 0.054, 0.071, 1, and 6 mJy
(i.e., those limiting magnitudes are better than 19.6, 19.3, 16.4, and 14.5 mag), respectively (Table \ref{tab:limmag}).
We use profile-fitting magnitudes for analyzing the ${\it WISE}$ photometric information (\citealt{2010AJ....140.1868W}).

\subsubsection{Clean samples} \label{subsubsec:CSAM}
In selecting the DOGs, we first made clean samples composed of only properly detected objects. 
The HSC S16A wide2 forced catalog contains 293,520,279 objects.
Based on the selection procedure of \cite{2015PASJ...67...86T}, we excluded objects with photometry that is not properly measured (see Figure \ref{fig:chart}).
Specifically, objects affected by bad pixels, cosmic rays, neighboring bright objects, saturated pixels, or the edge of charge-coupled devices (CCDs) were removed from the sample
by utilizing the following flags: flags\_pixel\_bad, flags\_pixel\_cr\_center, flags\_pixel\_bright\_object\_any, flags\_pixel\_ saturated\_center, and flags\_pixel\_edge.
Objects without a clean measurement of the centroid or cmodel flux in any bands were similarly excluded from the sample using the following two flags:
centroid\_sdss\_flags, and cmodel\_flux\_flags.
Furthermore, objects whose flux and shape are measured based on the $i$ band detection (i.e., $i$ band forced photometry) were selected using the flag: merge\_mesurement\_i, while objects which are not deblended and not unique were excluded from the sample using the flag: detect\_is\_primary.
For removing objects with unreliable photometry,
objects with $SN < 5$ in any bands were excluded from the sample. 
Objects with insufficient exposures in any of the HSC 5 bands were excluded from the sample (i.e., ${\rm{gcountinputs<3}}$, ${\rm{rcountinputs<3}}$, ${\rm{icountinputs<5}}$, ${\rm{zcountinputs<5}}$, or ${\rm{ycountinputs<5}}$).
The GAMA09H region includes objects with photometry that was not properly done due to too good seeing.
We excluded these objects from the HSC clean sample by adopting a criterion of ${\rm Decl.}<2.5\ {\rm deg}$.
We also excluded objects in some patches where the color sequence of Galactic stars (with $i_{\rm PSF} < 22$) showed a significant deviation from the expectation of the Gunn-Stryker stellar library (\citealt{1983ApJS...52..121G}); see \cite{2018PASJ...70S...8A} and \cite{2018PASJ...70S..34A} for more details. 
Consequently, 16,680,947 objects were left as the HSC clean sample.

The VIKING DR2 catalog contains 46,270,162 objects.
Following \cite{2015PASJ...67...86T},
we created the VIKING clean sample.
Specifically, objects which are not unique were excluded from the VIKING clean sample using a criterion of ``${\rm primary\_source}\ =\ 1$''.
Objects affected by significant noises were also excluded from the VIKING clean sample using a criterion of ``${\rm kspperrbits}\ =\ 0$''.
Further, objects with $SN<5$ in the $Ks$ band were excluded from the VIKING clean sample.
Consequently, 13,455,180 objects were left as the VIKING clean sample.

The ALLWISE catalog contains 747,634,026 objects.
We created the ALLWISE clean sample: objects with inadequate photometry because of significant noise were excluded from the ALLWISE clean sample by adopting the criteria of ``${\rm w4sat}\ =\ 0$'' and ``${\rm w4cc\_map}\ =\ 0$''.
Objects with $SN<3$ (that corresponds to $\sim$2 mJy) in the $W4$ band were also excluded from the ALLWISE clean sample.
Consequently, 9,439,990 objects were left as the ALLWISE clean sample.

\subsubsection{Crossmatching and selection} \label{subsubsec:CMADS}

We crossmatched the clean samples to select IR-bright DOGs.
The overlap region is $\sim 105\ {\rm{deg^2}}$ in total.
One problem with crossmatching the HSC objects with the ${\it WISE}$ objects is that the angular resolution of the HSC image is significantly different from the angular resolution of the ${\it WISE}$ image (the typical angular resolutions are $\sim 0\farcs6$ in the HSC $i$ band and $\sim 10''$ in the ${\it WISE}$ $W4$ band).
Thus, it is difficult to cross-identify the HSC objects and the ALLWISE objects, when one ALLWISE object has multiple candidates for its HSC counterpart.

In this study, we utilize the fact that DOGs show a very red optical-NIR color.
The VIKING angular resolution ($\sim 1''$) is close to the HSC angular resolution ($\sim 0\farcs6$).
Hence, we first joined the HSC clean sample with the VIKING clean sample, and adopted the same optical-NIR color cut as \cite{2015PASJ...67...86T}.
Next, we cross-identified the resulting ``red'' HSC objects with the ALLWISE objects.
Finally, we selected IR-bright DOGs by adopting the definition of DOGs.

We crossmatched the HSC clean sample (16,680,947) with the VIKING clean sample (13,455,180) using the nearest match with a search radius of 1.0 arcsec (see Figure 4 of \citealt{2015PASJ...67...86T}).
Consequently, we selected 1,534,327 objects.
However, if we performed the nearby match with the same search radius (i.e., selecting all objects within the search radius) between the HSC and the VIKING samples, we select 1,546,060 objects.
The difference in number between objects of the nearest match and the nearby match is 11,733 objects,
which corresponds to the number of possible miss-matched pairs.
However, we adopted the nearest match method because the frequency of such possible miss-matched pairs is negligible ($<1\%$).

We performed an optical-NIR color cut with a criterion adopted by \cite{2015PASJ...67...86T}:
\begin{eqnarray}
	(i - Ks)_{\rm AB} \geq 1.2\ {\rm{.}} \label{iks}
\end{eqnarray}
With this criterion, we selected 707,924 objects (hereafter the HSC-VIKING sample).

We crossmatched the HSC-VIKING sample (707,924) with the ALLWISE clean sample (9,439,990) using the nearest match with a search radius of 3.0 arcsec (see Figure 4 of \citealt{2015PASJ...67...86T}).
Consequently, we selected 1,915 objects (hereafter the HSC-${\it WISE}$ sample).
The number of matched objects with the nearby match is 1,980, which is almost the same as the number of the nearest-matched objects (the difference is $\sim3\%$).
Therefore, we decided to adopt the nearest-match method.

Finally, we adopted the following DOGs selection criterion:
\begin{eqnarray}
	(i\ - [22])_{\rm AB}\ \geq 7.0\ {\rm{.}} \label{dogs}
\end{eqnarray}
Consequently, we selected 571 IR-bright DOGs (see Figure \ref{fig:chart}).

\subsection{Classification of DOGs} \label{subsec: ClassD}

We classify our IR-bright DOG sample into two types (bump DOGs and PL DOGs; see \citealt{2008ApJ...677..943D}), based on the SED using the classification criterion of \cite{2015PASJ...67...86T}.

First, we assumed that each SED is described by a power law from NIR to MIR.
We fitted the MIR SED ($W2$, $W3$, and $W4$) by a power law and calculated the expected $Ks$ band flux described by the extrapolation from the MIR power-law fit($f_{Ks}^{fit}$).
We selected bump DOGs by adopting the following criterion:
\begin{eqnarray}
	\frac{f_{Ks}}{f_{Ks}^{fit}} > 3 {\rm ,} \label{fiks}
\end{eqnarray}
where $f_{Ks}$ is the observed $Ks$ band flux.
Then, we classified the remaining DOGs as PL DOGs.
We classified only 308 DOGs detected in all of the $W2$, $W3$, and $W4$ bands with $SN\geq2$ into bump DOG and PL DOG (see Table \ref{tab:ditect}).
We refer to the remaining 263 DOGs as unclassified DOGs.
Consequently, there are 51 bump DOGs and 257 PL DOGs in our sample (see Table \ref{tab:BP}).
\begin{deluxetable}{lr}
\tablecaption{MIR detection status of the IR-bright DOGs\label{tab:ditect}}
\tablehead{
	\colhead{Status}				& \colhead{Number of objects}	
}
\startdata
%\hline
	Detected in all of $W2$, $W3$, and $W4$ bands 					& 308  	\\ 
	Detected in the $W4$ band and								&		\\
	\ \ \ \ \ \ \ \ \ \ \ \ \ either of $W2$ or $W3$ band						& 216 	\\
	Detected only in the $W4$ band							& 47  \\ \hline 
	Total  												& 571   \\ \hline
\enddata
\end{deluxetable}

\begin{deluxetable}{lr}
\tablecaption{Result of the classification of DOGs \label{tab:BP}}
\tablehead{
	\colhead{Type}		& \colhead{Number of objects}	
}
\startdata
%\hline
      Bump DOGs 		& 51  				\\ 
      PL DOGs 			& 257  				\\ 
      Unclassified	 	& 263 				\\ \hline
      Total 				& 571  				\\ \hline
\enddata
\end{deluxetable}

\begin{figure}
	\begin{center}
	\includegraphics[width=8.5cm]{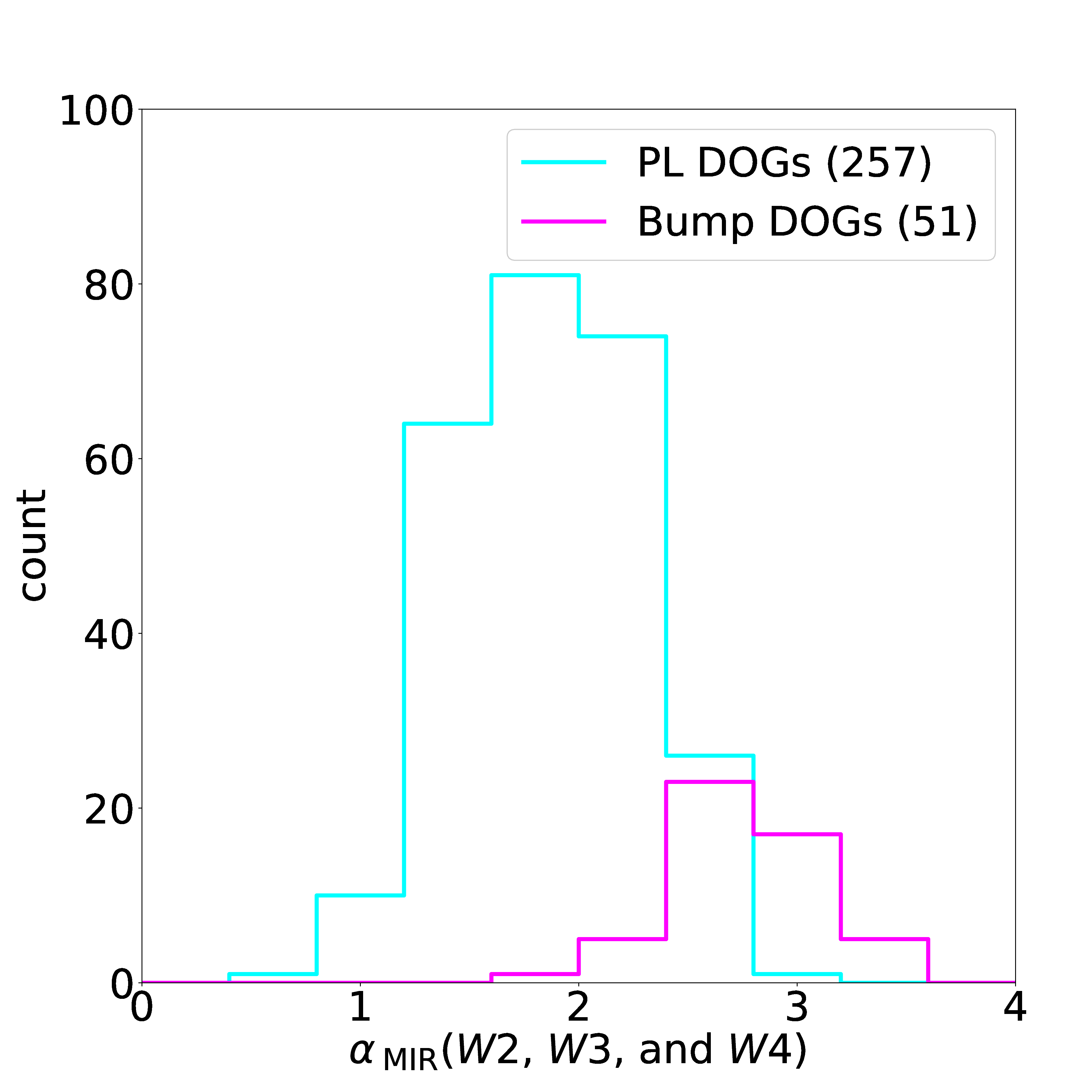}
	\end{center}
	\caption{The $\alpha_{\rm MIR}$ histogram. The magenta and cyan histograms represent the bump and PL DOGs, respectively. The numbers in parentheses indicate the number of objects.}
	\label{fig:AMIR} 
\end{figure}

The MIR spectral index in the power-law fit ($\alpha_{\rm MIR}$) is expressed as follows:
\begin{eqnarray}
	f_{\nu}\propto\lambda^{\alpha_{\rm MIR}} {\rm .}
\end{eqnarray}
Figure \ref{fig:AMIR} shows the frequency distribution of the MIR spectral index for bump and PL DOGs, whose mean and standard deviation are $2.75\pm0.31$ and $1.87\pm0.41$, respectively. 
This suggests that the SED in MIR is systematically steeper in bump DOGs than in PL DOGs.
However, a majority of the unclassified DOGs are expected to be bump DOGs, although a few of them might be low-luminosity PL DOGs (see Section \ref{subsec:BluDOGs}). We note that the contribution of polycyclic aromatic hydrocarbon (PAH) emission and/or silicate absorption to ${\it WISE}$ band fluxes would affect our classification. If we assume that the redshift of our DOGs is about one, the magnitude of $W3$ band is affected by the 6.2 ${\rm \mu m}$, 7.7 ${\rm \mu m}$, and 8.6 ${\rm \mu m}$ PAH emissions, and the magnitude of $W4$ band is affected by the 11.2 ${\rm \mu m}$ PAH emission and 9.7 ${\rm \mu m}$ silicate absorption. Therefore, a few PL DOGs may be SF-dominated DOGs.

%%%%%%%%%%%%%%%%%%%%%%%%%%%%

\section{Results} \label{sec:results}

\subsection{Comparison of the selection results of this study and previous studies}

The total survey area of this study is $\sim105\ {\rm{deg^2}}$, which is about 12 times wider than the survey area of \cite{2015PASJ...67...86T} ($\sim 9\ {\rm{deg^2}}$).
We selected 571 IR-bright DOGs (Figure \ref{fig:chart} and Appendix \ref{sec:app_cd}).
Our DOG sample is about 12 times larger than the \cite{2015PASJ...67...86T} DOG sample (48 objects).
The higher efficiency achieved in selecting DOGs in this study is likely because of improvements in the HSC pipeline, which detects faint objects more reliably.
However, the number of DOGs in this study is much smaller than the number of DOGs recently selected by \cite{2017ApJ...835...36T} using the HSC S15B catalog and the ALLWISE catalog (4,367 objects over $\sim 125\ {\rm deg^2}$).
This is partly because the survey area of our study is limited within the VIKING field, while \cite{2017ApJ...835...36T} utilized the full HSC survey region because they did not use optical-NIR color cut.
Even within the VIKING field, we probably failed to select some NIR-faint DOGs because of the insufficient sensitivity of the VIKING data.

\begin{figure}
	\begin{center}
	\includegraphics[width=9cm]{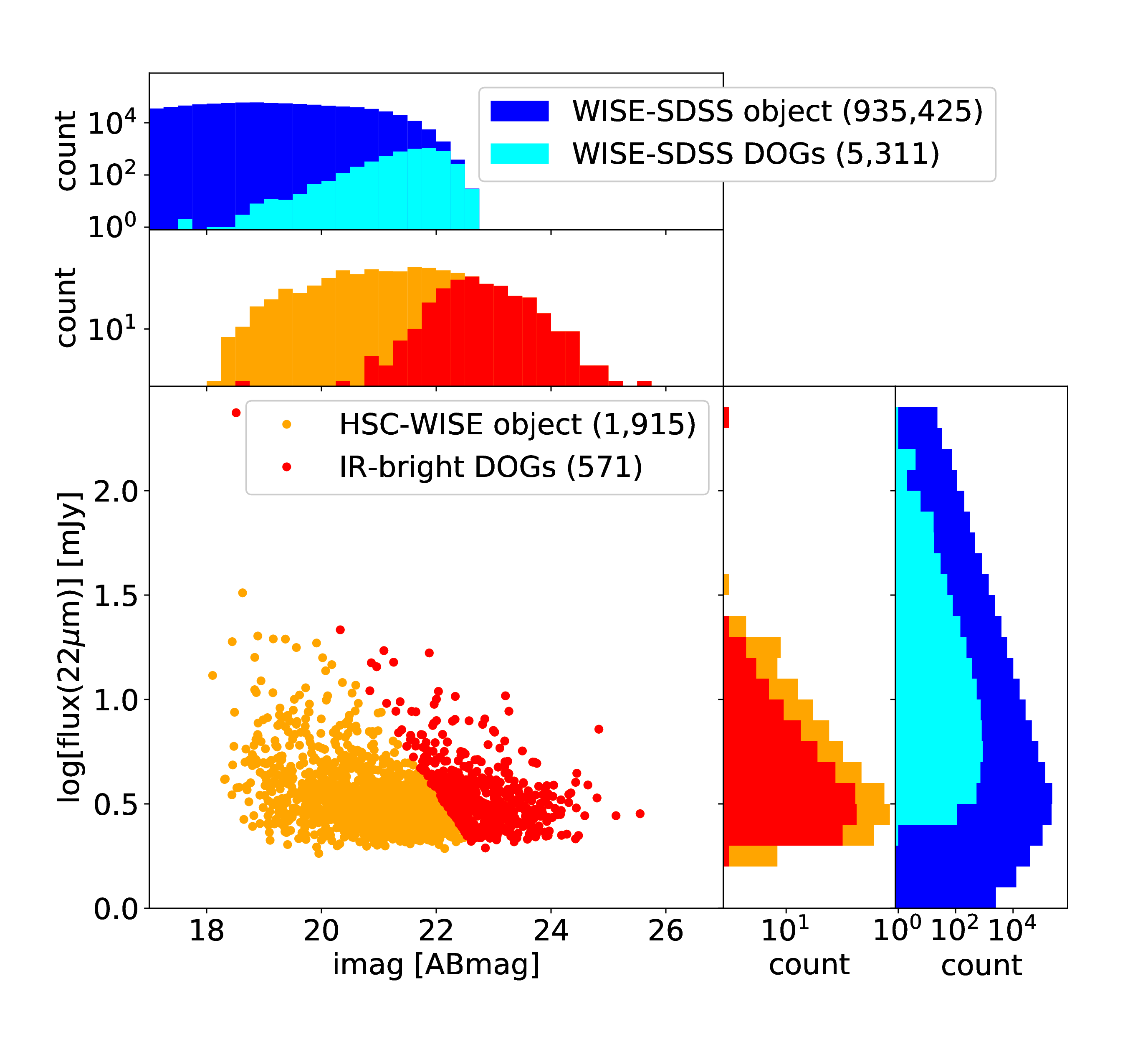}
	\end{center}
	\caption{Distribution of $i$ band magnitudes and 22 $\rm{\mu}$m fluxes in our sample. The histograms of $i$ band magnitude and 22 $\rm{\mu}$m flux are given at the top and right, respectively. The filled circles and histograms in orange and red represent the HSC-${\it WISE}$ objects and IR-bright DOGs, respectively. The blue histograms represent the ${\it WISE}$-SDSS objects, and the cyan histograms represent the ${\it WISE}$-SDSS DOGs (\citealt{2016ApJ...820...46T}). Consider that the $i$ band magnitudes of the ${\it WISE}$-SDSS sample and the ${\it WISE}$-SDSS DOGs in \cite{2016ApJ...820...46T} are not corrected for Galactic extinction. The numbers in parentheses indicate the number of galaxies in each class. }
	\label{fig:imag22} 
\end{figure}

Figure \ref{fig:imag22} shows the $i$ band vs. log[flux(22\ ${\rm \mu m}$)] diagram, where we compare the basic statistics of our DOG sample with those of a previous DOG sample. 
In comparison  with the \cite{2016ApJ...820...46T} DOG sample using the Sloan Digital Sky Survey (SDSS; \citealt{2000AJ....120.1579Y}),
this figure shows that most DOGs in our study are fainter in optical ($i\simeq 21-24$) than DOGs in \cite{2016ApJ...820...46T} ($i\simeq 19-22$),
suggesting that our DOG sample and that of \cite{2016ApJ...820...46T} are complementary.
However, the faintest limit in the $22\ {\rm \mu m}$ flux of our DOG sample is $\sim2$ mJy, 
which is shallower than the faint limit of the DOG sample of \cite{2017ApJ...835...36T} whose limit is $\sim1.5$ mJy (see Figure 2 in \citealt{2017ApJ...835...36T}).
This is because the survey area of \cite{2017ApJ...835...36T} includes northern fields (Decl. $\sim$ +40 deg) in the HSC footprint,
where the depth of the ${\it WISE}$ survey is slightly deeper than in the equatorial fields\footnote{http://wise2.ipac.caltech.edu/docs/release/allsky/}.

\subsection{Optical color distribution}

\subsubsection{Comparison of the optical color between IR-bright DOGs and other galaxy populations}

\begin{figure}
	\begin{center}
	\includegraphics[width=8.5cm]{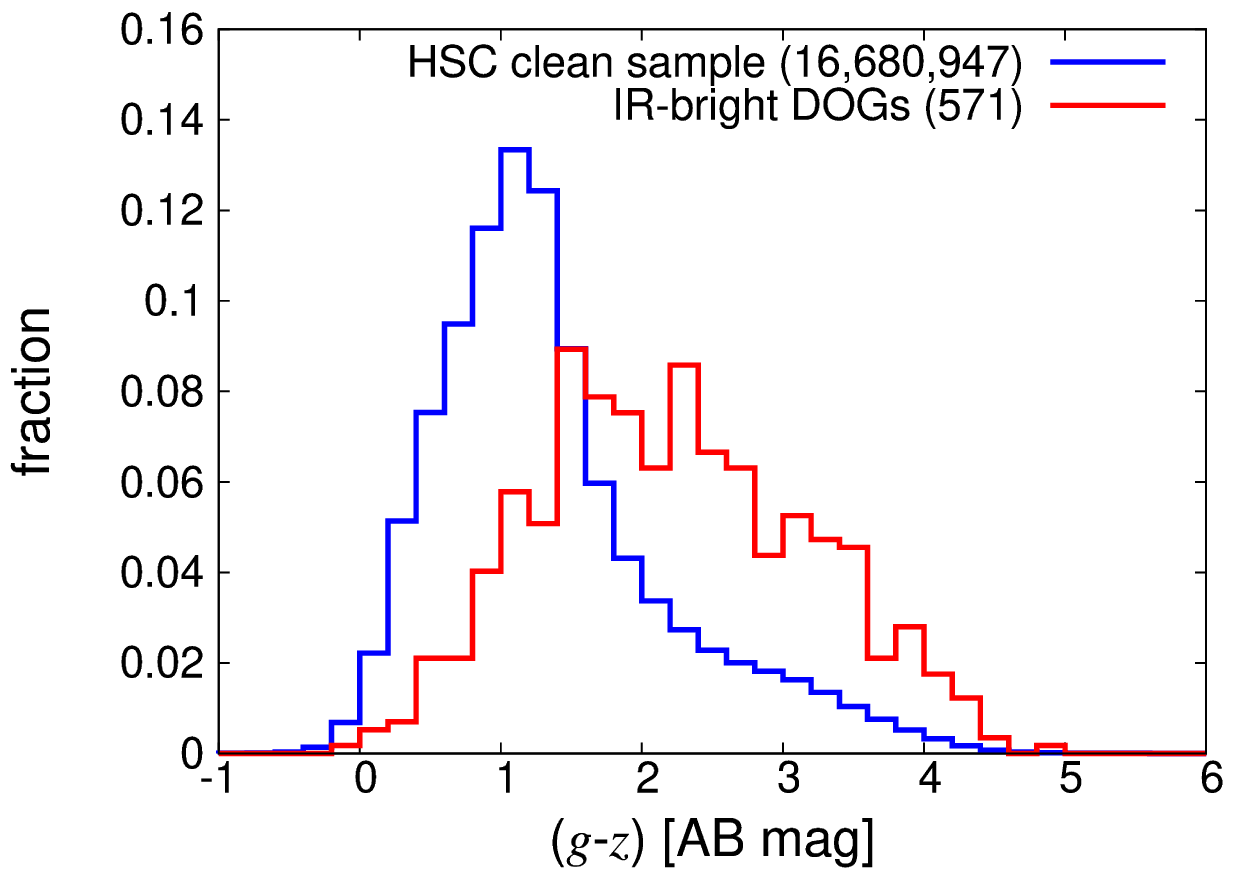}
	\end{center}
	\caption{The $(g-z)_{\rm AB}$ frequency distributions. The blue and the red histograms represent the HSC clean sample and the IR-bright DOGs, respectively. The numbers in parentheses indicate the number of objects.}
	\label{fig:histogramDOGs} 
\end{figure}

We investigated the $(g-z)_{\rm AB}$ color of the DOGs in comparison with that of other galaxy populations.
The $g$ and $z$ bands are used to cover the widest wavelength range possible in optical, while we avoid using shallow $y$ band data.
The average and the standard deviation of $(g-z)_{\rm AB}$ for our DOG sample is $2.21\pm0.95$.
We also investigated the $(g-z)_{\rm AB}$ color of the entire the HSC clean sample in comparison with our DOGs.
The average and the standard deviation of $(g-z)_{\rm AB}$ for the HSC clean sample is $1.36\pm0.83$.
We show the histograms of the $(g-z)_{\rm AB}$ color of our DOG sample and our HSC clean sample in Figure \ref{fig:histogramDOGs}.
This figure shows that the $(g-z)_{\rm AB}$ color of our DOG sample is significantly redder than the $(g-z)_{\rm AB}$ color of the HSC clean sample.
More surprising, the dispersion of the $(g-z)_{\rm AB}$ color of our DOG sample is much larger than that of the HSC clean sample.

\begin{deluxetable}{lrcc}
\tablecaption{The $(g-z)_{\rm AB}$ color of each population}
\tablehead{
	\colhead{Population}		& \colhead{$N_{\rm obj}$}	& \colhead{$(g-z)_{\rm AB}$}	& \colhead{Ref.}
}
\startdata
%\hline
	IR-bright DOGs 		& 571			& 2.21$\pm$0.95 	&			\\ 
	PL DOGs 				& 257			& 1.82$\pm$0.89 	&			\\ 
	Bump DOGs 			& 51				& 2.46$\pm$0.96  	&			\\ 
	HSC clean sample 		& 16,680,947		& 1.36$\pm$0.83  	&			\\ 
	ULIRG $(z < 0.5)$		& 96				& 1.34$\pm$0.48	& (1)			\\ 
	ULIRG $(z < 0.5)$		& 8				& 1.36$\pm$0.65	& (2)			\\  
	ULIRG $(0.5 < z < 1.5)$	& 75				& 0.44$\pm$0.61	& (2)			\\ 
	ULIRG $(1.5 < z < 2.5)$	& 62				& 0.36$\pm$0.26	& (2)			\\ 
	HyLIRG $(z < 0.5)$		& 1				& 1.29			& (1)			\\ 
	HyLIRG $(z < 0.5)$		& 2				& 1.36$\pm$0.24	& (2)			\\ 
	HyLIRG $(0.5 < z < 1.5)$	& 16				& 0.60$\pm$0.51	& (2)			\\ 
	HyLIRG $(1.5 < z < 2.5)$	& 14				& 0.34$\pm$0.20	& (2)			\\ 
	quasar $(z < 0.5)$		& 3,347			& 0.95$\pm$0.48	& (3)			\\ 
	quasar $(0.5 < z < 1.5)$	& 51,479			& 0.46$\pm$0.36	& (3)			\\ 
	quasar $(1.5 < z < 2.5)$	& 80,281			& 0.41$\pm$0.27	& (3)			\\ \hline
\enddata
\tablecomments{\newline (1) \cite{2014ApJ...797...54K} \newline (2) \cite{2013MNRAS.428.1958R} \newline (3) BOSS quasar catalog DR12 (\citealt{2017AandA...597A..79P}) \newline $N_{\rm obj}$ is the number of objects}
\label{tab:gzsam}
\end{deluxetable}

We compare the $(g-z)_{\rm AB}$ color of our DOG sample with that of other populations of galaxies (see Table \ref{tab:gzsam}).
\cite{2008ApJ...677..943D} reported that most DOGs also satisfy the criterion of ULIRGs (i.e., $12\leq \log{[L_{\rm IR}/L_{\odot}]}<13$).
Therefore, it is worth comparing the optical color of DOGs with such IR luminous populations of galaxies.
We investigated the optical color of ULIRGs and also hyper-luminous IR galaxies (HyLIRG; $13\leq \log{[L_{\rm IR}/L_{\odot}]}<14$; e.g., \citealt{2000MNRAS.316..885R}) selected by \cite{2014ApJ...797...54K} using the AKARI data (\citealt{2007PASJ...59S.369M}).
The IR galaxies studied by \cite{2014ApJ...797...54K} are mostly located in low-$z$ ($z<0.5$).
For this comparison, we converted the SDSS photometric magnitudes to the HSC photometric magnitudes by adopting the equations given by \cite{2018PASJ...70S..34A} as follows:
\begin{eqnarray}
	g_{{\rm{HSC}}}&=&g_{{\rm{SDSS}}}-0.074(g_{{\rm{SDSS}}}-r_{{\rm{SDSS}}})-0.011{\rm ,}\\
	z_{{\rm{HSC}}}&=&z_{{\rm{SDSS}}}+0.006(i_{{\rm{SDSS}}}-z_{{\rm{SDSS}}})-0.006{\rm .}
\end{eqnarray}
Consequently, the average and the standard deviation of $(g-z)_{\rm AB}$ in the low-$z$ ULIRGs are $1.34\pm0.48$,
and the $(g-z)_{\rm AB}$ color of 1 HyLIRG in the sample of \cite{2014ApJ...797...54K} is $1.29$ (see Table \ref{tab:gzsam}).
Clearly the optical color of low-$z$ IR galaxies is much bluer than that of our DOG sample.

\cite{2017ApJ...835...36T} reported that the redshift of IR-bright DOGs is typically $z\sim1-2$ (see also \citealt{2016ApJ...820...46T}).
Thus, we investigated the HSC $(g-z)_{\rm AB}$ color of higher-$z$ ULIRGs and higher-$z$ HyLIRGs at $z\sim1-2$ using the \cite{2013MNRAS.428.1958R} sample based on the data of SWIRE (\citealt{2003PASP..115..897L}).
We selected ULIRGs and HyLIRGs with a secure spec-$z$ and a secure IR luminosity derived through template fitting to the observed SED.
We then investigated the HSC $(g-z)_{\rm AB}$ color with the conversion equations by \cite{2018PASJ...70S..34A} for ULIRGs and HyLIRGs at $z < 0.5$, $0.5 < z < 1.5$, and $1.5 < z < 2.5$, respectively.
Consequently, the average of the HSC $(g-z)_{\rm AB}$ color for ULIRGs are $1.36\pm0.65$ $(z<0.5)$, $0.44\pm0.61$ $(0.5 < z < 1.5)$, and $0.36\pm0.26$ $(1.5 < z < 2.5)$,
while the average of the HSC $(g-z)_{\rm AB}$ color for HyLIRGs are $1.36\pm0.24$ $(z<0.5)$, $0.60\pm0.51$ $(0.5 < z < 1.5)$, and $0.34\pm0.20$ $(1.5 < z < 2.5)$.
Even at the redshift of $z\sim1-2$, where most of the IR-bright DOGs are expected to reside,
the optical color of ULIRGs and HyLIRGs show a much bluer color than that of DOGs.
The dispersion in the $(g-z)_{\rm AB}$ color of IR-bright DOGs is much larger than that of ULIRGs or HyLIRGs at similar redshift.

On the other hand, the DOGs are a candidate population of galaxies that are evolving to quasars through gas-rich mergers (\citealt{2009ApJ...700.1190D}; \citealt{2011ApJ...733...21B}).
Thus, it is interesting to also compare the optical color of DOGs with that of quasars.
We studied the $(g-z)_{\rm AB}$ color of quasars taken from the SDSS-III baryon oscillation spectroscopic survey (BOSS, \citealt{2013AJ....145...10D}) quasar catalog DR12 (\citealt{2017AandA...597A..79P}).
We selected objects with a reliable redshift (``${\rm ZWARNING}=0$''), a relatively small error in the derived redshift (``$|{\rm ERR\_ZPIPE}| <0.01$''), no features of broad absorption lines (BALs) (``${\rm BAL\_FLAG\_VI}=0$''), and a high confidence in the measured redshift (``$|{\rm (Z\_VI) - (Z\_PIPE)}|<0.05$'').
We also gave a constraint of $SN \geq 5$ for $g$, $r$, $i$, and $z$ bands of the BOSS quasars to obtain reliable optical colors.
The HSC $(g-z)_{\rm AB}$ color for the obtained quasar was calculated again using the conversion equations of \cite{2018PASJ...70S..34A}.
Consequently, the obtained averages of the HSC $(g-z)_{\rm AB}$ color for the quasar sample are $0.95\pm0.48$ $(z<0.5)$, $0.46\pm0.36$ $(0.5 < z < 1.5)$, and $0.41\pm0.27$ $(1.5 < z < 2.5)$.
The average of the HSC $(g-z)_{\rm AB}$ color for our DOG sample is much redder than the average of the HSC $(g-z)_{\rm AB}$ color of the BOSS quasar sample (Table \ref{tab:gzsam}).

\subsubsection{Fraction of PL DOGs as a function of the optical color} \label{subsubsec: fractionPLopt}

\begin{figure}
 \begin{center}
 \includegraphics[width=8.5cm]{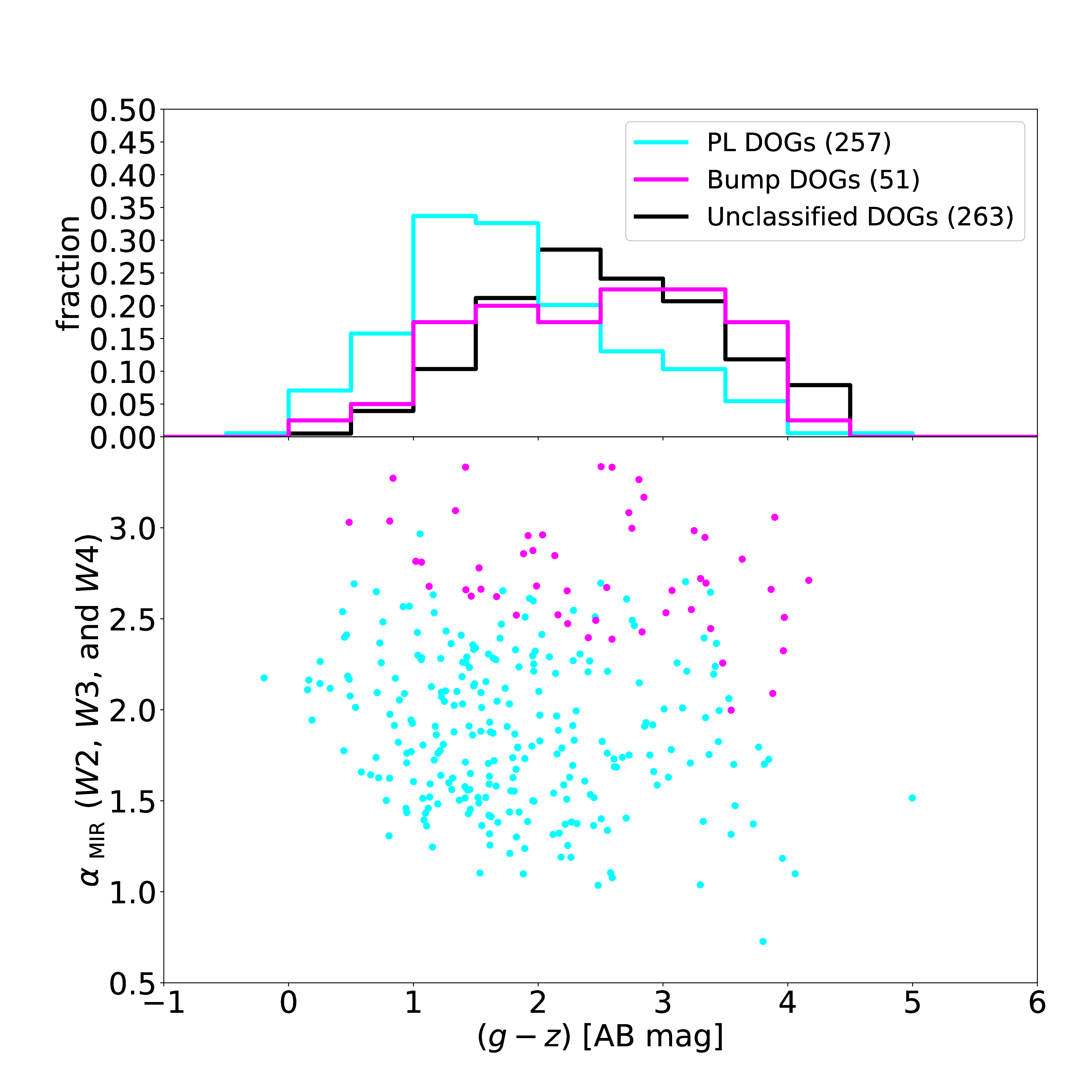}
 \end{center}
 \caption{Distribution of $(g-z)_{\rm AB}$ and $\alpha_{\rm MIR}$ of IR-bright DOGs. The histograms of $(g-z)_{\rm AB}$ frequency are given on the top. The cyan, magenta, and black histograms represent the PL, bump, and unclassified DOGs, respectively. The cyan and magenta filled circles represent the PL and bump DOGs, respectively. The numbers in parentheses indicate the number of objects.}
\label{fig:BUMP_PL_hist}
\end{figure}

\begin{figure}
 \begin{center}
 \includegraphics[width=8.5cm]{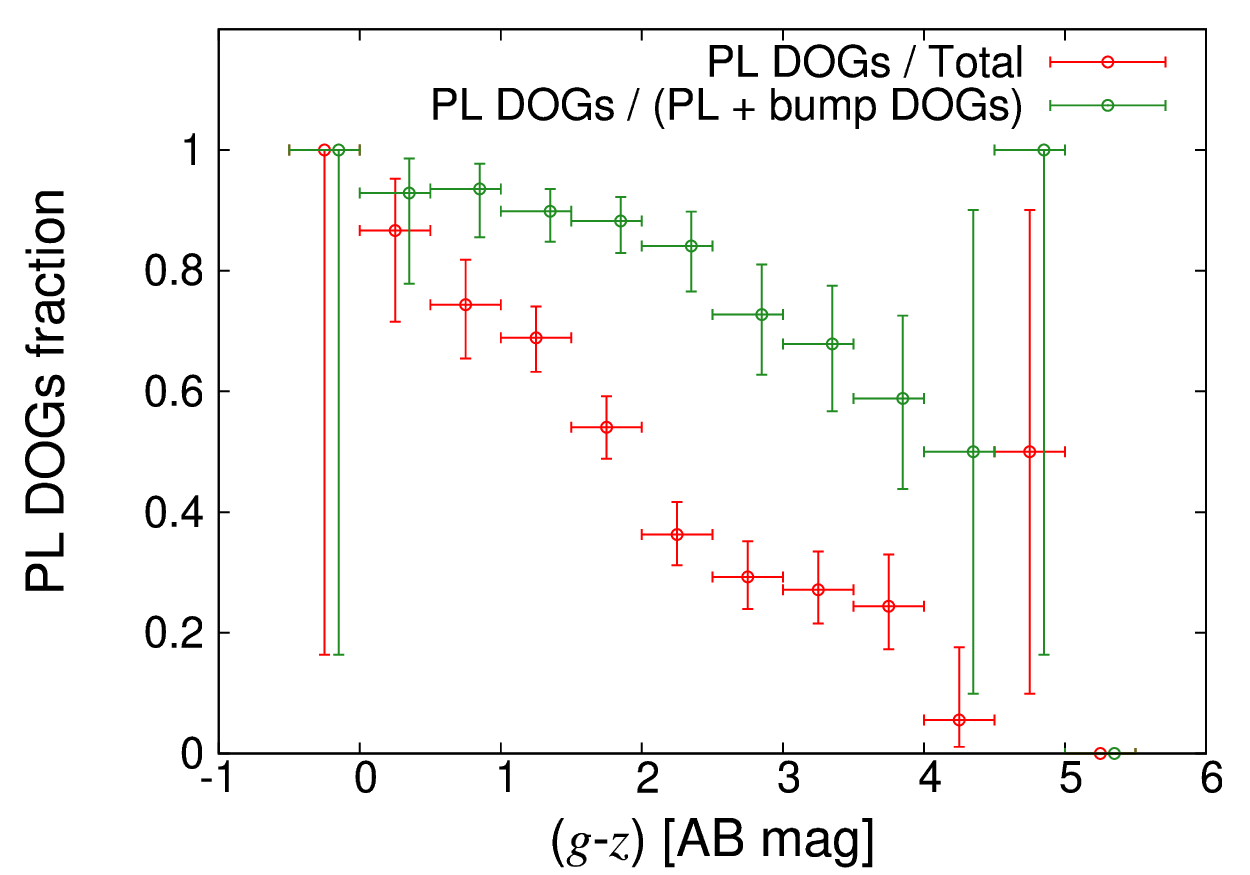}
 \end{center}
 \caption{The PL fraction as a function of the HSC $(g-z)_{\rm AB}$ color. The red and the green circles represent the PL DOGs fraction among total (i.e., including unclassified DOGs) and PL DOGs + bump DOGs (i.e., without unclassified DOGs), respectively. The error bar is given based on the bimodal statistics (see \citealt{1986ApJ...303..336G}). The green circles are shifted to $(g-z)_{\rm AB}+0.1$ for clarity.  }
\label{fig:BUMP_PL}
\end{figure}

To study the origin of the large dispersion in the HSC $(g-z)_{\rm AB}$ color shown in Figure \ref{fig:histogramDOGs}, 
we investigated the HSC $(g-z)_{\rm AB}$ color and $\alpha_{\rm MIR}$ distribution for PL DOGs and bump DOGs separately (Figure \ref{fig:BUMP_PL_hist}).
The top panel shows that the HSC $(g-z)_{\rm AB}$ color distribution of PL DOGs ($1.82\pm0.89$) is bluer than that of bump DOGs ($2.46\pm0.96$).
Figure \ref{fig:BUMP_PL_hist} also shows the HSC $(g-z)_{\rm AB}$ color distribution of unclassified DOGs ($2.54\pm0.87$)
that is very similar to that of bump DOGs.
This is consistent with the idea that a large fraction of unclassified DOGs are intrinsically bump DOGs, as already mentioned in Section \ref{subsec: ClassD}.
The lower panel shows that the two populations of DOGs are located in distinct regions in the $\alpha_{\rm MIR}$ v.s. $(g-z)_{\rm AB}$ plane.
Figure \ref{fig:BUMP_PL} shows the fraction of the PL DOGs among the sum of PL and bump DOGs, and that among the total population of DOGs (i.e., including also unclassified DOGs) as a function of the HSC $(g-z)_{\rm AB}$ color (see also Table \ref{tab:PBnum}).
We found a negative correlation between the HSC $(g-z)_{\rm AB}$ color and the fraction of PL DOGs, regardless of how we defined the fraction of the PL DOG.
This suggests that the PL DOGs and bump DOGs have a systematically different color in optical, which causes the large dispersion of the HSC $(g-z)_{\rm AB}$ color shown in Figure \ref{fig:histogramDOGs}.
This systematic difference in the HSC $(g-z)_{\rm AB}$ color between PL DOGs and bump DOGs is a major origin of the large dispersion of the HSC $(g-z)_{\rm AB}$ color of the IR-bright DOGs seen in Figure \ref{fig:histogramDOGs}.
We discuss this result in more detail in Section \ref{subsec:why_pl_blue}.

\begin{deluxetable}{ccccc}
\tablecaption{Number of PL, bump, and unclassified DOGs for each bin of the HSC $(g-z)_{\rm AB}$ color\label{tab:PBnum}}
\tablehead{
	\colhead{$(g-z)_{\rm AB}$}	& \colhead{PL DOGs}	& \colhead{Bump DOGs}	& \colhead{Unclassified}	& \colhead{Total}	
}
\startdata
%\hline
     -0.5 ... 0.0	& 1	& 0	& 0	& 1		\\ 
      0.0 ... 0.5	& 13	& 1	& 1	& 15		\\ 
      0.5 ... 1.0	& 29	& 2	& 8	& 39		\\ 
      1.0 ... 1.5	& 62	& 7	& 21	& 90		\\ 
      1.5 ... 2.0	& 60	& 8	& 43	& 111	\\ 
      2.0 ... 2.5	& 37	& 7	& 58	& 102	\\ 
      2.5 ... 3.0	& 24	& 9	& 49	& 82		\\ 
      3.0 ... 3.5	& 19	& 9	& 42	& 70		\\ 
      3.5 ... 4.0	& 10	& 7	& 24	& 41		\\ 
      4.0 ... 4.5	& 1	& 1	& 16	& 18		\\ 
      4.5 ... 5.0	& 1	& 0	& 1	& 2		\\ 
      5.0 ... 5.5	& 0	& 0	& 0	& 0		\\ \hline
\enddata
\end{deluxetable}

\subsection{Redshift Distribution of IR-bright DOGs} \label{subsec:rdoirdog}

\begin{figure}
 \begin{center}
 \includegraphics[width=8.5cm]{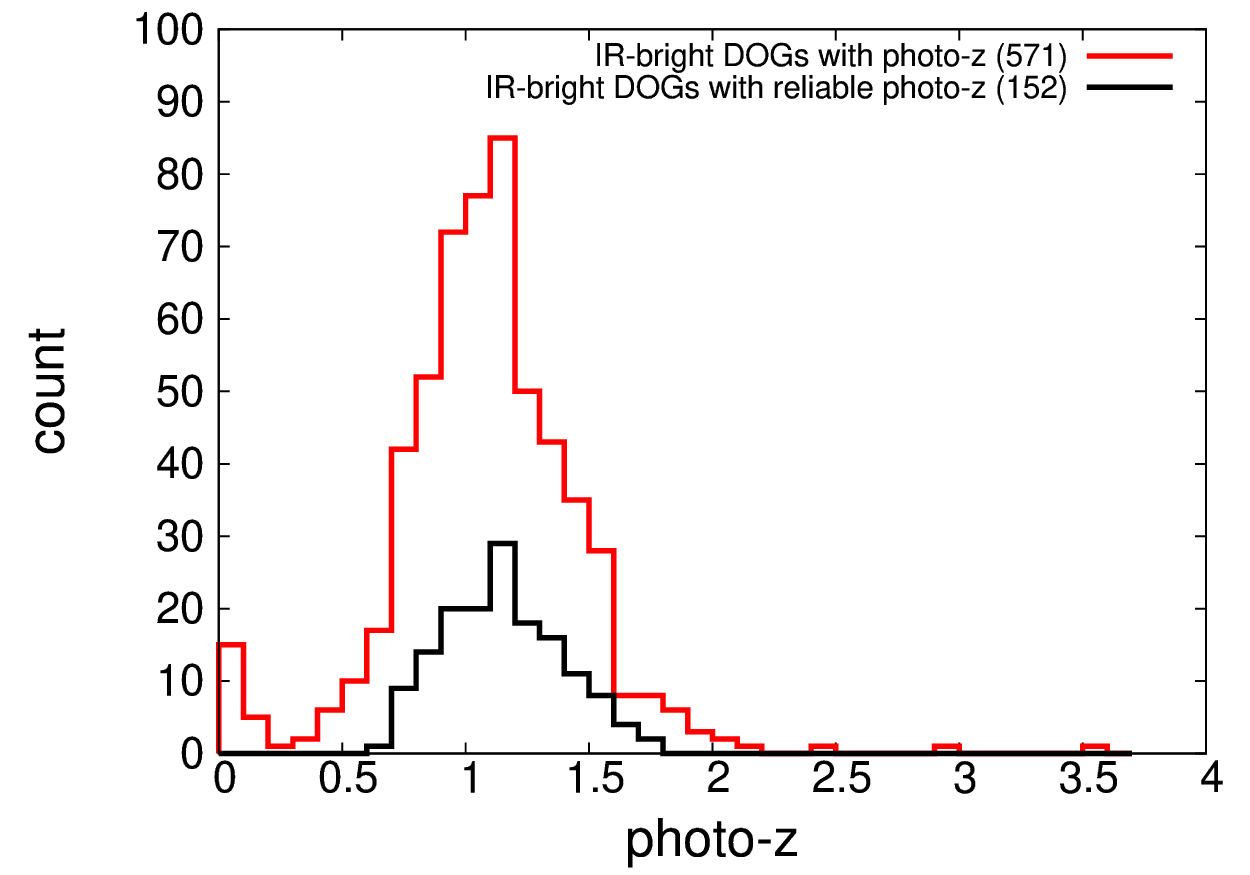}
 \end{center}
 \caption{Photometric redshift distribution of our DOG sample. The red line represents the photometric redshift distribution of 571 DOGs, while the black line represents the distribution of the reliable photometric redshift of 152 DOGs.}
 \label{fig:photoz} 
\end{figure}

\begin{figure}
 \begin{center}
 \includegraphics[width=8.5cm]{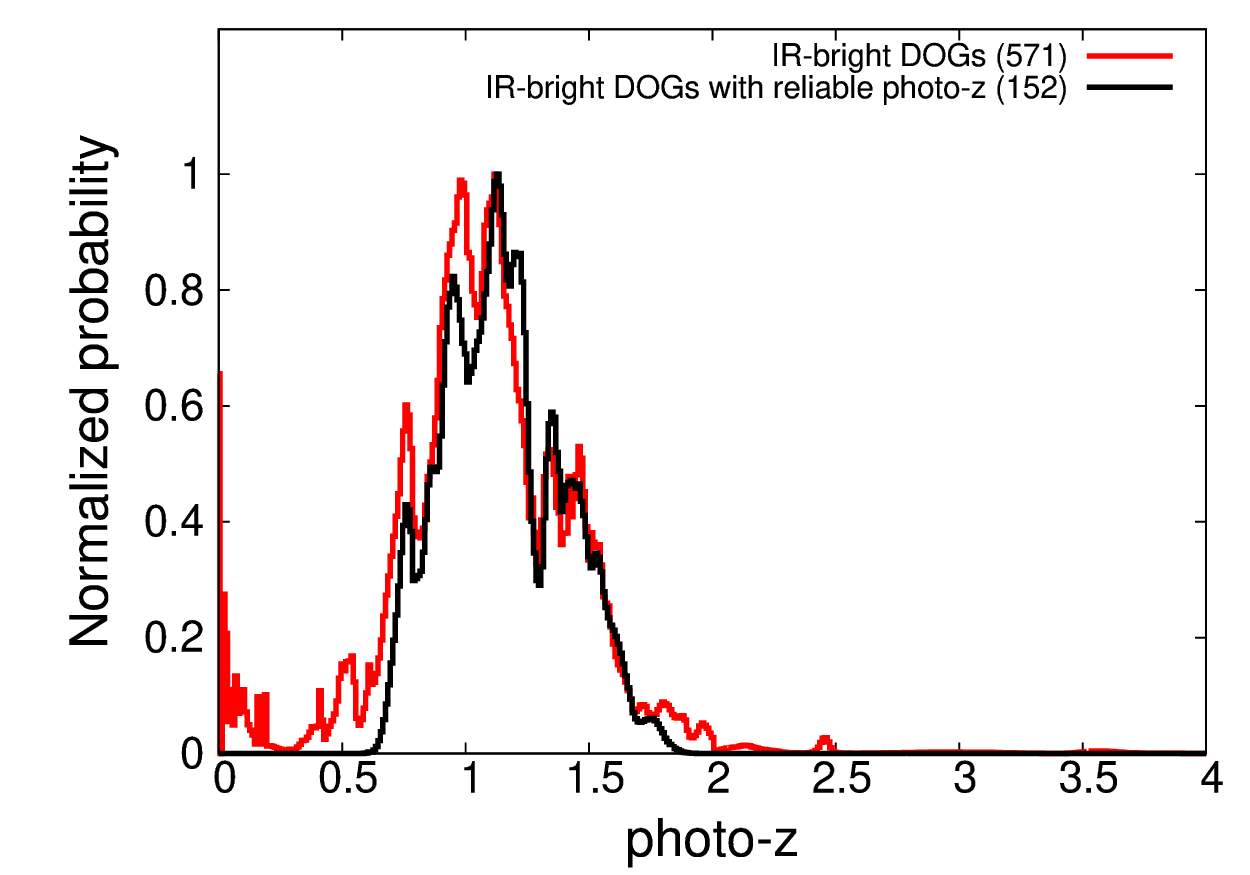}
 \end{center}
\caption{Average probability distribution function (PDF) of the photometric redshift of our DOG sample. The red line represents the average PDF of 571 DOGs, while the black line represents the average PDF of 152 DOGs with the reliable photometric redshift. The histograms are normalized by the peak of each histogram.}
 \label{fig:ALLPDF} 
\end{figure}

\begin{figure}
 \begin{center}
 \includegraphics[width=8.5cm]{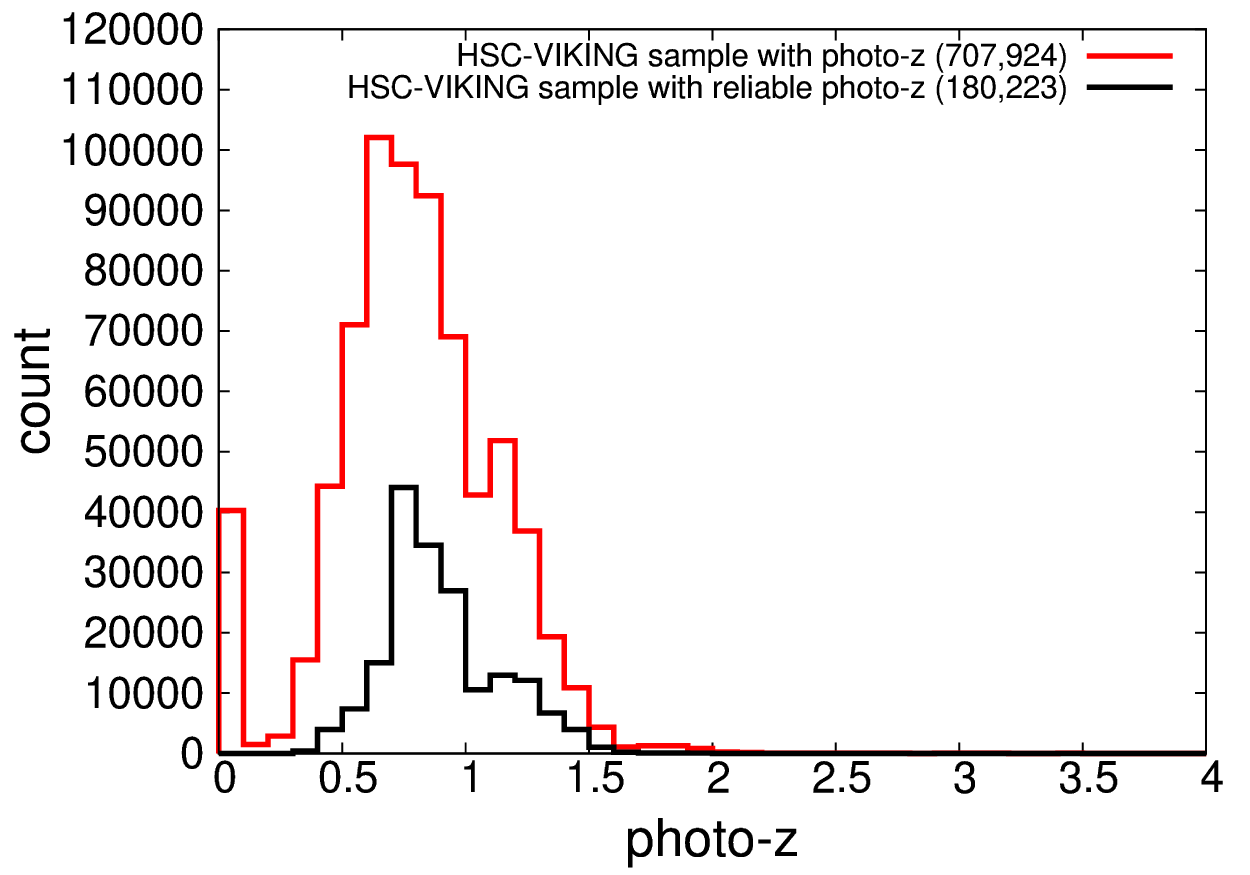}
 \end{center}
\caption{Photometric redshift distribution of our HSC-VIKING sample. The red line represents the photometric redshift distribution of 707,924 HSC-VIKING sample, while the black line represents the distribution of the reliable photometric redshift of 180,223 HSC-VIKING sample.}
 \label{fig:HSC_VIKING_photoz} 
\end{figure}

We obtained the photometric redshift of our DOG sample derived with a photo-$z$ code (MIZUKI: \citealt{2015ApJ...801...20T}; \citealt{2018PASJ...70S...9T})  using five-band HSC photometry.
This photo-$z$ code adopts the SED fitting technique where the spectral templates of galaxies by \cite{2003MNRAS.344.1000B} are used.
\cite{2017ApJ...835...36T} investigated the redshift distribution of IR-bright DOGs using a sample with a ``reliable $z_{\rm ph}$'' 
that is defined by the following criteria:
\begin{eqnarray}
	\chi^2_z < 1.5\\
	{\rm and}\nonumber\\
	\sigma_z/z < 0.05, \label{reliable_criteria}
\end{eqnarray}
where $\chi^2_z$ is the reduced $\chi^2$ for the template fitting and $\sigma_z$ is the estimated uncertainly in the derived $z_{\rm ph}$.
In this study, we adopted the same criteria as \cite{2017ApJ...835...36T} for defining the reliable photometric redshift, and investigated its distribution.
Consequently, we obtained 152 DOGs with a reliable photometric redshift (Figure \ref{fig:photoz}).
The averages and standard deviations of the photometric redshift for all the IR-bright DOGs (i.e., including non-reliable photometric redshift) and those for IR-bright DOGs with a reliable photometric redshift are $z_{\rm ph} = 1.08\pm0.38$ and $1.15\pm0.24$, respectively.
This reliable photometric redshift ($z_{\rm ph}=1.15\pm0.24$) is consistent with the reliable photometric redshift of IR-bright DOGs studied in \cite{2017ApJ...835...36T} ($z_{\rm ph}=1.19\pm0.30$).
We also investigated the distribution of the probability distribution function (hereafter PDF) of the photometric redshift for our DOGs sample (Figure \ref{fig:ALLPDF}).
The weighted mean of PDF ($z_{\rm PDF}$) of the photometric redshift for all the IR-bright DOGs and that for IR-bright DOGs with a reliable photometric redshift are $z_{\rm PDF}=1.08$ and $z_{\rm PDF}=1.15$.
There is a peak around $z_{\rm ph} \sim 0$ in $z_{\rm PDF}$ of all IR-bright DOGs, but this peak is not seen in $z_{\rm PDF}$ of DOGs with a reliable $z_{\rm ph}$ (as seen also in Figure \ref{fig:photoz}).
The $z_{\rm PDF}$ of IR-bright DOGs is therefore consistent with the photo-$z$ distribution shown in Figure \ref{fig:photoz}.
Note that the photo-$z$ distribution of our IR-bright DOG sample is not strongly affected by the photo-$z$ distribution of the parent sample.
Figure \ref{fig:HSC_VIKING_photoz} shows the frequency distributions of photo-$z$ and ``reliable'' photo-$z$ for the HSC-VIKING sample (see Figure \ref{fig:chart}),
whose averages and the standard deviations are $z_{\rm ph}=0.78\pm0.34$ and $0.90\pm0.23$ respectively.
The peak and the entire shape are systematically different between the IR-bright DOGs and HSC-VIKING samples.


\begin{figure}
 \begin{center}
 \includegraphics[width=8.5cm]{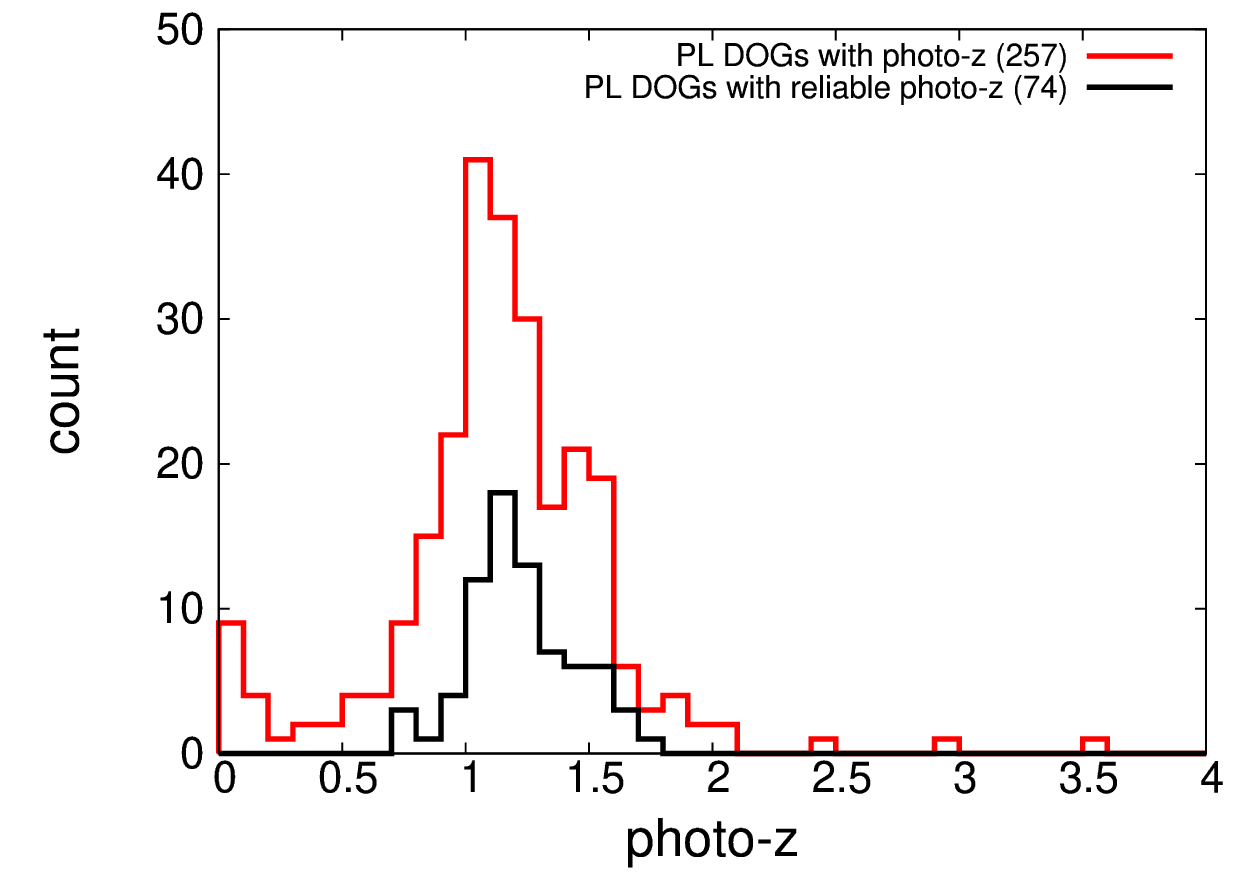}
 \end{center}
 \caption{Photometric redshift distribution of our PL DOG sample. The red line represents the photometric redshift distribution of 257 PL DOGs, while the black line represents the distribution of the reliable photometric redshift of 74 PL DOGs.}
 \label{fig:plphotoz} 
\end{figure}

\begin{figure}
 \begin{center}
 \includegraphics[width=8.5cm]{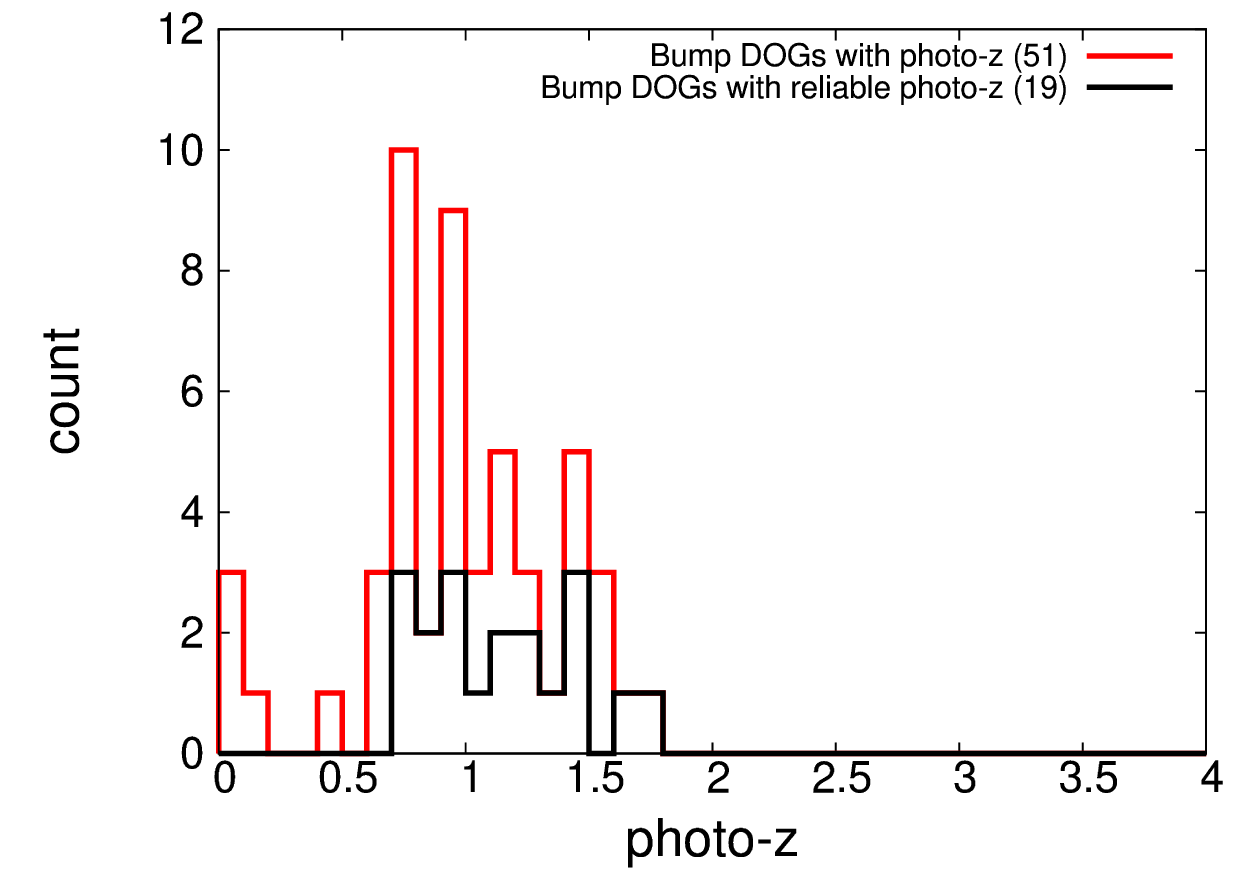}
 \end{center}
  \caption{Photometric redshift distribution of our bump DOG sample. The red line represents the photometric redshift distribution of 51 bump DOGs, while the black line represents the distribution of the reliable photometric redshift of 19 bump DOGs.}
 \label{fig:bumpphotoz} 
\end{figure}

\begin{figure}
 \begin{center}
  \includegraphics[width=8.5cm]{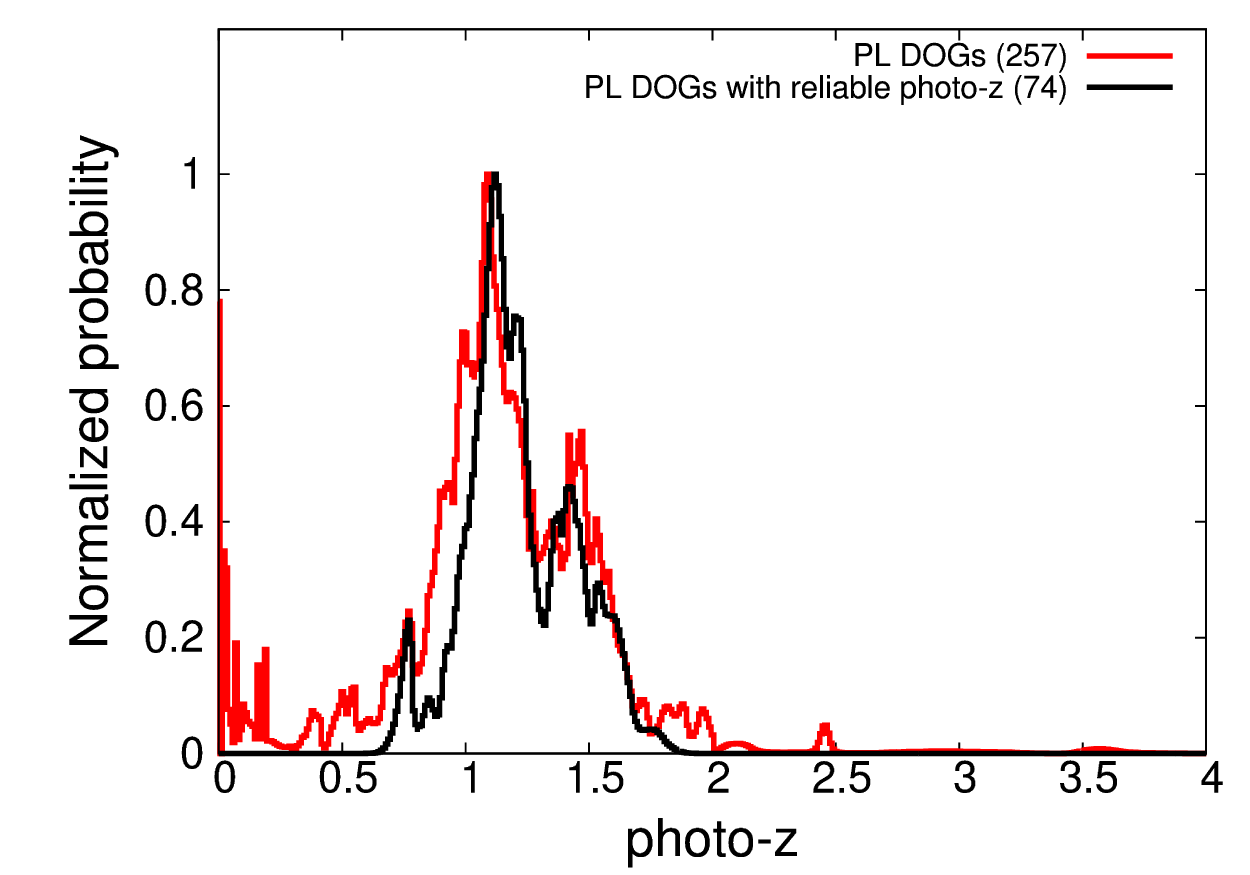}
 \end{center}
 \caption{PDF of the photometric redshift of PL DOG sample. The red line represents the photo-$z$ PDF of 257 PL DOGs, while the black line represents the photo-$z$ PDF of 74 PL DOGs with the reliable photometric redshift. The histograms are normalized by the peak of each histogram.}
 \label{fig:PLPDF} 
\end{figure}

\begin{figure}
 \begin{center}
 \includegraphics[width=8.5cm]{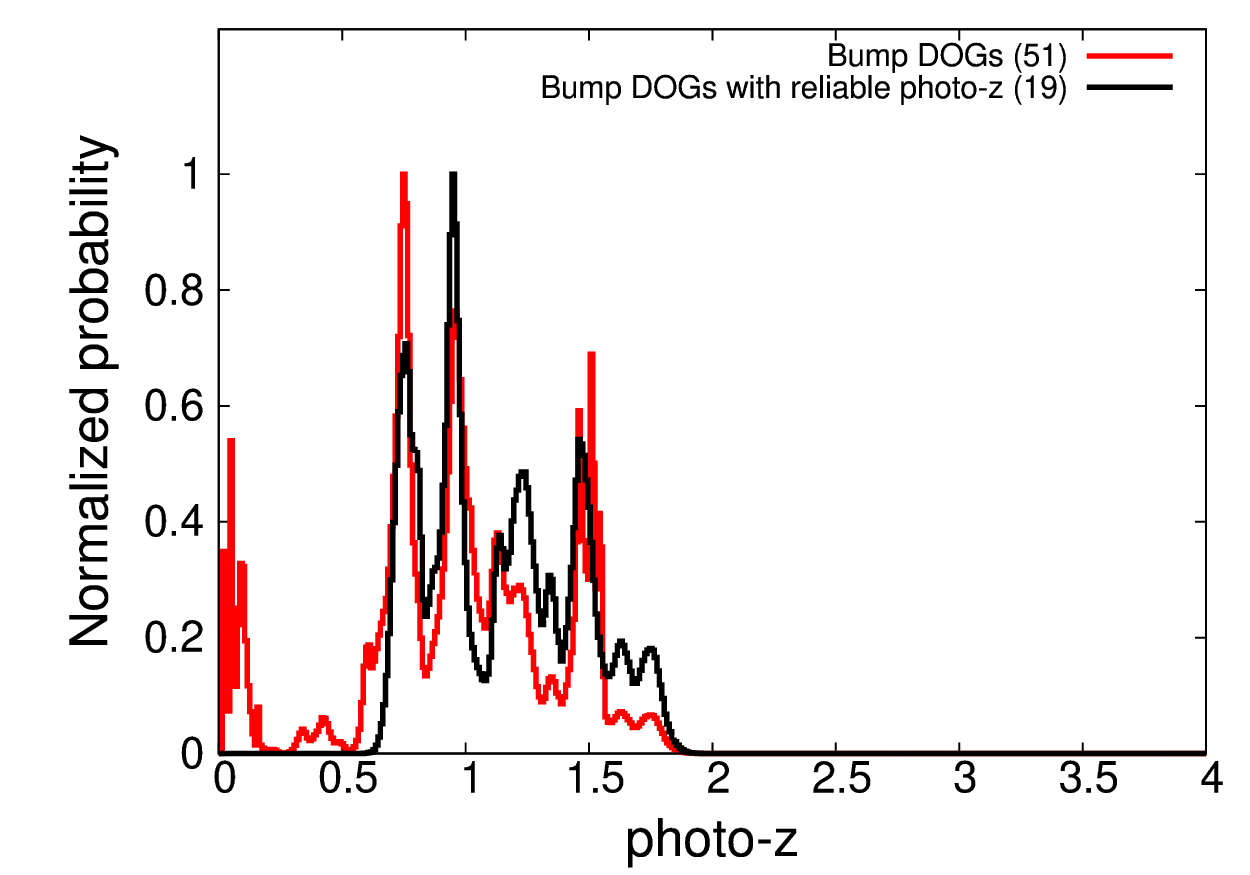}
 \end{center}
 \caption{PDF of the photometric redshift of bump DOG sample. The red line represents the photo-$z$ PDF of 51 bump DOGs, while the black line represents the photo-$z$ PDF of 19 bump DOGs with the reliable photometric redshift. The histograms are normalized by the peak of each histogram.}
 \label{fig:BumpPDF} 
\end{figure}

As we are now studying the optical color in the observed frame,
it may introduce systematic uncertainties if the redshift distribution of PL DOGs and that of bump DOGs is systematically different.
Therefore, we investigate the distribution of the photometric redshift of PL DOGs and that of bump DOGs separately (Figures \ref{fig:plphotoz} and \ref{fig:bumpphotoz}).
The average and the standard deviation of the photometric redshift for all the PL DOGs (i.e., including non-reliable photometric redshift) and those for PL DOGs with a reliable photometric redshift are $z_{\rm ph} = 1.14\pm0.44$ and $1.22\pm0.22$, respectively.
The average and the standard deviation of the photometric redshift for all the bump DOGs and those for bump DOGs with a reliable photometric redshift are $z_{\rm ph} = 0.97\pm0.40$ and $1.14\pm0.31$, respectively.
We thus conclude that there is no significant different in the photometric redshift between PL DOGs and bump DOGs in our IR-bright DOG sample.
We also investigate the PDF of PL and bump DOGs (see Figures \ref{fig:PLPDF} and \ref{fig:BumpPDF}).
The weighted means of the photo-$z$ PDF for all the PL DOGs, PL DOGs with reliable photometric redshift, all the bump DOGs, and bump DOGs with reliable photometric redshift are $z_{\rm PDF}=1.13$, $z_{\rm PDF}=1.22$, $z_{\rm PDF}=0.97$, and $z_{\rm PDF}=1.14$, respectively.
The $z_{\rm PDF}$ of PL and bump DOGs is therefore consistent with the $z_{\rm ph}$ distribution.

\subsection{Optical color-color diagram}

\begin{figure}
 \includegraphics[width=9cm]{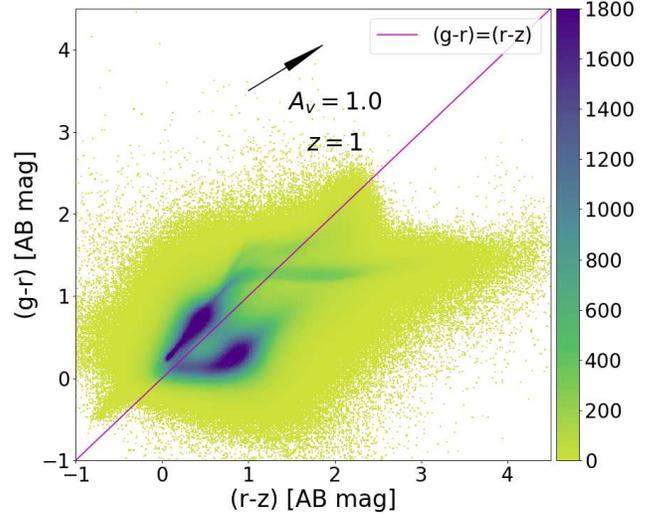}
  \caption{The $(g-r)_{\rm AB}$ vs. $(r-z)_{\rm AB}$ color-color diagram with the HSC clean sample. The color of the dots denote the number density of the HSC clean sample within $0.01\times0.01\ {\rm mag^2}$. A black arrow denotes the reddening vector at $z=1$, adopting the extinction curve of  \cite{2000ApJ...533..682C}.}
 \label{fig:HSCgrrz} 
\end{figure}

\begin{figure}
 \includegraphics[width=8.7cm]{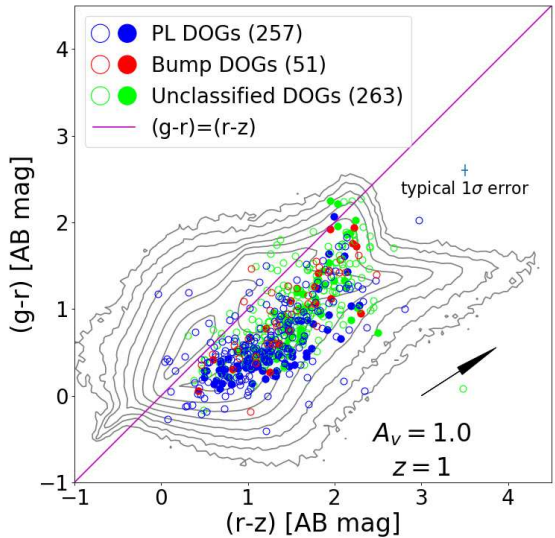}
 \caption{The $(g-r)_{\rm AB}$ vs. $(r-z)_{\rm AB}$ color-color diagram with our DOG sample (shown by colored circles). The blue, red, and green open circles represent our PL DOGs, bump DOGs, and unclassified DOGs, respectively. The filled circles of those colors represent those DOGs with reliable photometric redshift. The gray contour represents the number density of the HSC clean sample within $0.05\times0.05\ {\rm {mag^2}}$ in the logarithmic scale. The number in parentheses indicates the number of objects in each class of DOGs. The cyan cross denotes the typical 1 $\sigma$ error of the HSC color for DOGs. The black arrow is the same as in Figure \ref{fig:HSCgrrz}.}
\label{fig:2dogs}
\end{figure}

In the previous subsection, it is shown that the redshift distribution of our DOG sample is mostly distributed around $z\sim1$.
At $z\sim1$, the $(g-r)_{\rm AB}$ color corresponds to the rest-UV slope at a shorter wavelength than the 4000 {\AA} break 
while the $(r-z)_{\rm AB}$ color corresponds to the amplitude of the 4000 {\AA} break.
Therefore, we here investigate the color-color diagram with the HSC $(g-r)_{\rm AB}$ vs. $(r-z)_{\rm AB}$ colors.
Note that we do not use $y$ band data because the limiting depth of the HSC $y$ band data is relatively shallow with respect to the other HSC bands.
Figure \ref{fig:HSCgrrz} shows the $(g-r)_{\rm AB}$ vs. $(r-z)_{\rm AB}$ color-color diagram with the HSC clean sample.
The color distribution of the HSC clean sample shows 2 sequences, which correspond to the stellar sequence (upper one) and the galaxy sequence (lower one) respectively.
As a rough separator for these two sequences, we also show a line of $(g-r)_{\rm AB}=(r-z)_{\rm AB}$.
We plot our DOG sample on the same HSC $(g-r)_{\rm AB}$ vs. $(r-z)_{\rm AB}$ color-color diagram (Figure \ref{fig:2dogs}).
Most DOGs in our sample is located in the lower-right side from the separator, i.e., $(g-r)_{\rm AB}<(r-z)_{\rm AB}$.
In the upper-left side from the separator, $\sim6$\% of DOGs in our sample are populated and they may be Galactic stars. In addition to this, our DOGs sample located in the lower-right side from the separator may be also contaminated by late-type Galactic stars because the stellar sequence seen at $(g-r)_{\rm AB}\sim1.2$ extends to the lower-right side from the separator (see Figure \ref{fig:HSCgrrz}).
We will discuss the optical color properties of DOGs in our sample more in Section \ref{subsec:waht_pop_dogs}.


%%%%%%%%%%%%%%%%%%%%%%%%%%%%
\section{Discussion} \label{sec:disscussion}

\subsection{Selection effects} \label{subsec:sel_eff}

First, we discuss the influence of the adopted order for $(i-Ks)_{\rm AB}$ and $(i-[22])_{\rm AB}$ criteria in the DOGs selection. 
To assess the effect of the order of each selection criterion, we here try to start the DOGs selection by applying the $(i-[22])_{\rm AB}$ instead of $(i-Ks)_{\rm AB}$. We cross-match the HSC clean sample (16,680,947) with the ALLWISE clean sample (9,439,990) by the nearest match with a search radius of 3.0 arcsec, and select 9,539 objects. By performing the $(i-[22])_{\rm AB}$ color cut, we obtain 1,604 objects. Next, we cross-match this sample (1,604) with the VIKING clean sample (13,455,180) by nearest match with a search radius of 1.0 arcsec, and select 528 objects. By performing the $(i-Ks)_{\rm AB}$ color cut, we obtain 521 objects. The number of the selected DOGs (521) is less than the number of the originally selected DOGs (571; see Figure \ref{fig:chart}). However, if we adopt this selection order, we may fail to select DOGs when we match the HSC clean sample and ALLWISE clean sample, because non-DOGs objects may be selected by the ``nearest'' criterion. Actually, if we perform the ``nearby match'' (i.e., selecting all objects within the search radius) instead of the ``nearest match'', 12,600 objects are selected. When we remove the cases where more than 1 HSC source is found within 3.0 arcsec from the ${\it WISE}$ sources, we obtain 6,923 objects. The difference among the nearest matching sample (9,539) and the above sample (6,923) between the HSC and the ALLWISE clean sample is 2,616 objects. Thus, it is possible that the nearest matching sample (9,539) contains miss-matched sources up to 2,616 objects, and the replacement of the order of the $(i-Ks)_{\rm AB}$ with $(i-[22])_{\rm AB}$ significantly increases such miss match. Therefore, in this study, we first apply the criteria of $(i-Ks)_{\rm AB}$, and then apply the criteria of $(i-[22])_{\rm AB}$.

In Section \ref{sec:results} we reported the rest-frame optical properties of IR-bright DOGs and also their dependences on the spectral type (bump DOGs or PL DOGs).
However, the selection procedure described in Section \ref{sec:data} may introduce some selection effects that could affect the statistical properties of rest-frame optical properties of IR-bright DOGs.
Therefore, hereafter in this subsection, we discuss possible selection effects especially by examining those related to the detection at the HSC optical bands and to the optical-NIR color cut.

\begin{figure}
 \begin{center}
 \includegraphics[width=8.5cm]{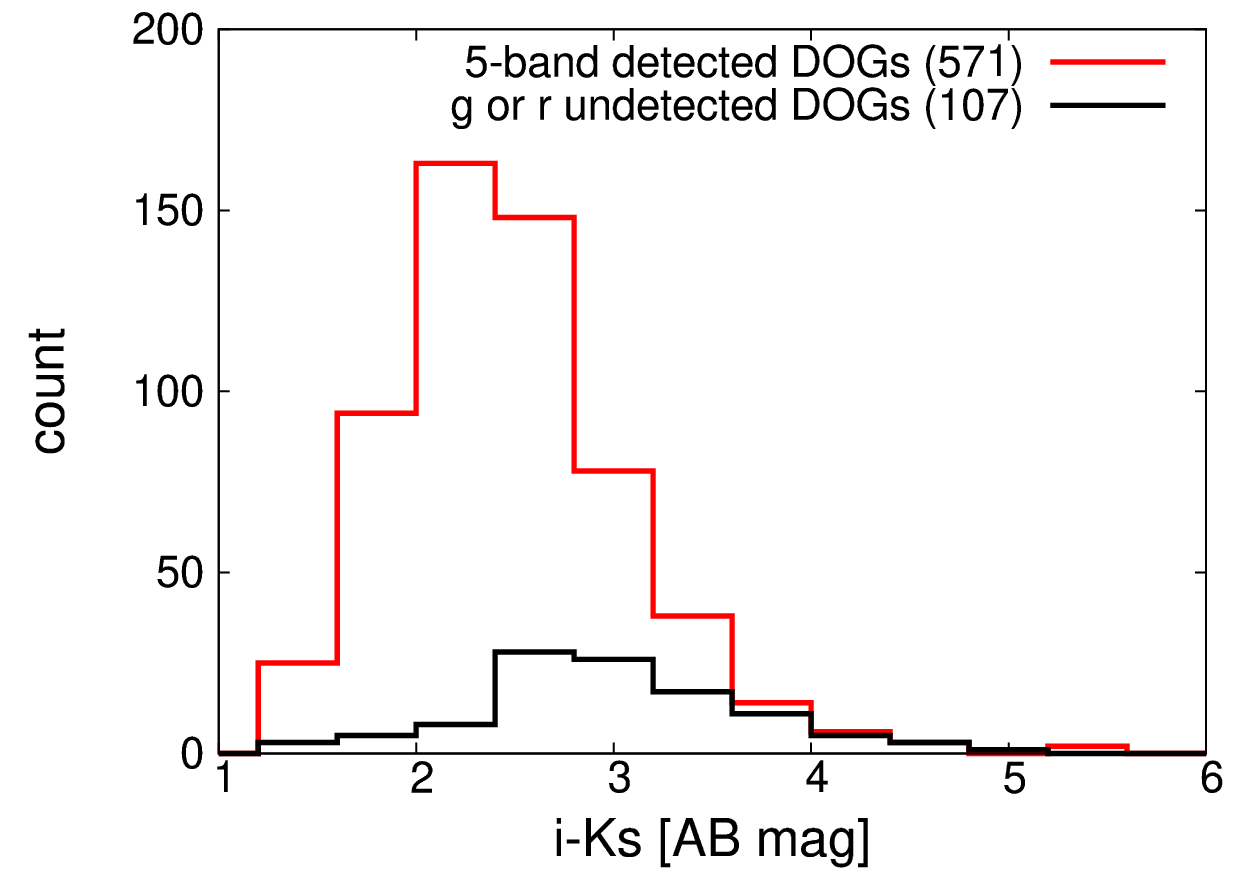}
 \end{center}
\caption{The $(i-Ks)_{\rm AB}$ histograms. The red and the black lines represent the five-band detected DOG sample and the $g$-or-$r$-undetected DOG sample, respectively. Numbers in parentheses indicate the number of objects.}
\label{fig:5izyhist}
\end{figure}

One possible origin of the selection effect is the criterion for the HSC detection; in our selection procedure, a significant detection at all of the five HSC bands is required.
This may result in losing DOGs with a very red color in optical, as such DOGs do not show detectable fluxes in blue optical bands.
To check this effect, we define ``$izy$-detected DOGs'' for which we do not require the detection in the HSC $g$ and $r$ bands 
(but the remaining criteria are the same as the main IR-bright DOGs).
There are 673 $izy$-detected DOGs, among which 107 DOGs are undetected in $g$ band and/or $r$ band (hereafter ``$g$-or-$r$-undetected DOGs'').
Such $g$-or-$r$-undetected DOGs are not included in our original sample (``five-band detected DOGs'').
Table \ref{tab:gruDOGsbpu} shows the classification result of the $g$-or-$r$-undetected DOGs.
Figure \ref{fig:5izyhist} shows the $(i-Ks)_{\rm AB}$ histograms for five-band detected (i.e., original) DOG sample and for $g$-or-$r$-undetected DOG sample, respectively.
The color of $(i-Ks)_{\rm AB}$ for five-band detected DOG sample and for $g$-or-$r$-undetected DOG sample are $2.48\pm0.60$ and $3.01\pm0.70$, respectively.
Because $g$-or-$r$-undetected DOGs show a redder $(i-Ks)_{\rm AB}$ color than five-band detected DOGs, our original selection criterion requiring five-band detection in the HSC image results in losing DOGs with a relatively red color in the optical-NIR range.
Note that most of $g$-or-$r$-undetected DOGs categorized as unclassified DOGs are expected to be bump DOGs (see Section \ref{subsec: ClassD}).
Therefore, even by considering the effect of detection in the HSC optical bands, we still selectively see DOGs with a blue optical color in the PL DOG sample rather than in the bump DOG sample.
 
\begin{deluxetable}{lr}
\tablecaption{Result of the classification of $g$-or-$r$-undetected DOGs \label{tab:gruDOGsbpu}}
\tablehead{
	\colhead{Type}		& \colhead{Number of objects}	
}
\startdata
%\hline
      Bump DOGs 		& 6  					\\ 
      PL DOGs 			& 26  				\\ 
      Unclassified	 	& 75 					\\ \hline
      Total 				& 107  				\\ \hline
\enddata
%\tablecomments{}
\end{deluxetable}

\begin{figure}
 \begin{center}
 \includegraphics[width=8.5cm]{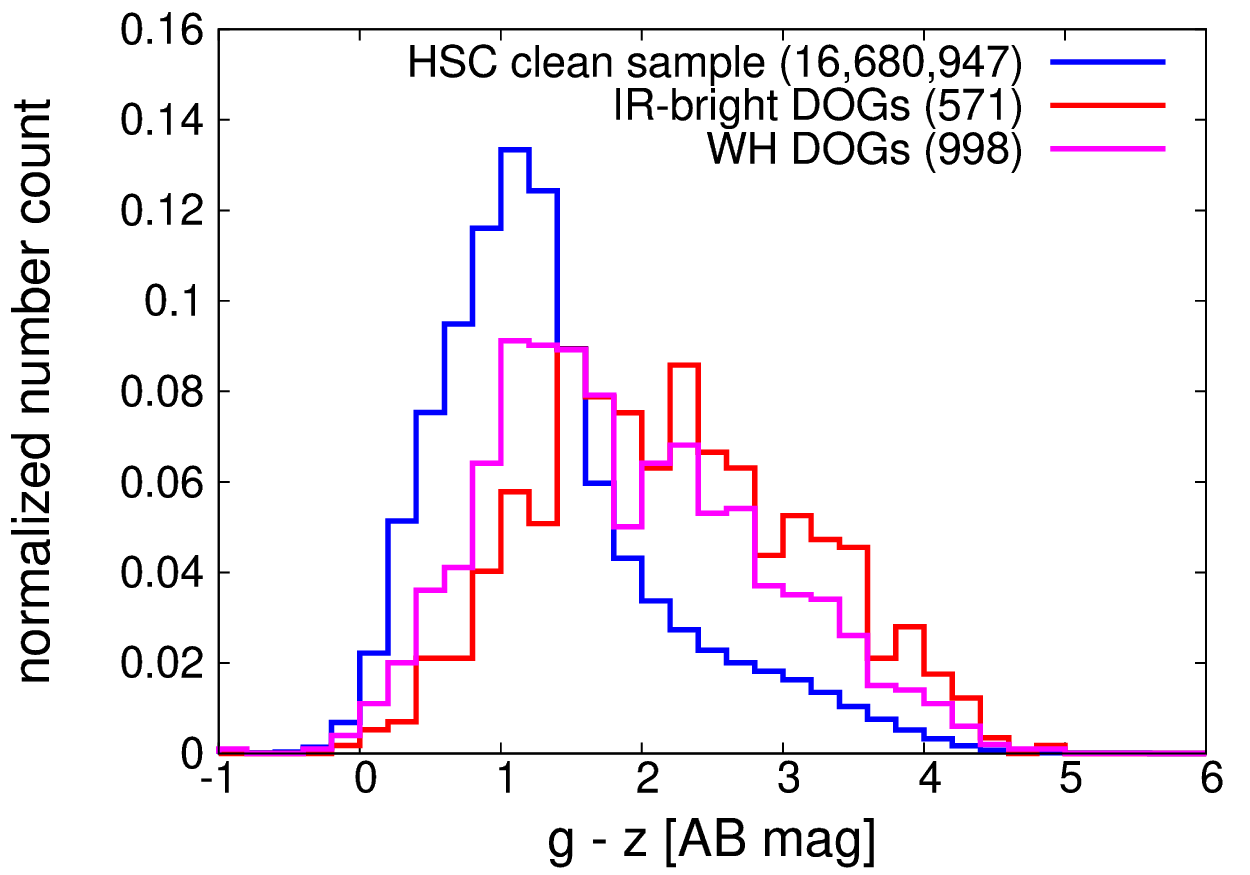}
 \end{center}
 \caption{The $(g-z)_{\rm AB}$ frequency distribution. The blue, red, and magenta histograms represent the HSC clean sample, IR-bright DOGs, and WH DOGs, respectively. The numbers in parentheses indicate the number of objects.}
  \label{fig:WH_DOGs} 
\end{figure}

Another possible origin of the observational bias is the criterion of $(i-Ks)_{\rm AB}\geq1.2$.
This criterion is introduced to reduce the mis-match between the ${\it WISE}$ source and the HSC source (see Section \ref{subsubsec:CMADS}).
Here the adopted criterion was originally introduced by \cite{2015PASJ...67...86T}, by showing that all of DOGs in \cite{2012ApJ...744..150B} satisfy this criterion.
However, the DOG sample of \cite{2012ApJ...744..150B}  was selected from the NOAO Deep Wide-Field Survey (NDWFS) Bo\"otes field (see \citealt{2008ApJ...677..943D}) and the selected DOGs are systematically fainter ($F_{24}\sim0.3-1$ mJy) than our IR-bright DOGs ($F_{22}\sim4-10$ mJy; see Figure \ref{fig:imag22}).
Therefore, it is not clear whether the optical-IR color properties of IR-bright DOGs are the same as IR-fainter DOGs of \cite{2012ApJ...744..150B}; in other words, we may lose some IR-bright DOGs with a relatively blue optical-IR color by introducing the $i-Ks$ criterion.
To examine this possible selection effect, we construct a sample of DOGs without adopting the $i-Ks$ criterion.
Here we cross-match the HSC clean sample with the ALLWISE clean sample.
Then we remove the cases where more than 1 HSC sources are found within 3 arcsec from the ${\it WISE}$ source when we search for the optical counterpart of ${\it WISE}$ sources to avoid mis-matches.
Through the HSC-${\it WISE}$ DOG selection criteria, we obtain 998 DOGs ${\it WISE}$-HSC DOGs (hereafter WH DOG sample).
Figure \ref{fig:WH_DOGs} shows $(g-z)_{\rm AB}$ color histograms for the WH DOGs, the original IR-bright DOGs, and the HSC clean sample.
The $(g-z)_{\rm AB}$ color of the WH DOGs is $1.88\pm0.96$,
which is systematically bluer than the original IR-bright DOGs ($2.21\pm0.95$; see Table \ref{tab:gzsam}).
On the other hand, we focus on the case where more than one HSC source is found within 3 arcsec from the ${\it WISE}$ source.
Such a case may correspond to DOGs in a merger phase. 
The number of ${\it WISE}$-HSC DOGs in this case is 606, and their $(g-z)_{\rm AB}$ color is $1.88\pm0.92$, which is almost the same as the color of the WH DOGs with only one optical counterpart within 3 arcsec.
These considerations suggest that we are losing DOGs with a relatively blue optical-IR color in our IR-bright DOG selection procedure.
In the following sections we used the original IR-bright DOG sample to investigate the optical color properties of IR-bright DOGs by keeping this observational bias against relatively blue DOGs in mind.

\subsection{Why do PL DOGs show a bluer optical color?} \label{subsec:why_pl_blue}

\begin{figure}
 \begin{center}
 \includegraphics[width=9.5cm]{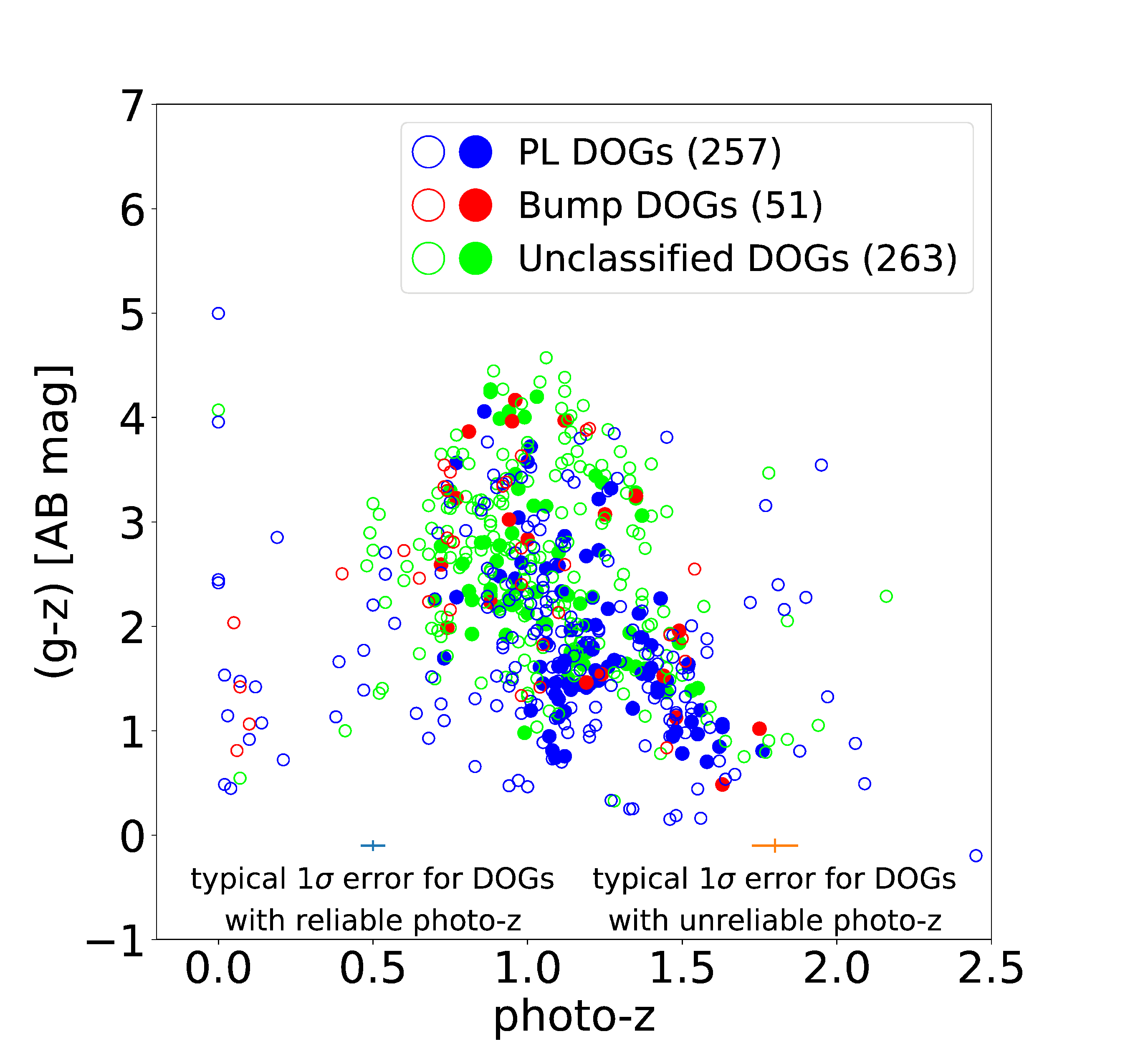}
 \end{center}
 \caption{Distribution of photo-$z$ and $(g-z)_{\rm AB}$ color of our DOG sample. The blue, red, and green circles represent our PL DOGs, bump DOGs, and unclassified DOGs, respectively. The filled and open circles represent DOGs with reliable and unreliable photometric redshift, respectively. The numbers in parentheses indicate the number of objects. The cyan and orange crosses shown at the bottom of the panel denote the typical 1 $\sigma$ error for DOGs with reliable photo-$z$ and unreliable photo-$z$, respectively.}
  \label{fig:DOGs_gz_photoz} 
\end{figure}

\begin{figure}
 \begin{center}
  \includegraphics[width=9cm]{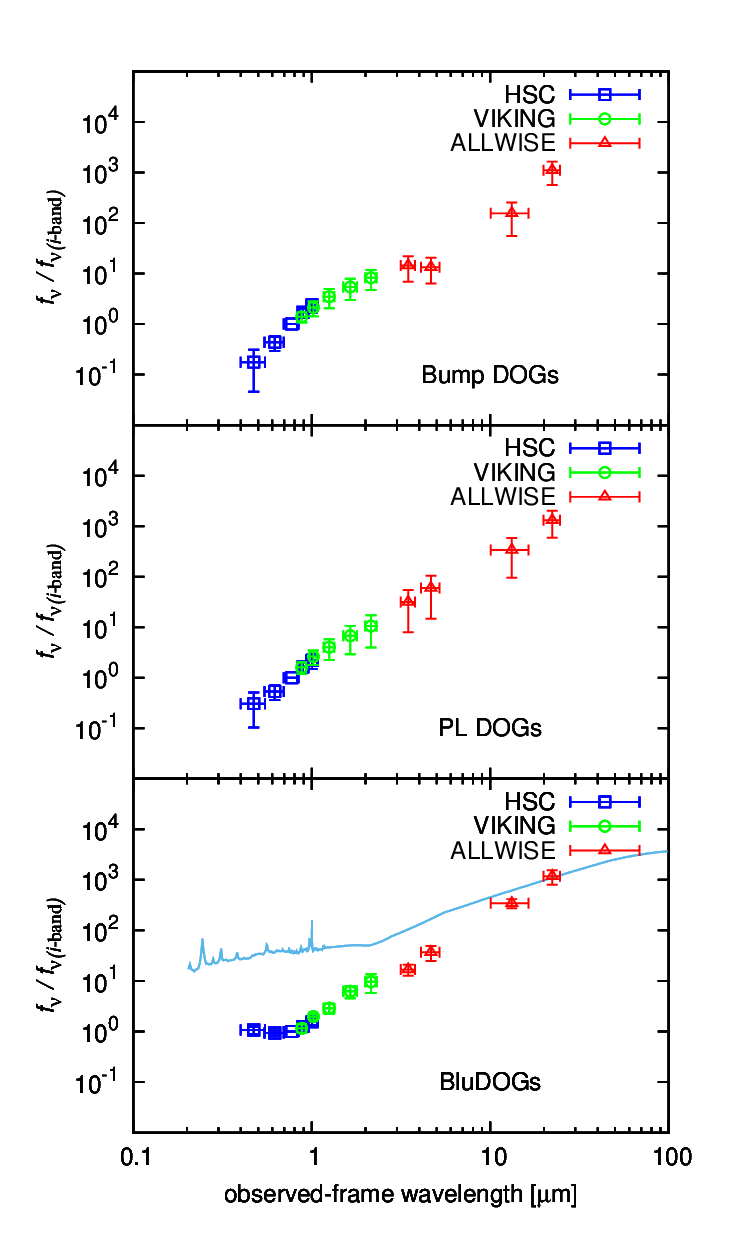}
 \end{center}
 \caption{Mean SEDs of bump DOGs, PL DOGs, and BluDOGs. The top, center, and bottom panels show the mean SEDs of bump DOGs, PL DOGs, and BluDOGs, respectively. The blue square, the green circle, and the red triangle represent the HSC bands, the VIKING bands, and the ALLWISE bands, respectively. The error bar in the vertical direction denotes the standard deviation. The means of $i$ band flux for bump DOGs, PL DOGs, and BluDOGs are $3.38\times10^{-6}\ {\rm{Jy}}$, $3.16\times10^{-6}\ {\rm{Jy}}$, and  $3.09\times10^{-6}\ {\rm{Jy}}$, respectively. The cyan line represents a type 1 quasar template (\citealt{2007ApJ...663...81P}), which is shifted to redshift $z$=1, and is scaled up to the 22 ${\rm \mu m}$ flux of the average SED for BluDOGs.}
\label{fig:BUMPave}
\end{figure}

Here we discuss why PL DOGs show a bluer optical color as presented in Figures \ref{fig:BUMP_PL_hist} and \ref{fig:BUMP_PL}.
First, we show that the observed $(g-z)_{\rm AB}$ color is not sensitive to redshift (Figure \ref{fig:DOGs_gz_photoz}), suggesting that a slight systematic difference in the redshift distribution among sub-classes of DOGs does not affect the $(g-z)_{\rm AB}$ color significantly.
For the sake of discussing the origin of the bluer color of PL DOGs, we show the mean SEDs of bump DOGs and PL DOGs in Figure \ref{fig:BUMPave},
where the SEDs are normalized by the $i$ band flux and averaged on the logarithmic scale (i.e., the geometric mean).
The normalized mean flux of each band is given in Table \ref{tab:aveplbump}.
In the top panel of Figure \ref{fig:BUMPave}, the mean SED of bump DOGs clearly shows the bump feature that is dominated by the stellar emission (\citealt{2009ApJ...700.1190D}; \citealt{2011ApJ...733...21B}).
The optical bands consist of the shorter-wavelength side of the SED, suggesting that the $(g-z)_{\rm AB}$ color is determined by the reddened stellar spectrum.
However, the mean SED of PL DOGs is consistent with a single power law from MIR to optical,
strongly suggesting that the optical emission is dominated by the unreddened AGN emission (see the center panel of Figure \ref{fig:BUMPave}).
Therefore, the origin of the optical emission is different for bump DOGs and PL DOGs,
which creates a systematic difference in the $(g-z)_{\rm AB}$ color between the two populations of DOGs.

\cite{2008ApJ...677..943D} proposed an evolutionary scenario, in which gas-rich major mergers cause the SF-dominated phase (bump DOGs) and then subsequently, the AGN-dominated phase (PL DOGs) appears.
Our results and the scenario of \cite{2008ApJ...677..943D} suggest that the optical emission of DOGs is redder in the early phase after the major merger, and the subsequent enhanced AGN emission overlays the stellar emission in the later phase, resulting in a blue optical color.

\begin{deluxetable}{cccc}
\tablecaption{$i$ band normalized mean flux of bump DOGs, PL DOGs, and BluDOGs\label{tab:aveplbump}}
\tablehead{
	\colhead{Band}	& \colhead{Bump DOGs}	& \colhead{PL DOGs}	& \colhead{BluDOGs}\\
	\colhead{}	& \colhead{$N_{\rm obj}=51$}	& \colhead{$N_{\rm obj}=257$}	& \colhead{$N_{\rm obj}=8$}
}
\startdata
%\hline\textcolor{blue}{
      $g$	& $0.176\pm0.131$	& $0.307\pm0.203$	& $1.07\pm0.18$	\\
      $r$	& $0.435\pm0.141$	& $0.537\pm0.175$	& $0.940\pm0.106$	\\
      $i$	& $1.00$			& $1.00$			& $1.00$			\\
      $z$	& $1.70\pm0.34$	& $1.64\pm0.37$	& $1.27\pm0.15$	\\
      $y$	& $2.37\pm0.70$	& $2.21\pm0.71$	& $1.55\pm0.21$	\\
      $Z$	& $1.39\pm0.31$	& $1.61\pm0.43$	& $1.16\pm0.22$	\\
      $Y$	& $2.12\pm0.70$	& $2.63\pm0.83$	& $1.97\pm0.25$	\\
      $J$	& $3.49\pm1.42$	& $4.02\pm1.76$	& $2.90\pm0.65$	\\
      $H$	& $5.41\pm2.43$	& $6.80\pm3.89$	& $6.24\pm1.67$	\\
      $Ks$	& $8.25\pm3.50$	& $10.6\pm6.6$	& $9.68\pm3.90$	\\
      $W1$	& $14.4\pm7.5$	& $31.5\pm23.5$	& $16.8\pm4.1$	\\
      $W2$	& $13.4\pm7.0$	& $59.7\pm45.0$	& $37.1\pm12.1$	\\
      $W3$	& $155\pm100$	& $339\pm243$	& $345\pm72$		\\
      $W4$	& $1110\pm540$	& $1310\pm720$	& $1180\pm380$	\\ \hline
\enddata
%\tablecomments{}
\end{deluxetable}

\subsection{DOGs in the optical color-color diagram} \label{subsec:waht_pop_dogs}

\begin{figure}
 \begin{center}
  \includegraphics[width=8.5cm]{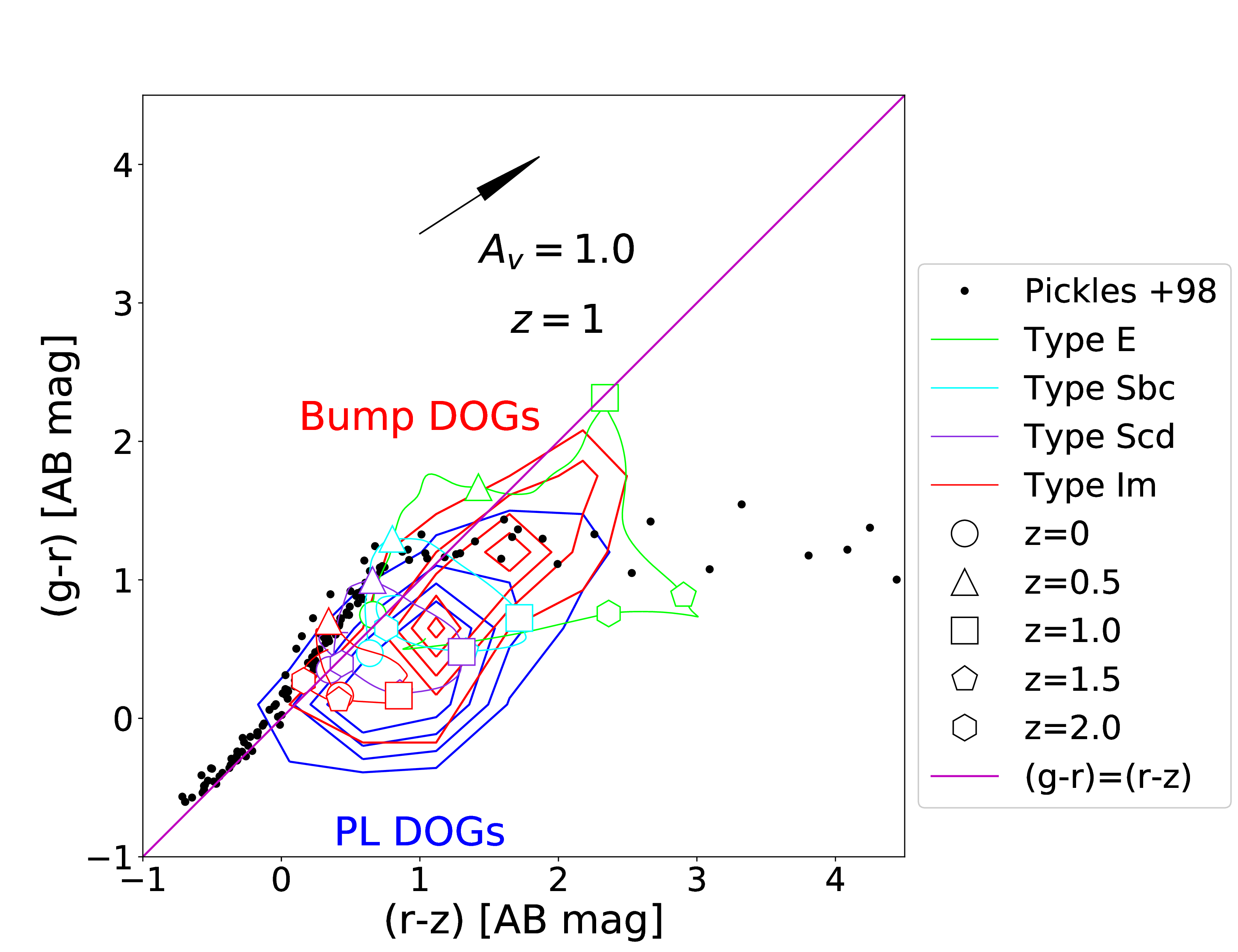}
 \end{center}
 \caption{The $(g-r)_{\rm AB}$ vs. $(r-z)_{\rm AB}$ color-color diagram with star templates and galaxy templates. The black filled circles represent the star templates of \cite{1998PASP..110..863P}. The green, the cyan, the purple, and the red lines represent the type E, Sbc, Scd, and Im galaxy templates of \cite{1980ApJS...43..393C}, respectively. The symbols represent indications of redshift (circle, triangle, square, pentagon, and hexagon denote  $z=0$, $z=0.5$, $z=1.0$, $z=1.5$, and $z=2.0$, respectively). The blue and red contours represent the number density of the PL DOGs and the bump DOGs for each $0.5\times0.5\ {\rm mag^2}$ bin.}
\label{fig:star_gal}
\end{figure}

To understand the optical color distribution of DOGs shown in Figure \ref{fig:2dogs},
we compare the observed color distribution of DOGs and some galaxy spectral templates in Figure \ref{fig:star_gal}.
We used the E, Sbc, Scd, and Im galaxy templates of \cite{1980ApJS...43..393C}.
This figure suggests that the optical color of bump DOGs is understood through the reddened spectrum of star-forming galaxies at $0.5\lesssim z \lesssim1.5$,
which is consistent with the picture shown in Section \ref{subsec:why_pl_blue}, and also with the distribution of the photometric redshift (Figure \ref{fig:bumpphotoz}).
The optical color distribution of PL DOGs can be understood by the additional bluer AGN continuum emission on the reddened stellar emission, as discussed in Section \ref{subsec:why_pl_blue}.

\subsection{Blue-excess DOGs} \label{subsec:BluDOGs}

\begin{figure}
\begin{center}
 \includegraphics[width=8.5cm]{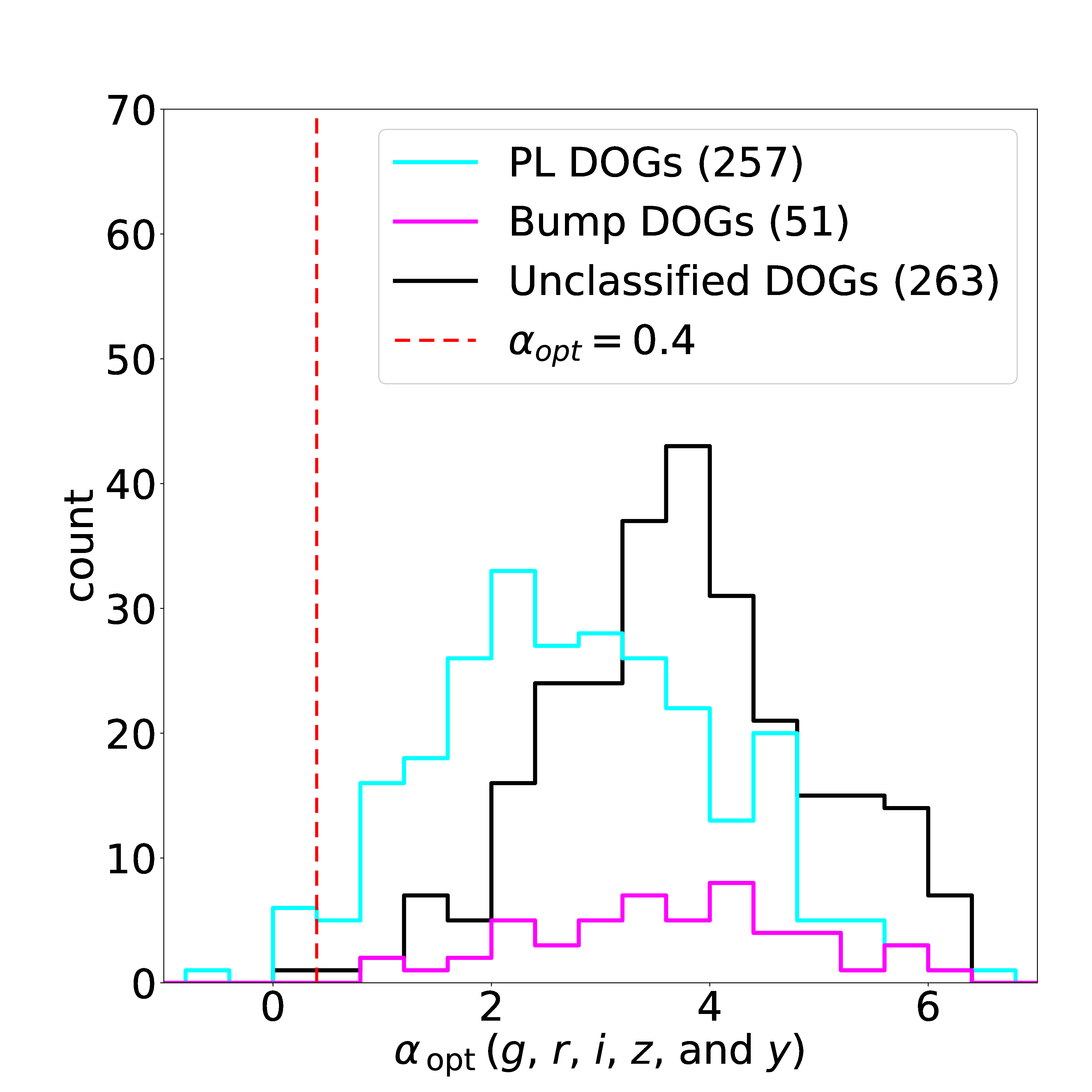}
 \end{center}
 \caption{The $\alpha_{\rm opt}$ histogram. The magenta and the cyan histograms represent the bump DOGs and the PL DOGs. The red dashed line represents $\alpha_{\rm opt} = 0.4$. The numbers in parentheses indicate the number of objects.}
\label{fig:Aoptfisto}
\end{figure}

\begin{figure}
\begin{center}
\includegraphics[width=8.5 cm]{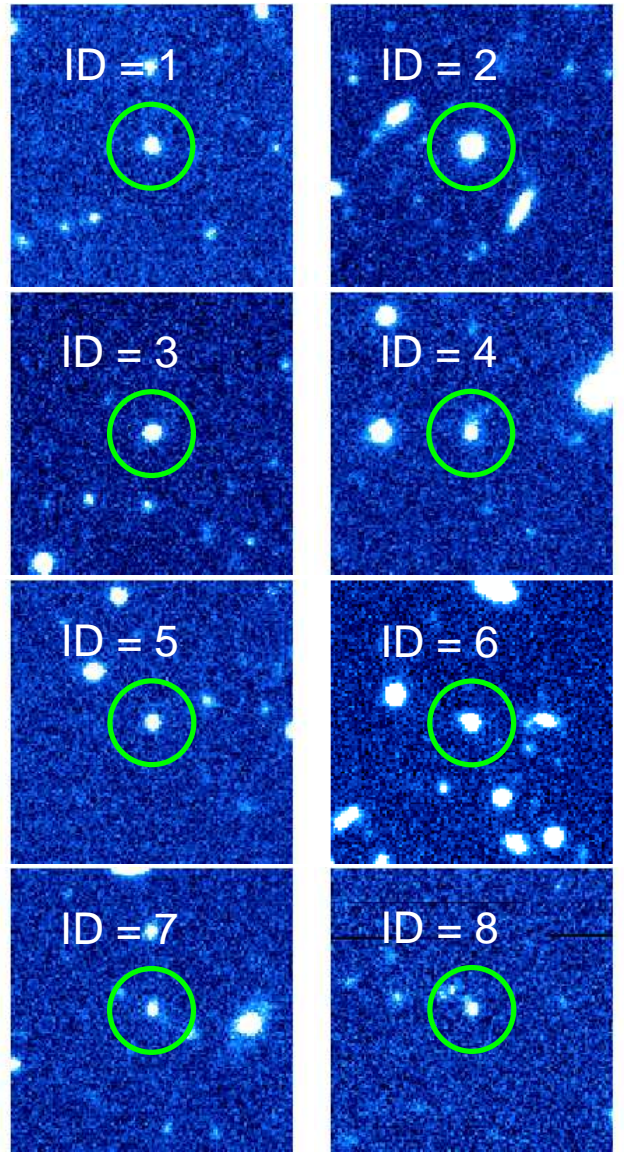}
 \end{center}
 \caption{HSC $i$ band images of BluDOGs. The size of each image is 20'' $\times$ 20''. The radius of green circles is 3''.}
\label{fig:image_BLU}
\end{figure}

\begin{figure}
\begin{center}
\includegraphics[width=8.5 cm]{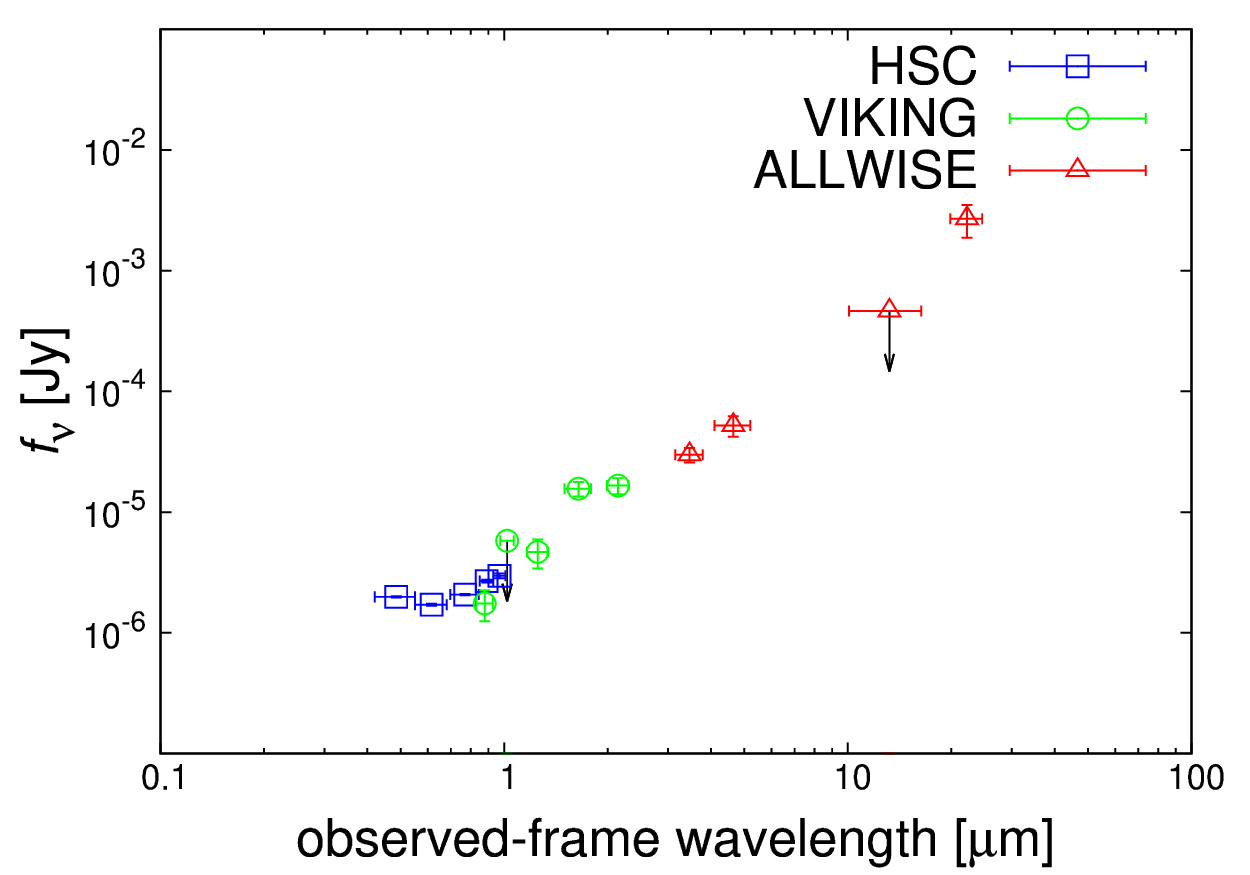}
\end{center}
\caption{SED of the unclassified BluDOG (ID=4). Symbols are the same as in Figure \ref{fig:BUMPave}.}
\label{fig:BLU_un_sed}
\end{figure}

In Section \ref{subsubsec: fractionPLopt}, we showed that PL DOGs have systematically bluer $(g-z)_{\rm AB}$ color than bump DOGs.
Figure \ref{fig:BUMP_PL_hist} shows that a part of PL DOGs show very blue $(g-z)_{\rm AB}$ color, such as $(g-z)_{\rm AB}\sim0.5$.
This is surprising because this color is similar to optically selected BOSS quasars at similar redshift (Table \ref{tab:gzsam}).
To investigate these ``blue-excess DOGs'' (hereafter BluDOGs) in detail, we quantified the blueness of DOGs as follows:
First, we assumed that the optical emission covered by HSC (from $g$ band to $y$ band) is described by a power law.
Then the optical spectral index in the power-law fit ($\alpha_{\rm opt}$) is defined as follow:
\begin{eqnarray}
	f_{\nu}\propto\lambda^{\alpha_{\rm opt}} {\rm .}
\end{eqnarray}
Figure \ref{fig:Aoptfisto} shows the histogram of $\alpha_{\rm opt}$ for bump DOGs and PL DOGs.
This figure shows that PL DOGs have flatter optical SED than bump DOGs;
the averages and standard deviations of $\alpha_{\rm opt}$ for bump DOGs and PL DOGs are $3.60\pm1.26$ and $2.84\pm1.31$, respectively.
Finally, we selected BluDOGs by adopting the following criterion:
\begin{eqnarray}
	\alpha_{\rm opt}<0.4{\rm .}
\end{eqnarray}
Consequently, we select 8 BluDOGs.
This corresponds to $\sim1\%\ (=8/571)$ among the entire IR-bright DOGs sample. This fraction is uncertain because some likely observational biases (discussed in Section \ref{subsec:sel_eff}) are not taken into account.
Regarding the nature of BluDOGs, we examined possible neighboring foreground or background objects around BluDOGs that may cause a very blue color to be observed.
Figure \ref{fig:image_BLU} shows the $i$ band images of BluDOGs.
No BluDOGs have bright objects within 3 arcsec.
Therefore, the photometric properties of the eight BluDOGs are not significantly affected by neighboring foreground or background objects.
However, Figure \ref{fig:Aoptfisto} shows that our BluDOGs sample contains one unclassified DOG.
It is noteworthy that most unclassified DOGs are thought to be bump DOGs as discussed in Section \ref{subsubsec: fractionPLopt}.
To investigate the nature of this unclassified BluDOG (ID=4), we show the SED of this unclassified BluDOG in Figure \ref{fig:BLU_un_sed}.
By comparing the SED of the unclassified DOG with the average SED of bump and PL DOGs (Figure \ref{fig:BUMPave}), the SED of the unclassified DOG looks very similar to the average SED of PL DOGs.
Therefore, we conclude that this unclassified BluDOG is also a PL DOG intrinsically. We thus conclude that all the BluDOGs discovered in this study are PL DOGs.

\begin{figure}
\begin{center}
  \includegraphics[width=9 cm]{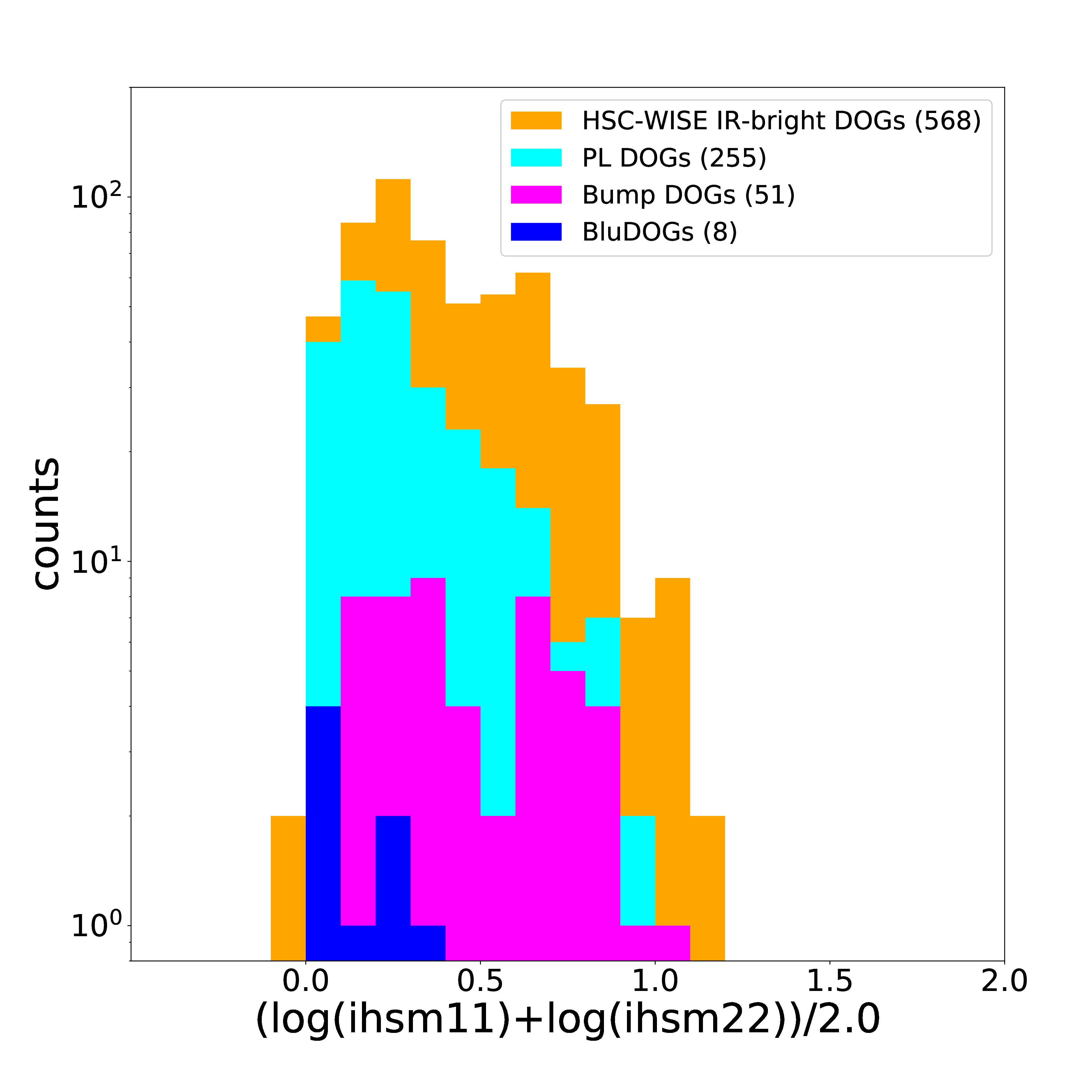}
 \end{center}
 \caption{Histogram for the geometric mean of ihsm11 and ihsm22. The orange, cyan, magenta, and blue histograms represent the HSC-${\it WISE}$ IR-bright DOGs, PL DOGs, Bump DOGs, and BluDOGs. The number in parentheses indicates the number of objects.}
\label{fig:dogs_shape_i}
\end{figure}

To investigate the morphological shape of BluDOGs, we examined the adaptive moment in the HSC-SSP database (\citealt{2018PASJ...70S..34A}; see also \citealt{2003MNRAS.343..459H}). 
The adaptive moment is an indicator of the spatial extension of HSC sources, and is calculated by the algorithm described in \cite{2003MNRAS.343..459H}. In this study, we adopted the adaptive moments of ``ishape\_hsm\_moments\_11'', ``ishape\_hsm\_psf-moments\_11'', ``ishape\_hsm\_moments\_22'', and ``ishape\_-hsm\_psfmoments\_22'' based on $i$ band images, because the image quality (e.g., the spatial resolution) of $i$ band HSC images is better than that of HSC images obtained in the other photometric bands (see \citealt{2018PASJ...70S...8A}). ``ishape\_hsm\_moments\_11'' and ``ishape\_hsm\_-moments\_22'' represent the second-order adaptive moments of sources, while ``ishape\_hsm\_psf-moments\_11'' and ``ishape\_hsm\_psfmoments\_22'' represent the second-order adaptive moments of point spread functions at each source position. The suffixes of 11 and 22 represent each perpendicular axis in each image (see, e.g., \citealt{2018ApJ...866..140Y}). Using this data, we calculated the following values:
\begin{eqnarray}
	{\rm ihsm11} &=& \frac{\rm ishape\_hsm\_moments\_11}{\rm ishape\_hsm\_psfmoments\_11} \nonumber\\
	{\rm ihsm22} &=& \frac{\rm ishape\_hsm\_moments\_22}{\rm ishape\_hsm\_psfmoments\_22},
\end{eqnarray}
where ihsm11 and ihsm22 are expected to be close to one if a source has a PSF-shape (see \citealt{2018PASJ...70S..34A}). 
We adopted the geometric mean of ihsm11 and ihsm22, and Figure \ref{fig:dogs_shape_i} shows the histogram for the calculated mean of ihsm11 and ihsm22. 
Three objects (two PL DOGs without optically blue excess and one unclassified DOG) do not have these values. Therefore, we removed them from this statistical analysis because the statistical properties of the 571 DOGs are not affected significantly by removing the three objects.
The logarithmic means and standard deviations of the geometric mean value for all IR-bright DOGs, PL DOGs, Bump DOGs, and BluDOGs are $0.40\pm0.25$, $0.31\pm0.21$, $0.46\pm0.25$, and $0.15\pm0.11$, respectively. This suggests that the fraction of point source is higher in PL DOGs than in Bump DOGs, and higher in BluDOGs than in PL DOGs. Specifically, most BluDOGs seems to be consistent with point sources.

To investigate the SED of BluDOGs, we derived the mean SED (the geometric mean flux for each band) of BluDOGs in the same manner as described in Section \ref{subsec:why_pl_blue}.
The mean SED of BluDOGs shows an extreme blue excess as expected, at $\lambda_{\rm obs}<1{\rm \mu m}$ (see Table \ref{tab:aveplbump} and the bottom panel of Figure \ref{fig:BUMPave}).
We also show a template spectrum of type 1 quasars at $z=1$ (\citealt{2007ApJ...663...81P}) to Figure \ref{fig:BUMPave}. The similarity in the spectral index at the blue part between BluDOGs and type 1 quasars suggests that the origin of the blue excess in BluDOGs is the unobscured AGN emission; specifically, the leaked AGN emission or scattered AGN emission.

Such a blue excess has also been reported for some Hot DOGs (\citealt{2016ApJ...819..111A}).
\cite{2016ApJ...819..111A} argued that the Hot DOGs with a blue excess can be explained by the leaked AGN emission, and thus the BluDOGs in our sample seem to be a very similar population to the blue-excess Hot DOGs.
This is an interesting idea because these populations of DOGs may correspond to the transition phase from the dust-obscured phase to the UV-bright quasar phase in the major-merger scenario of \cite{2008ApJ...677..943D} (see also \citealt{2008ApJS..175..356H}).
We examined whether the BluDOGs in our sample also satisfy the criteria for Hot DOGs (\citealt{2012ApJ...755..173E}; \citealt{2012ApJ...756...96W}), which are:
\begin{eqnarray}
	W1_{\rm Vega} &>& 17.4, W4_{\rm Vega} < 7.7,\ \&\ (W2 - W4)_{\rm Vega} > 8.2,\nonumber\\ \\
	&&{\rm or}\nonumber\\
	W1_{\rm Vega} &>& 17.4, W3_{\rm Vega} < 10.6,\ \&\ (W2 - W3)_{\rm Vega} > 5.3.\nonumber\\
\end{eqnarray}
We found that only one object among 571 IR-bright DOGs satisfies the criteria for Hot DOGs, and no BluDOGs satisfy the criteria for Hot DOGs.
Therefore, the combination of the HSC and ${\it WISE}$ offers a complementary path to the Hot DOG criteria for identifying very interesting objects in terms of the co-evolution between galaxies and SMBHs.
In \cite{2016ApJ...819..111A}, the blue excess of Hot DOGs is thought to be a leaking AGN emission scattered into our line of sight, because the X-ray observation of blue-excess Hot DOGs shows that their hydrogen column density is $N_{\rm H} \sim 6\times10^{23}\ {\rm cm^{-2}}$. By observing our BluDOGs with X-ray, we can understand the origin of the blue excess which is either a directly leaking AGN emission or a scattered AGN emission.

On the evolutionary link between BluDOGs and quasars, it is interesting to compare the bolometric luminosity ($L_{\rm bol}$) of BluDOGs and quasars. To compare $L_{\rm bol}$ of quasars with $L_{\rm bol}$ of BluDOGs and our entire DOGs, we estimated the $L_{\rm bol}$ of DOGs using the conversion factor of the infrared luminosity ($L_{\rm IR}$) to $L_{\rm bol}$, and the average of the $L_{\rm IR}$ over $\nu^{\rm obs} L_{\nu}^{\rm obs}$ ratio at $\nu = 22 {\rm \mu m}$. For estimating the $L_{\rm IR}$ from $\nu^{\rm obs} L_{\nu}^{\rm obs}$, we averaged the $L_{\rm IR}/\nu^{\rm obs} L_{\nu}^{\rm obs}$ of the PL and bump DOGs within the redshift range of $0.5<z<2.0$ in \cite{2012AJ....143..125M}, and obtained $7.5\pm1.2$ and $10.7\pm3.0$, respectively. The average of the estimated $\log (L_{\rm IR}/L_{\odot})$ for all IR-bright DOGs, PL DOGs, bump DOGs, and BluDOGs are $12.8\pm0.5$, $12.8\pm0.6$, $12.7\pm0.7$, and $13.1\pm0.3$, while the average of the estimated $\log (L_{\rm IR}/L_{\odot})$ for IR-bright DOGs, PL DOGs, and bump DOGs with reliable photo-$z$ are (see Section \ref{subsec:rdoirdog}) $12.9\pm0.2$, $12.9\pm0.2$, and $13.0\pm0.3$, respectively (see Table \ref{tab:estimatedll}). There are no BluDOGs with reliable photo-$z$. By using the conversion factor of $L_{\rm bol} = 1.4\times L_{\rm IR}$ (\citealt{2017ApJ...835...36T}), the average of the estimated $\log (L_{\rm bol}/L_{\odot})$ for all IR-bright DOGs, PL DOGs, bump DOGs, and BluDOGs are $12.9\pm0.5$, $12.9\pm0.6$, $12.8\pm0.7$, and $13.3\pm0.3$, while the average of the estimated $\log (L_{\rm bol}/L_{\odot})$ for IR-bright DOGs, PL DOGs, and bump DOGs with reliable photo-$z$ are $13.0\pm0.2$, $13.1\pm0.2$, and $13.1\pm0.3$, respectively (see Table \ref{tab:estimatedll}). By comparing the $L_{\rm bol}$ of our BluDOGs and entire DOGs with the characteristic luminosity of the quasar luminosity function at redshift $\sim$1 in \cite{2015ApJ...810...74A}, we understand that the $L_{\rm bol}$ of our BluDOGs and DOGs is roughly consistent with the characteristic $L_{\rm bol}$ for quasars at a redshift of about one. This suggests that DOGs harbor an obscured AGN whose bolometric luminosity is statistically comparable to typical quasars.

\begin{deluxetable}{ccccc}
\tablecaption{Estimated $L_{\rm IR}$ and $L_{\rm bol}$ for IR-bright DOGs, PL DOGs, bump DOGs, and BluDOGs\label{tab:estimatedll}}
\tablehead{
	\colhead{Sample}	& \multicolumn{2}{c}{$\log (L_{\rm IR}/L_{\odot})$}	& \multicolumn{2}{c}{$\log (L_{\rm bol}/L_{\odot})$}\\
					& \colhead{All}			& \colhead{\shortstack{Reliable\\photo-$z$}}	& \colhead{All}			& \colhead{\shortstack{Reliable\\photo-$z$}}}
\startdata
%\hline
      All DOGs 			& $12.8\pm0.5$		& $12.9\pm0.2$		& $12.9\pm0.5$	& $13.0\pm0.2$	\\
      PL DOGs			& $12.8\pm0.6$		& $12.9\pm0.2$		& $12.9\pm0.6$	& $13.1\pm0.2$	\\
      Bump DOGs		& $12.7\pm0.7$		& $13.0\pm0.3$		& $12.8\pm0.7$	& $13.1\pm0.3$	\\
      BluDOGs			& $13.1\pm0.3$		& \nodata				& $13.3\pm0.3$	& \nodata			\\
      \hline
\enddata
%\tablecomments{}
\end{deluxetable}

Finally, we discuss the lifetime of BluDOGs. In \cite{2010MNRAS.407.1701N}, the lifetime of DOGs at redshift $\sim$2 is inferred to be $\sim$70 Myr (the lifetimes of bump DOGs and PL DOGs are estimated to be 30 Myr and 40 Myr, respectively). This inferred lifetime of DOGs is consistent with the lifetime of the IR-bright DOGs ($\sim$50 Myr) estimated by \cite{2017ApJ...835...36T}. Assuming the lifetime of DOGs to be $\sim$70 Myr, the lifetime of BluDOGs is naively estimated to be $\sim$1 Myr, given the fact that the abundance of BluDOGs among entire DOGs is roughly 1\%. This timescale seems too short if BluDOGs are the only population corresponding to galaxies in a blowing-out phase. Therefore, we speculate that the BluDOGs occupy only a small fraction of the entire population of galaxies experiencing the blowing-out of the optically thick dusty interstellar medium (ISM). Another likely population corresponding to galaxies in the blowing-out phase is the extremely red quasar studied by \cite{2015MNRAS.453.3932R} (see also Section \ref{sec:intro}), because their SED shows some features of quasars at optical, and also a very red color between optical and mid-IR. However, the relation between the BluDOG and the extremely red quasar is unclear. Both the BluDOGs and the extremely red quasars should be spectroscopically studied in a statistical manner for revealing the nature of galaxies at the blowing-out phase in the so-called gas-rich major merger scenario.

%%%%%%%%%%%%%%%%%%%%%%%%%%%%%
\section{CONCLUSION} \label{sec:conclusion}

We selected 571 IR-bright DOGs using the HSC S16A catalog, the VIKING DR2 catalog, and the ALLWISE catalog.
The main results from the statistical analysis of the IR-bright DOGs are as follows:
\begin{enumerate}
\item The $(g-z)_{\rm AB}$ color distribution of DOGs is significantly redder than that of ULIRGs, HyLIRGs, and quasars at similar redshift with a large dispersion.
\item The $(g-z)_{\rm AB}$ color of the PL DOGs are bluer than that of the bump DOGs. This is explained by the reddened stellar optical emission seen in bump DOGs, but it is overlaid by the power-law emission in the PL DOGs.
\item We identified eight IR-bright DOGs showing an extreme blue excess (BluDOGs). This population seems to be similar to the blue-excess hot DOGs reported by \cite{2016ApJ...819..111A}, and is likely a very interesting population corresponding to the phase of transition from the dust-obscured phase to the optically thin quasar phase in major-merger scenarios.
\end{enumerate}

%%%%%%%%%%%%%%%%%%%%%%%%%%%%%
\acknowledgments
The authors gratefully acknowledge the anonymous referee for a careful reading of the manuscript and very helpful comments.
%HSC
This study is based on data collected at the Subaru Telescope and retrieved from the HSC data archive system, which is operated by the Subaru Telescope and Astronomy Data Center at the National Astronomical Observatory of Japan.
The Hyper Suprime-Cam (HSC) collaboration includes the astronomical communities of Japan and Taiwan, and Princeton University. 
The HSC instrumentation and software were developed by the National Astronomical Observatory of Japan (NAOJ), the Kavli Institute for the Physics and Mathematics of the Universe (Kavli IPMU), the University of Tokyo, the High Energy Accelerator Research Organization (KEK), the Academia Sinica Institute for Astronomy and Astrophysics in Taiwan (ASIAA), and Princeton University. 
Funding was contributed by the FIRST program from Japanese Cabinet Office, the Ministry of Education, Culture, Sports, Science and Technology (MEXT), the Japan Society for the Promotion of Science (JSPS), Japan Science and Technology Agency (JST), the Toray Science Foundation, NAOJ, Kavli IPMU, KEK, ASIAA, and Princeton University.
The Pan-STARRS1 Surveys (PS1) have been made possible through contributions of the Institute for Astronomy, the University of Hawaii, the Pan-STARRS Project Office, the Max-Planck Society and its participating institutes, the Max Planck Institute for Astronomy, Heidelberg and the Max Planck Institute for Extraterrestrial Physics, Garching, The Johns Hopkins University, Durham University, the University of Edinburgh, Queen's University Belfast, the Harvard-Smithsonian Center for Astrophysics, the Las Cumbres Observatory Global Telescope Network Incorporated, the National Central University of Taiwan, the Space Telescope Science Institute, the National Aeronautics and Space Administration under Grant No. NNX08AR22G issued through the Planetary Science Division of the NASA Science Mission Directorate, the National Science Foundation under Grant No. AST-1238877, the University of Maryland, and Eotvos Lorand University (ELTE).
This paper makes use of software developed for the Large Synoptic Survey Telescope. 
We thank the LSST Project for making their code available as free software at http://dm.lsstcorp.org.
%%
%VIKING
This publication has made use of data from the VIKING survey from VISTA at the ESO Paranal Observatory, programme ID 179.A-2004. 
Data processing has been contributed by the VISTA Data Flow System at CASU, Cambridge and WFAU, Edinburgh.
%%
%WISE
This publication makes use of data products from the Wide-field Infrared Survey Explorer, which is a joint project of the University of California, Los Angeles, and the Jet Propulsion Laboratory/California Institute of Technology, funded by the National Aeronautics and Space Administration.
This research was supported by a grant from the Hayakawa Satio Fund awarded by the Astronomical Society of Japan.
This study was financially supported by the Japan Society for the Promotion of Science (JSPS) KAKENHI 16H01101, 16H03958, 17H01114 (T.Nagao), 18J01050 (Y.Toba) 15H02070, and 16K05296 (Y.Terashima).
%%%%%%%%%%%%%%%%%%%%%%%%%%%%

%%%%%%%%%%%%%%%%%%%%%%%%%%%%
\bibliography{Ref}{}

\begin{thebibliography}{}
\expandafter\ifx\csname natexlab\endcsname\relax\def\natexlab#1{#1}\fi
\providecommand{\url}[1]{\href{#1}{#1}}

\bibitem[{{Abazajian} {et~al.}(2004){Abazajian}, {Adelman-McCarthy},
  {Ag{\"u}eros}, {Allam}, {Anderson}, {Anderson}, {Annis}, {Bahcall}, {Baldry},
  {Bastian}, {Berlind}, {Bernardi}, {Blanton}, {Bochanski}, {Boroski},
  {Briggs}, {Brinkmann}, {Brunner}, {Budav{\'a}ri}, {Carey}, {Carliles},
  {Castander}, {Connolly}, {Csabai}, {Doi}, {Dong}, {Eisenstein}, {Evans},
  {Fan}, {Finkbeiner}, {Friedman}, {Frieman}, {Fukugita}, {Gal}, {Gillespie},
  {Glazebrook}, {Gray}, {Grebel}, {Gunn}, {Gurbani}, {Hall}, {Hamabe},
  {Harris}, {Harris}, {Harvanek}, {Heckman}, {Hendry}, {Hennessy}, {Hindsley},
  {Hogan}, {Hogg}, {Holmgren}, {Ichikawa}, {Ichikawa}, {Ivezi{\'c}}, {Jester},
  {Johnston}, {Jorgensen}, {Kent}, {Kleinman}, {Knapp}, {Kniazev}, {Kron},
  {Krzesinski}, {Kunszt}, {Kuropatkin}, {Lamb}, {Lampeitl}, {Lee}, {Leger},
  {Li}, {Lin}, {Loh}, {Long}, {Loveday}, {Lupton}, {Malik}, {Margon},
  {Matsubara}, {McGehee}, {McKay}, {Meiksin}, {Munn}, {Nakajima}, {Nash},
  {Neilsen}, {Newberg}, {Newman}, {Nichol}, {Nicinski}, {Nieto-Santisteban},
  {Nitta}, {Okamura}, {O'Mullane}, {Ostriker}, {Owen}, {Padmanabhan},
  {Peoples}, {Pier}, {Pope}, {Quinn}, {Richards}, {Richmond}, {Rix}, {Rockosi},
  {Schlegel}, {Schneider}, {Scranton}, {Sekiguchi}, {Seljak}, {Sergey},
  {Sesar}, {Sheldon}, {Shimasaku}, {Siegmund}, {Silvestri}, {Smith}, {Smol{\v
  c}i{\'c}}, {Snedden}, {Stebbins}, {Stoughton}, {Strauss}, {SubbaRao},
  {Szalay}, {Szapudi}, {Szkody}, {Szokoly}, {Tegmark}, {Teodoro}, {Thakar},
  {Tremonti}, {Tucker}, {Uomoto}, {Vanden Berk}, {Vandenberg}, {Vogeley},
  {Voges}, {Vogt}, {Walkowicz}, {Wang}, {Weinberg}, {West}, {White}, {Wilhite},
  {Xu}, {Yanny}, {Yasuda}, {Yip}, {Yocum}, {York}, {Zehavi}, {Zibetti}, \&
  {Zucker}}]{2004AJ....128..502A}
{Abazajian}, K., {Adelman-McCarthy}, J.~K., {Ag{\"u}eros}, M.~A., {et~al.}
  2004, \aj, 128, 502

\bibitem[{{Aihara} {et~al.}(2018{\natexlab{a}}){Aihara}, {Arimoto},
  {Armstrong}, {Arnouts}, {Bahcall}, {Bickerton}, {Bosch}, {Bundy}, {Capak},
  {Chan}, {Chiba}, {Coupon}, {Egami}, {Enoki}, {Finet}, {Fujimori}, {Fujimoto},
  {Furusawa}, {Furusawa}, {Goto}, {Goulding}, {Greco}, {Greene}, {Gunn},
  {Hamana}, {Harikane}, {Hashimoto}, {Hattori}, {Hayashi}, {Hayashi},
  {He{\l}miniak}, {Higuchi}, {Hikage}, {Ho}, {Hsieh}, {Huang}, {Huang},
  {Ikeda}, {Imanishi}, {Inoue}, {Iwasawa}, {Iwata}, {Jaelani}, {Jian},
  {Kamata}, {Karoji}, {Kashikawa}, {Katayama}, {Kawanomoto}, {Kayo}, {Koda},
  {Koike}, {Kojima}, {Komiyama}, {Konno}, {Koshida}, {Koyama}, {Kusakabe},
  {Leauthaud}, {Lee}, {Lin}, {Lin}, {Lupton}, {Mandelbaum}, {Matsuoka},
  {Medezinski}, {Mineo}, {Miyama}, {Miyatake}, {Miyazaki}, {Momose}, {More},
  {More}, {Moritani}, {Moriya}, {Morokuma}, {Mukae}, {Murata}, {Murayama},
  {Nagao}, {Nakata}, {Niida}, {Niikura}, {Nishizawa}, {Obuchi}, {Oguri},
  {Oishi}, {Okabe}, {Okamoto}, {Okura}, {Ono}, {Onodera}, {Onoue}, {Osato},
  {Ouchi}, {Price}, {Pyo}, {Sako}, {Sawicki}, {Shibuya}, {Shimasaku},
  {Shimono}, {Shirasaki}, {Silverman}, {Simet}, {Speagle}, {Spergel},
  {Strauss}, {Sugahara}, {Sugiyama}, {Suto}, {Suyu}, {Suzuki}, {Tait},
  {Takada}, {Takata}, {Tamura}, {Tanaka}, {Tanaka}, {Tanaka}, {Tanaka},
  {Terai}, {Terashima}, {Toba}, {Tominaga}, {Toshikawa}, {Turner}, {Uchida},
  {Uchiyama}, {Umetsu}, {Uraguchi}, {Urata}, {Usuda}, {Utsumi}, {Wang}, {Wang},
  {Wong}, {Yabe}, {Yamada}, {Yamanoi}, {Yasuda}, {Yeh}, {Yonehara}, \&
  {Yuma}}]{2018PASJ...70S...4A}
{Aihara}, H., {Arimoto}, N., {Armstrong}, R., {et~al.} 2018{\natexlab{a}},
  \pasj, 70, S4

\bibitem[{{Aihara} {et~al.}(2018{\natexlab{b}}){Aihara}, {Armstrong},
  {Bickerton}, {Bosch}, {Coupon}, {Furusawa}, {Hayashi}, {Ikeda}, {Kamata},
  {Karoji}, {Kawanomoto}, {Koike}, {Komiyama}, {Lang}, {Lupton}, {Mineo},
  {Miyatake}, {Miyazaki}, {Morokuma}, {Obuchi}, {Oishi}, {Okura}, {Price},
  {Takata}, {Tanaka}, {Tanaka}, {Tanaka}, {Uchida}, {Uraguchi}, {Utsumi},
  {Wang}, {Yamada}, {Yamanoi}, {Yasuda}, {Arimoto}, {Chiba}, {Finet},
  {Fujimori}, {Fujimoto}, {Furusawa}, {Goto}, {Goulding}, {Gunn}, {Harikane},
  {Hattori}, {Hayashi}, {He{\l}miniak}, {Higuchi}, {Hikage}, {Ho}, {Hsieh},
  {Huang}, {Huang}, {Imanishi}, {Iwata}, {Jaelani}, {Jian}, {Kashikawa},
  {Katayama}, {Kojima}, {Konno}, {Koshida}, {Kusakabe}, {Leauthaud}, {Lee},
  {Lin}, {Lin}, {Mandelbaum}, {Matsuoka}, {Medezinski}, {Miyama}, {Momose},
  {More}, {More}, {Mukae}, {Murata}, {Murayama}, {Nagao}, {Nakata}, {Niida},
  {Niikura}, {Nishizawa}, {Oguri}, {Okabe}, {Ono}, {Onodera}, {Onoue}, {Ouchi},
  {Pyo}, {Shibuya}, {Shimasaku}, {Simet}, {Speagle}, {Spergel}, {Strauss},
  {Sugahara}, {Sugiyama}, {Suto}, {Suzuki}, {Tait}, {Takada}, {Terai}, {Toba},
  {Turner}, {Uchiyama}, {Umetsu}, {Urata}, {Usuda}, {Yeh}, \&
  {Yuma}}]{2018PASJ...70S...8A}
{Aihara}, H., {Armstrong}, R., {Bickerton}, S., {et~al.} 2018{\natexlab{b}},
  \pasj, 70, S8

\bibitem[{{Akiyama} {et~al.}(2018){Akiyama}, {He}, {Ikeda}, {Niida}, {Nagao},
  {Bosch}, {Coupon}, {Enoki}, {Imanishi}, {Kashikawa}, {Kawaguchi}, {Komiyama},
  {Lee}, {Matsuoka}, {Miyazaki}, {Nishizawa}, {Oguri}, {Ono}, {Onoue}, {Ouchi},
  {Schulze}, {Silverman}, {Tanaka}, {Tanaka}, {Terashima}, {Toba}, \&
  {Ueda}}]{2018PASJ...70S..34A}
{Akiyama}, M., {He}, W., {Ikeda}, H., {et~al.} 2018, \pasj, 70, S34

\bibitem[{{Arnaboldi} {et~al.}(2007){Arnaboldi}, {Neeser}, {Parker}, {Rosati},
  {Lombardi}, {Dietrich}, \& {Hummel}}]{2007Msngr.127...28A}
{Arnaboldi}, M., {Neeser}, M.~J., {Parker}, L.~C., {et~al.} 2007, The
  Messenger, 127

\bibitem[{{Assef} {et~al.}(2016){Assef}, {Walton}, {Brightman}, {Stern},
  {Alexander}, {Bauer}, {Blain}, {Diaz-Santos}, {Eisenhardt}, {Finkelstein},
  {Hickox}, {Tsai}, \& {Wu}}]{2016ApJ...819..111A}
{Assef}, R.~J., {Walton}, D.~J., {Brightman}, M., {et~al.} 2016, \apj, 819, 111

\bibitem[{{Aversa} {et~al.}(2015){Aversa}, {Lapi}, {de Zotti}, {Shankar}, \&
  {Danese}}]{2015ApJ...810...74A}
{Aversa}, R., {Lapi}, A., {de Zotti}, G., {Shankar}, F., \& {Danese}, L. 2015,
  \apj, 810, 74

\bibitem[{{Axelrod} {et~al.}(2010){Axelrod}, {Kantor}, {Lupton}, \&
  {Pierfederici}}]{2010SPIE.7740E..15A}
{Axelrod}, T., {Kantor}, J., {Lupton}, R.~H., \& {Pierfederici}, F. 2010, in
  \procspie, Vol. 7740, Software and Cyberinfrastructure for Astronomy, 774015

\bibitem[{{Bosch} {et~al.}(2018){Bosch}, {Armstrong}, {Bickerton}, {Furusawa},
  {Ikeda}, {Koike}, {Lupton}, {Mineo}, {Price}, {Takata}, {Tanaka}, {Yasuda},
  {AlSayyad}, {Becker}, {Coulton}, {Coupon}, {Garmilla}, {Huang}, {Krughoff},
  {Lang}, {Leauthaud}, {Lim}, {Lust}, {MacArthur}, {Mandelbaum}, {Miyatake},
  {Miyazaki}, {Murata}, {More}, {Okura}, {Owen}, {Swinbank}, {Strauss},
  {Yamada}, \& {Yamanoi}}]{2018PASJ...70S...5B}
{Bosch}, J., {Armstrong}, R., {Bickerton}, S., {et~al.} 2018, \pasj, 70, S5

\bibitem[{{Bruzual} \& {Charlot}(2003)}]{2003MNRAS.344.1000B}
{Bruzual}, G., \& {Charlot}, S. 2003, \mnras, 344, 1000

\bibitem[{{Bussmann} {et~al.}(2009){Bussmann}, {Dey}, {Borys}, {Desai},
  {Jannuzi}, {Le Floc'h}, {Melbourne}, {Sheth}, \&
  {Soifer}}]{2009ApJ...705..184B}
{Bussmann}, R.~S., {Dey}, A., {Borys}, C., {et~al.} 2009, \apj, 705, 184

\bibitem[{{Bussmann} {et~al.}(2011){Bussmann}, {Dey}, {Lotz}, {Armus}, {Brown},
  {Desai}, {Eisenhardt}, {Higdon}, {Higdon}, {Jannuzi}, {Le Floc'h},
  {Melbourne}, {Soifer}, \& {Weedman}}]{2011ApJ...733...21B}
{Bussmann}, R.~S., {Dey}, A., {Lotz}, J., {et~al.} 2011, \apj, 733, 21

\bibitem[{{Bussmann} {et~al.}(2012){Bussmann}, {Dey}, {Armus}, {Brown},
  {Desai}, {Gonzalez}, {Jannuzi}, {Melbourne}, \&
  {Soifer}}]{2012ApJ...744..150B}
{Bussmann}, R.~S., {Dey}, A., {Armus}, L., {et~al.} 2012, \apj, 744, 150

\bibitem[{{Calzetti} {et~al.}(2000){Calzetti}, {Armus}, {Bohlin}, {Kinney},
  {Koornneef}, \& {Storchi-Bergmann}}]{2000ApJ...533..682C}
{Calzetti}, D., {Armus}, L., {Bohlin}, R.~C., {et~al.} 2000, \apj, 533, 682

\bibitem[{{Coleman} {et~al.}(1980){Coleman}, {Wu}, \&
  {Weedman}}]{1980ApJS...43..393C}
{Coleman}, G.~D., {Wu}, C.-C., \& {Weedman}, D.~W. 1980, \apjs, 43, 393

\bibitem[{{Cutri} \& {et al.}(2014)}]{2014yCat.2328....0C}
{Cutri}, R.~M., \& {et al.} 2014, VizieR Online Data Catalog, 2328

\bibitem[{{Dalton} {et~al.}(2006){Dalton}, {Caldwell}, {Ward}, {Whalley},
  {Woodhouse}, {Edeson}, {Clark}, {Beard}, {Gallie}, {Todd}, {Strachan},
  {Bezawada}, {Sutherland}, \& {Emerson}}]{2006SPIE.6269E..0XD}
{Dalton}, G.~B., {Caldwell}, M., {Ward}, A.~K., {et~al.} 2006, in \procspie,
  Vol. 6269, Society of Photo-Optical Instrumentation Engineers (SPIE)
  Conference Series, 62690X

\bibitem[{{Davies} {et~al.}(2007){Davies}, {M{\"u}ller S{\'a}nchez}, {Genzel},
  {Tacconi}, {Hicks}, {Friedrich}, \& {Sternberg}}]{2007ApJ...671.1388D}
{Davies}, R.~I., {M{\"u}ller S{\'a}nchez}, F., {Genzel}, R., {et~al.} 2007,
  \apj, 671, 1388

\bibitem[{{Dawson} {et~al.}(2013){Dawson}, {Schlegel}, {Ahn}, {Anderson},
  {Aubourg}, {Bailey}, {Barkhouser}, {Bautista}, {Beifiori}, {Berlind},
  {Bhardwaj}, {Bizyaev}, {Blake}, {Blanton}, {Blomqvist}, {Bolton}, {Borde},
  {Bovy}, {Brandt}, {Brewington}, {Brinkmann}, {Brown}, {Brownstein}, {Bundy},
  {Busca}, {Carithers}, {Carnero}, {Carr}, {Chen}, {Comparat}, {Connolly},
  {Cope}, {Croft}, {Cuesta}, {da Costa}, {Davenport}, {Delubac}, {de Putter},
  {Dhital}, {Ealet}, {Ebelke}, {Eisenstein}, {Escoffier}, {Fan}, {Filiz Ak},
  {Finley}, {Font-Ribera}, {G{\'e}nova-Santos}, {Gunn}, {Guo}, {Haggard},
  {Hall}, {Hamilton}, {Harris}, {Harris}, {Ho}, {Hogg}, {Holder}, {Honscheid},
  {Huehnerhoff}, {Jordan}, {Jordan}, {Kauffmann}, {Kazin}, {Kirkby}, {Klaene},
  {Kneib}, {Le Goff}, {Lee}, {Long}, {Loomis}, {Lundgren}, {Lupton}, {Maia},
  {Makler}, {Malanushenko}, {Malanushenko}, {Mandelbaum}, {Manera}, {Maraston},
  {Margala}, {Masters}, {McBride}, {McDonald}, {McGreer}, {McMahon}, {Mena},
  {Miralda-Escud{\'e}}, {Montero-Dorta}, {Montesano}, {Muna}, {Myers},
  {Naugle}, {Nichol}, {Noterdaeme}, {Nuza}, {Olmstead}, {Oravetz}, {Oravetz},
  {Owen}, {Padmanabhan}, {Palanque-Delabrouille}, {Pan}, {Parejko},
  {P{\^a}ris}, {Percival}, {P{\'e}rez-Fournon}, {P{\'e}rez-R{\`a}fols},
  {Petitjean}, {Pfaffenberger}, {Pforr}, {Pieri}, {Prada}, {Price-Whelan},
  {Raddick}, {Rebolo}, {Rich}, {Richards}, {Rockosi}, {Roe}, {Ross}, {Ross},
  {Rossi}, {Rubi{\~n}o-Martin}, {Samushia}, {S{\'a}nchez}, {Sayres}, {Schmidt},
  {Schneider}, {Sc{\'o}ccola}, {Seo}, {Shelden}, {Sheldon}, {Shen}, {Shu},
  {Slosar}, {Smee}, {Snedden}, {Stauffer}, {Steele}, {Strauss}, {Streblyanska},
  {Suzuki}, {Swanson}, {Tal}, {Tanaka}, {Thomas}, {Tinker}, {Tojeiro},
  {Tremonti}, {Vargas Maga{\~n}a}, {Verde}, {Viel}, {Wake}, {Watson}, {Weaver},
  {Weinberg}, {Weiner}, {West}, {White}, {Wood-Vasey}, {Yeche}, {Zehavi},
  {Zhao}, \& {Zheng}}]{2013AJ....145...10D}
{Dawson}, K.~S., {Schlegel}, D.~J., {Ahn}, C.~P., {et~al.} 2013, \aj, 145, 10

\bibitem[{{Desai} {et~al.}(2009){Desai}, {Soifer}, {Dey}, {Le Floc'h}, {Armus},
  {Brand}, {Brown}, {Brodwin}, {Jannuzi}, {Houck}, {Weedman}, {Ashby},
  {Gonzalez}, {Huang}, {Smith}, {Teplitz}, {Willner}, \&
  {Melbourne}}]{2009ApJ...700.1190D}
{Desai}, V., {Soifer}, B.~T., {Dey}, A., {et~al.} 2009, \apj, 700, 1190

\bibitem[{{Dey} {et~al.}(2008){Dey}, {Soifer}, {Desai}, {Brand}, {Le Floc'h},
  {Brown}, {Jannuzi}, {Armus}, {Bussmann}, {Brodwin}, {Bian}, {Eisenhardt},
  {Higdon}, {Weedman}, \& {Willner}}]{2008ApJ...677..943D}
{Dey}, A., {Soifer}, B.~T., {Desai}, V., {et~al.} 2008, \apj, 677, 943

\bibitem[{{Driver} {et~al.}(2009){Driver}, {Norberg}, {Baldry}, {Bamford},
  {Hopkins}, {Liske}, {Loveday}, {Peacock}, {Hill}, {Kelvin}, {Robotham},
  {Cross}, {Parkinson}, {Prescott}, {Conselice}, {Dunne}, {Brough}, {Jones},
  {Sharp}, {van Kampen}, {Oliver}, {Roseboom}, {Bland-Hawthorn}, {Croom},
  {Ellis}, {Cameron}, {Cole}, {Frenk}, {Couch}, {Graham}, {Proctor}, {De
  Propris}, {Doyle}, {Edmondson}, {Nichol}, {Thomas}, {Eales}, {Jarvis},
  {Kuijken}, {Lahav}, {Madore}, {Seibert}, {Meyer}, {Staveley-Smith},
  {Phillipps}, {Popescu}, {Sansom}, {Sutherland}, {Tuffs}, \&
  {Warren}}]{2009A&G....50e..12D}
{Driver}, S.~P., {Norberg}, P., {Baldry}, I.~K., {et~al.} 2009, Astronomy and
  Geophysics, 50, 5.12

\bibitem[{{Driver} {et~al.}(2011){Driver}, {Hill}, {Kelvin}, {Robotham},
  {Liske}, {Norberg}, {Baldry}, {Bamford}, {Hopkins}, {Loveday}, {Peacock},
  {Andrae}, {Bland-Hawthorn}, {Brough}, {Brown}, {Cameron}, {Ching}, {Colless},
  {Conselice}, {Croom}, {Cross}, {de Propris}, {Dye}, {Drinkwater}, {Ellis},
  {Graham}, {Grootes}, {Gunawardhana}, {Jones}, {van Kampen}, {Maraston},
  {Nichol}, {Parkinson}, {Phillipps}, {Pimbblet}, {Popescu}, {Prescott},
  {Roseboom}, {Sadler}, {Sansom}, {Sharp}, {Smith}, {Taylor}, {Thomas},
  {Tuffs}, {Wijesinghe}, {Dunne}, {Frenk}, {Jarvis}, {Madore}, {Meyer},
  {Seibert}, {Staveley-Smith}, {Sutherland}, \& {Warren}}]{2011MNRAS.413..971D}
{Driver}, S.~P., {Hill}, D.~T., {Kelvin}, L.~S., {et~al.} 2011, \mnras, 413,
  971

\bibitem[{{Eisenhardt} {et~al.}(2012){Eisenhardt}, {Wu}, {Tsai}, {Assef},
  {Benford}, {Blain}, {Bridge}, {Condon}, {Cushing}, {Cutri}, {Evans},
  {Gelino}, {Griffith}, {Grillmair}, {Jarrett}, {Lonsdale}, {Masci}, {Mason},
  {Petty}, {Sayers}, {Stanford}, {Stern}, {Wright}, \&
  {Yan}}]{2012ApJ...755..173E}
{Eisenhardt}, P.~R.~M., {Wu}, J., {Tsai}, C.-W., {et~al.} 2012, \apj, 755, 173

\bibitem[{{Fiore} {et~al.}(2008){Fiore}, {Grazian}, {Santini}, {Puccetti},
  {Brusa}, {Feruglio}, {Fontana}, {Giallongo}, {Comastri}, {Gruppioni},
  {Pozzi}, {Zamorani}, \& {Vignali}}]{2008ApJ...672...94F}
{Fiore}, F., {Grazian}, A., {Santini}, P., {et~al.} 2008, \apj, 672, 94

\bibitem[{{Furusawa} {et~al.}(2018){Furusawa}, {Koike}, {Takata}, {Okura},
  {Miyatake}, {Lupton}, {Bickerton}, {Price}, {Bosch}, {Yasuda}, {Mineo},
  {Yamada}, {Miyazaki}, {Nakata}, {Koshida}, {Komiyama}, {Utsumi},
  {Kawanomoto}, {Jeschke}, {Noumaru}, {Schubert}, {Iwata}, {Finet},
  {Fujiyoshi}, {Tajitsu}, {Terai}, \& {Lee}}]{2018PASJ...70S...3F}
{Furusawa}, H., {Koike}, M., {Takata}, T., {et~al.} 2018, \pasj, 70, S3

\bibitem[{{Gehrels}(1986)}]{1986ApJ...303..336G}
{Gehrels}, N. 1986, \apj, 303, 336

\bibitem[{{Gonz{\'a}lez-Fern{\'a}ndez}
  {et~al.}(2018){Gonz{\'a}lez-Fern{\'a}ndez}, {Hodgkin}, {Irwin},
  {Gonz{\'a}lez-Solares}, {Koposov}, {Lewis}, {Emerson}, {Hewett}, {Yolda{\c
  s}}, \& {Riello}}]{2018MNRAS.474.5459G}
{Gonz{\'a}lez-Fern{\'a}ndez}, C., {Hodgkin}, S.~T., {Irwin}, M.~J., {et~al.}
  2018, \mnras, 474, 5459

\bibitem[{{G{\"u}ltekin} {et~al.}(2009){G{\"u}ltekin}, {Richstone}, {Gebhardt},
  {Lauer}, {Tremaine}, {Aller}, {Bender}, {Dressler}, {Faber}, {Filippenko},
  {Green}, {Ho}, {Kormendy}, {Magorrian}, {Pinkney}, \&
  {Siopis}}]{2009ApJ...698..198G}
{G{\"u}ltekin}, K., {Richstone}, D.~O., {Gebhardt}, K., {et~al.} 2009, \apj,
  698, 198

\bibitem[{{Gunn} \& {Stryker}(1983)}]{1983ApJS...52..121G}
{Gunn}, J.~E., \& {Stryker}, L.~L. 1983, \apjs, 52, 121

\bibitem[{{Hirata} \& {Seljak}(2003)}]{2003MNRAS.343..459H}
{Hirata}, C., \& {Seljak}, U. 2003, \mnras, 343, 459

\bibitem[{{Hopkins}(2012)}]{2012MNRAS.420L...8H}
{Hopkins}, P.~F. 2012, \mnras, 420, L8

\bibitem[{{Hopkins} {et~al.}(2006){Hopkins}, {Hernquist}, {Cox}, {Di Matteo},
  {Robertson}, \& {Springel}}]{2006ApJS..163....1H}
{Hopkins}, P.~F., {Hernquist}, L., {Cox}, T.~J., {et~al.} 2006, \apjs, 163, 1

\bibitem[{{Hopkins} {et~al.}(2008){Hopkins}, {Hernquist}, {Cox}, \& {Kere{\v
  s}}}]{2008ApJS..175..356H}
{Hopkins}, P.~F., {Hernquist}, L., {Cox}, T.~J., \& {Kere{\v s}}, D. 2008,
  \apjs, 175, 356

\bibitem[{{Ivezic} {et~al.}(2008){Ivezic}, {Tyson}, {Abel}, {Acosta},
  {Allsman}, {AlSayyad}, {Anderson}, {Andrew}, {Angel}, {Angeli}, {Ansari},
  {Antilogus}, {Arndt}, {Astier}, {Aubourg}, {Axelrod}, {Bard}, {Barr},
  {Barrau}, {Bartlett}, {Bauman}, {Beaumont}, {Becker}, {Becla}, {Beldica},
  {Bellavia}, {Blanc}, {Blandford}, {Bloom}, {Bogart}, {Borne}, {Bosch},
  {Boutigny}, {Brandt}, {Brown}, {Bullock}, {Burchat}, {Burke}, {Cagnoli},
  {Calabrese}, {Chandrasekharan}, {Chesley}, {Cheu}, {Chiang}, {Claver},
  {Connolly}, {Cook}, {Cooray}, {Covey}, {Cribbs}, {Cui}, {Cutri}, {Daubard},
  {Daues}, {Delgado}, {Digel}, {Doherty}, {Dubois}, {Dubois-Felsmann},
  {Durech}, {Eracleous}, {Ferguson}, {Frank}, {Freemon}, {Gangler}, {Gawiser},
  {Geary}, {Gee}, {Geha}, {Gibson}, {Gilmore}, {Glanzman}, {Goodenow},
  {Gressler}, {Gris}, {Guyonnet}, {Hascall}, {Haupt}, {Hernandez}, {Hogan},
  {Huang}, {Huffer}, {Innes}, {Jacoby}, {Jain}, {Jee}, {Jernigan},
  {Jevremovic}, {Johns}, {Jones}, {Juramy-Gilles}, {Juric}, {Kahn}, {Kalirai},
  {Kallivayalil}, {Kalmbach}, {Kantor}, {Kasliwal}, {Kessler}, {Kirkby},
  {Knox}, {Kotov}, {Krabbendam}, {Krughoff}, {Kubanek}, {Kuczewski},
  {Kulkarni}, {Lambert}, {Le Guillou}, {Levine}, {Liang}, {Lim}, {Lintott},
  {Lupton}, {Mahabal}, {Marshall}, {Marshall}, {May}, {McKercher}, {Migliore},
  {Miller}, {Mills}, {Monet}, {Moniez}, {Neill}, {Nief}, {Nomerotski},
  {Nordby}, {O'Connor}, {Oliver}, {Olivier}, {Olsen}, {Ortiz}, {Owen}, {Pain},
  {Peterson}, {Petry}, {Pierfederici}, {Pietrowicz}, {Pike}, {Pinto}, {Plante},
  {Plate}, {Price}, {Prouza}, {Radeka}, {Rajagopal}, {Rasmussen}, {Regnault},
  {Ridgway}, {Ritz}, {Rosing}, {Roucelle}, {Rumore}, {Russo}, {Saha},
  {Sassolas}, {Schalk}, {Schindler}, {Schneider}, {Schumacher}, {Sebag},
  {Sembroski}, {Seppala}, {Shipsey}, {Silvestri}, {Smith}, {Smith}, {Strauss},
  {Stubbs}, {Sweeney}, {Szalay}, {Takacs}, {Thaler}, {Van Berg}, {Vanden Berk},
  {Vetter}, {Virieux}, {Xin}, {Walkowicz}, {Walter}, {Wang}, {Warner},
  {Willman}, {Wittman}, {Wolff}, {Wood-Vasey}, {Yoachim}, {Zhan}, \& {for the
  LSST Collaboration}}]{2008arXiv0805.2366I}
{Ivezic}, Z., {Tyson}, J.~A., {Abel}, B., {et~al.} 2008, ArXiv e-prints,
  arXiv:0805.2366

\bibitem[{{Jarrett} {et~al.}(2011){Jarrett}, {Cohen}, {Masci}, {Wright},
  {Stern}, {Benford}, {Blain}, {Carey}, {Cutri}, {Eisenhardt}, {Lonsdale},
  {Mainzer}, {Marsh}, {Padgett}, {Petty}, {Ressler}, {Skrutskie}, {Stanford},
  {Surace}, {Tsai}, {Wheelock}, \& {Yan}}]{2011ApJ...735..112J}
{Jarrett}, T.~H., {Cohen}, M., {Masci}, F., {et~al.} 2011, \apj, 735, 112

\bibitem[{{Kilerci Eser} {et~al.}(2014){Kilerci Eser}, {Goto}, \&
  {Doi}}]{2014ApJ...797...54K}
{Kilerci Eser}, E., {Goto}, T., \& {Doi}, Y. 2014, \apj, 797, 54

\bibitem[{{Komiyama} {et~al.}(2018){Komiyama}, {Obuchi}, {Nakaya}, {Kamata},
  {Kawanomoto}, {Utsumi}, {Miyazaki}, {Uraguchi}, {Furusawa}, {Morokuma},
  {Uchida}, {Miyatake}, {Mineo}, {Fujimori}, {Aihara}, {Karoji}, {Gunn}, \&
  {Wang}}]{2018PASJ...70S...2K}
{Komiyama}, Y., {Obuchi}, Y., {Nakaya}, H., {et~al.} 2018, \pasj, 70, S2

\bibitem[{{Kormendy} \& {Ho}(2013)}]{2013ARA&A..51..511K}
{Kormendy}, J., \& {Ho}, L.~C. 2013, \araa, 51, 511

\bibitem[{{Lonsdale} {et~al.}(2003){Lonsdale}, {Smith}, {Rowan-Robinson},
  {Surace}, {Shupe}, {Xu}, {Oliver}, {Padgett}, {Fang}, {Conrow},
  {Franceschini}, {Gautier}, {Griffin}, {Hacking}, {Masci}, {Morrison},
  {O'Linger}, {Owen}, {P{\'e}rez-Fournon}, {Pierre}, {Puetter}, {Stacey},
  {Castro}, {Polletta}, {Farrah}, {Jarrett}, {Frayer}, {Siana}, {Babbedge},
  {Dye}, {Fox}, {Gonzalez-Solares}, {Salaman}, {Berta}, {Condon}, {Dole}, \&
  {Serjeant}}]{2003PASP..115..897L}
{Lonsdale}, C.~J., {Smith}, H.~E., {Rowan-Robinson}, M., {et~al.} 2003, \pasp,
  115, 897

\bibitem[{{LSST Science Collaboration} {et~al.}(2009){LSST Science
  Collaboration}, {Abell}, {Allison}, {Anderson}, {Andrew}, {Angel}, {Armus},
  {Arnett}, {Asztalos}, {Axelrod}, \& et~al.}]{2009arXiv0912.0201L}
{LSST Science Collaboration}, {Abell}, P.~A., {Allison}, J., {et~al.} 2009,
  ArXiv e-prints, arXiv:0912.0201

\bibitem[{{Lupton} {et~al.}(2001){Lupton}, {Gunn}, {Ivezi{\'c}}, {Knapp}, \&
  {Kent}}]{2001ASPC..238..269L}
{Lupton}, R., {Gunn}, J.~E., {Ivezi{\'c}}, Z., {Knapp}, G.~R., \& {Kent}, S.
  2001, in Astronomical Society of the Pacific Conference Series, Vol. 238,
  Astronomical Data Analysis Software and Systems X, ed. F.~R. {Harnden}, Jr.,
  F.~A. {Primini}, \& H.~E. {Payne}, 269

\bibitem[{{Madau} \& {Dickinson}(2014)}]{2014ARA&A..52..415M}
{Madau}, P., \& {Dickinson}, M. 2014, \araa, 52, 415

\bibitem[{{Magnier} {et~al.}(2013){Magnier}, {Schlafly}, {Finkbeiner}, {Juric},
  {Tonry}, {Burgett}, {Chambers}, {Flewelling}, {Kaiser}, {Kudritzki},
  {Morgan}, {Price}, {Sweeney}, \& {Stubbs}}]{2013ApJS..205...20M}
{Magnier}, E.~A., {Schlafly}, E., {Finkbeiner}, D., {et~al.} 2013, \apjs, 205,
  20

\bibitem[{{Magorrian} {et~al.}(1998){Magorrian}, {Tremaine}, {Richstone},
  {Bender}, {Bower}, {Dressler}, {Faber}, {Gebhardt}, {Green}, {Grillmair},
  {Kormendy}, \& {Lauer}}]{1998AJ....115.2285M}
{Magorrian}, J., {Tremaine}, S., {Richstone}, D., {et~al.} 1998, \aj, 115, 2285

\bibitem[{{Marconi} \& {Hunt}(2003)}]{2003ApJ...589L..21M}
{Marconi}, A., \& {Hunt}, L.~K. 2003, \apjl, 589, L21

\bibitem[{{Matsuoka} {et~al.}(2017){Matsuoka}, {Nagao}, {Maiolino}, {Marconi},
  {Park}, \& {Taniguchi}}]{2017A&A...608A..90M}
{Matsuoka}, K., {Nagao}, T., {Maiolino}, R., {et~al.} 2017, \aap, 608, A90

\bibitem[{{Melbourne} {et~al.}(2012){Melbourne}, {Soifer}, {Desai}, {Pope},
  {Armus}, {Dey}, {Bussmann}, {Jannuzi}, \& {Alberts}}]{2012AJ....143..125M}
{Melbourne}, J., {Soifer}, B.~T., {Desai}, V., {et~al.} 2012, \aj, 143, 125

\bibitem[{{Miyazaki} {et~al.}(2012){Miyazaki}, {Komiyama}, {Nakaya}, {Kamata},
  {Doi}, {Hamana}, {Karoji}, {Furusawa}, {Kawanomoto}, {Morokuma}, {Ishizuka},
  {Nariai}, {Tanaka}, {Uraguchi}, {Utsumi}, {Obuchi}, {Okura}, {Oguri},
  {Takata}, {Tomono}, {Kurakami}, {Namikawa}, {Usuda}, {Yamanoi}, {Terai},
  {Uekiyo}, {Yamada}, {Koike}, {Aihara}, {Fujimori}, {Mineo}, {Miyatake},
  {Yasuda}, {Nishizawa}, {Saito}, {Tanaka}, {Uchida}, {Katayama}, {Wang},
  {Chen}, {Lupton}, {Loomis}, {Bickerton}, {Price}, {Gunn}, {Suzuki},
  {Miyazaki}, {Muramatsu}, {Yamamoto}, {Endo}, {Ezaki}, {Itoh}, {Miwa},
  {Yokota}, {Matsuda}, {Ebinuma}, \& {Takeshi}}]{2012SPIE.8446E..0ZM}
{Miyazaki}, S., {Komiyama}, Y., {Nakaya}, H., {et~al.} 2012, in \procspie, Vol.
  8446, Ground-based and Airborne Instrumentation for Astronomy IV, 84460Z

\bibitem[{{Miyazaki} {et~al.}(2018){Miyazaki}, {Komiyama}, {Kawanomoto}, {Doi},
  {Furusawa}, {Hamana}, {Hayashi}, {Ikeda}, {Kamata}, {Karoji}, {Koike},
  {Kurakami}, {Miyama}, {Morokuma}, {Nakata}, {Namikawa}, {Nakaya}, {Nariai},
  {Obuchi}, {Oishi}, {Okada}, {Okura}, {Tait}, {Takata}, {Tanaka}, {Tanaka},
  {Terai}, {Tomono}, {Uraguchi}, {Usuda}, {Utsumi}, {Yamada}, {Yamanoi},
  {Aihara}, {Fujimori}, {Mineo}, {Miyatake}, {Oguri}, {Uchida}, {Tanaka},
  {Yasuda}, {Takada}, {Murayama}, {Nishizawa}, {Sugiyama}, {Chiba}, {Futamase},
  {Wang}, {Chen}, {Ho}, {Liaw}, {Chiu}, {Ho}, {Lai}, {Lee}, {Jeng}, {Iwamura},
  {Armstrong}, {Bickerton}, {Bosch}, {Gunn}, {Lupton}, {Loomis}, {Price},
  {Smith}, {Strauss}, {Turner}, {Suzuki}, {Miyazaki}, {Muramatsu}, {Yamamoto},
  {Endo}, {Ezaki}, {Ito}, {Kawaguchi}, {Sofuku}, {Taniike}, {Akutsu}, {Dojo},
  {Kasumi}, {Matsuda}, {Imoto}, {Miwa}, {Suzuki}, {Takeshi}, \&
  {Yokota}}]{2018PASJ...70S...1M}
{Miyazaki}, S., {Komiyama}, Y., {Kawanomoto}, S., {et~al.} 2018, \pasj, 70, S1

\bibitem[{{Murakami} {et~al.}(2007){Murakami}, {Baba}, {Barthel}, {Clements},
  {Cohen}, {Doi}, {Enya}, {Figueredo}, {Fujishiro}, {Fujiwara}, {Fujiwara},
  {Garcia-Lario}, {Goto}, {Hasegawa}, {Hibi}, {Hirao}, {Hiromoto}, {Hong},
  {Imai}, {Ishigaki}, {Ishiguro}, {Ishihara}, {Ita}, {Jeong}, {Jeong},
  {Kaneda}, {Kataza}, {Kawada}, {Kawai}, {Kawamura}, {Kessler}, {Kester},
  {Kii}, {Kim}, {Kim}, {Kobayashi}, {Koo}, {Kwon}, {Lee}, {Lorente}, {Makiuti},
  {Matsuhara}, {Matsumoto}, {Matsuo}, {Matsuura}, {M{\"U}ller}, {Murakami},
  {Nagata}, {Nakagawa}, {Naoi}, {Narita}, {Noda}, {Oh}, {Ohnishi}, {Ohyama},
  {Okada}, {Okuda}, {Oliver}, {Onaka}, {Ootsubo}, {Oyabu}, {Pak}, {Park},
  {Pearson}, {Rowan-Robinson}, {Saito}, {Sakon}, {Salama}, {Sato}, {Savage},
  {Serjeant}, {Shibai}, {Shirahata}, {Sohn}, {Suzuki}, {Takagi}, {Takahashi},
  {Tanab{\'E}}, {Takeuchi}, {Takita}, {Thomson}, {Uemizu}, {Ueno}, {Usui},
  {Verdugo}, {Wada}, {Wang}, {Watabe}, {Watarai}, {White}, {Yamamura},
  {Yamauchi}, \& {Yasuda}}]{2007PASJ...59S.369M}
{Murakami}, H., {Baba}, H., {Barthel}, P., {et~al.} 2007, \pasj, 59, S369

\bibitem[{{Narayanan} {et~al.}(2010){Narayanan}, {Dey}, {Hayward}, {Cox},
  {Bussmann}, {Brodwin}, {Jonsson}, {Hopkins}, {Groves}, {Younger}, \&
  {Hernquist}}]{2010MNRAS.407.1701N}
{Narayanan}, D., {Dey}, A., {Hayward}, C.~C., {et~al.} 2010, \mnras, 407, 1701

\bibitem[{{P{\^a}ris} {et~al.}(2017){P{\^a}ris}, {Petitjean}, {Ross}, {Myers},
  {Aubourg}, {Streblyanska}, {Bailey}, {Armengaud}, {Palanque-Delabrouille},
  {Y{\`e}che}, {Hamann}, {Strauss}, {Albareti}, {Bovy}, {Bizyaev}, {Niel
  Brandt}, {Brusa}, {Buchner}, {Comparat}, {Croft}, {Dwelly}, {Fan},
  {Font-Ribera}, {Ge}, {Georgakakis}, {Hall}, {Jiang}, {Kinemuchi},
  {Malanushenko}, {Malanushenko}, {McMahon}, {Menzel}, {Merloni}, {Nandra},
  {Noterdaeme}, {Oravetz}, {Pan}, {Pieri}, {Prada}, {Salvato}, {Schlegel},
  {Schneider}, {Simmons}, {Viel}, {Weinberg}, \& {Zhu}}]{2017AandA...597A..79P}
{P{\^a}ris}, I., {Petitjean}, P., {Ross}, N.~P., {et~al.} 2017, \aap, 597, A79

\bibitem[{{Pickles}(1998)}]{1998PASP..110..863P}
{Pickles}, A.~J. 1998, \pasp, 110, 863

\bibitem[{{Pierre} {et~al.}(2016){Pierre}, {Pacaud}, {Adami}, {Alis},
  {Altieri}, {Baran}, {Benoist}, {Birkinshaw}, {Bongiorno}, {Bremer}, {Brusa},
  {Butler}, {Ciliegi}, {Chiappetti}, {Clerc}, {Corasaniti}, {Coupon}, {De
  Breuck}, {Democles}, {Desai}, {Delhaize}, {Devriendt}, {Dubois}, {Eckert},
  {Elyiv}, {Ettori}, {Evrard}, {Faccioli}, {Farahi}, {Ferrari}, {Finet},
  {Fotopoulou}, {Fourmanoit}, {Gandhi}, {Gastaldello}, {Gastaud},
  {Georgantopoulos}, {Giles}, {Guennou}, {Guglielmo}, {Horellou}, {Husband},
  {Huynh}, {Iovino}, {Kilbinger}, {Koulouridis}, {Lavoie}, {Le Brun}, {Le
  Fevre}, {Lidman}, {Lieu}, {Lin}, {Mantz}, {Maughan}, {Maurogordato},
  {McCarthy}, {McGee}, {Melin}, {Melnyk}, {Menanteau}, {Novak}, {Paltani},
  {Plionis}, {Poggianti}, {Pomarede}, {Pompei}, {Ponman}, {Ramos-Ceja},
  {Ranalli}, {Rapetti}, {Raychaudury}, {Reiprich}, {Rottgering}, {Rozo},
  {Rykoff}, {Sadibekova}, {Santos}, {Sauvageot}, {Schimd}, {Sereno}, {Smith},
  {Smol{\v c}i{\'c}}, {Snowden}, {Spergel}, {Stanford}, {Surdej}, {Valageas},
  {Valotti}, {Valtchanov}, {Vignali}, {Willis}, \&
  {Ziparo}}]{2016A&A...592A...1P}
{Pierre}, M., {Pacaud}, F., {Adami}, C., {et~al.} 2016, \aap, 592, A1

\bibitem[{{Polletta} {et~al.}(2007){Polletta}, {Tajer}, {Maraschi},
  {Trinchieri}, {Lonsdale}, {Chiappetti}, {Andreon}, {Pierre}, {Le F{\`e}vre},
  {Zamorani}, {Maccagni}, {Garcet}, {Surdej}, {Franceschini}, {Alloin},
  {Shupe}, {Surace}, {Fang}, {Rowan-Robinson}, {Smith}, \&
  {Tresse}}]{2007ApJ...663...81P}
{Polletta}, M., {Tajer}, M., {Maraschi}, L., {et~al.} 2007, \apj, 663, 81

\bibitem[{{Richards} {et~al.}(2006){Richards}, {Strauss}, {Fan}, {Hall},
  {Jester}, {Schneider}, {Vanden Berk}, {Stoughton}, {Anderson}, {Brunner},
  {Gray}, {Gunn}, {Ivezi{\'c}}, {Kirkland}, {Knapp}, {Loveday}, {Meiksin},
  {Pope}, {Szalay}, {Thakar}, {Yanny}, {York}, {Barentine}, {Brewington},
  {Brinkmann}, {Fukugita}, {Harvanek}, {Kent}, {Kleinman}, {Krzesi{\'n}ski},
  {Long}, {Lupton}, {Nash}, {Neilsen}, {Nitta}, {Schlegel}, \&
  {Snedden}}]{2006AJ....131.2766R}
{Richards}, G.~T., {Strauss}, M.~A., {Fan}, X., {et~al.} 2006, \aj, 131, 2766

\bibitem[{{Ross} {et~al.}(2015){Ross}, {Hamann}, {Zakamska}, {Richards},
  {Villforth}, {Strauss}, {Greene}, {Alexandroff}, {Brandt}, {Liu}, {Myers},
  {P{\^a}ris}, \& {Schneider}}]{2015MNRAS.453.3932R}
{Ross}, N.~P., {Hamann}, F., {Zakamska}, N.~L., {et~al.} 2015, \mnras, 453,
  3932

\bibitem[{{Rowan-Robinson}(2000)}]{2000MNRAS.316..885R}
{Rowan-Robinson}, M. 2000, \mnras, 316, 885

\bibitem[{{Rowan-Robinson} {et~al.}(2013){Rowan-Robinson}, {Gonzalez-Solares},
  {Vaccari}, \& {Marchetti}}]{2013MNRAS.428.1958R}
{Rowan-Robinson}, M., {Gonzalez-Solares}, E., {Vaccari}, M., \& {Marchetti}, L.
  2013, \mnras, 428, 1958

\bibitem[{{Sanders} \& {Mirabel}(1996)}]{1996ARA&A..34..749S}
{Sanders}, D.~B., \& {Mirabel}, I.~F. 1996, \araa, 34, 749

\bibitem[{{Sanders} {et~al.}(1988){Sanders}, {Soifer}, {Elias}, {Madore},
  {Matthews}, {Neugebauer}, \& {Scoville}}]{1988ApJ...325...74S}
{Sanders}, D.~B., {Soifer}, B.~T., {Elias}, J.~H., {et~al.} 1988, \apj, 325, 74

\bibitem[{{Schlafly} {et~al.}(2012){Schlafly}, {Finkbeiner}, {Juri{\'c}},
  {Magnier}, {Burgett}, {Chambers}, {Grav}, {Hodapp}, {Kaiser}, {Kudritzki},
  {Martin}, {Morgan}, {Price}, {Rix}, {Stubbs}, {Tonry}, \&
  {Wainscoat}}]{2012ApJ...756..158S}
{Schlafly}, E.~F., {Finkbeiner}, D.~P., {Juri{\'c}}, M., {et~al.} 2012, \apj,
  756, 158

\bibitem[{{Schlegel} {et~al.}(1998){Schlegel}, {Finkbeiner}, \&
  {Davis}}]{1998ApJ...500..525S}
{Schlegel}, D.~J., {Finkbeiner}, D.~P., \& {Davis}, M. 1998, \apj, 500, 525

\bibitem[{{Tanaka}(2015)}]{2015ApJ...801...20T}
{Tanaka}, M. 2015, \apj, 801, 20

\bibitem[{{Tanaka} {et~al.}(2018){Tanaka}, {Coupon}, {Hsieh}, {Mineo},
  {Nishizawa}, {Speagle}, {Furusawa}, {Miyazaki}, \&
  {Murayama}}]{2018PASJ...70S...9T}
{Tanaka}, M., {Coupon}, J., {Hsieh}, B.-C., {et~al.} 2018, \pasj, 70, S9

\bibitem[{{Toba} \& {Nagao}(2016)}]{2016ApJ...820...46T}
{Toba}, Y., \& {Nagao}, T. 2016, \apj, 820, 46

\bibitem[{{Toba} {et~al.}(2015){Toba}, {Nagao}, {Strauss}, {Aoki}, {Goto},
  {Imanishi}, {Kawaguchi}, {Terashima}, {Ueda}, {Bosch}, {Bundy}, {Doi},
  {Inami}, {Komiyama}, {Lupton}, {Matsuhara}, {Matsuoka}, {Miyazaki},
  {Morokuma}, {Nakata}, {Oi}, {Onoue}, {Oyabu}, {Price}, {Tait}, {Takata},
  {Tanaka}, {Terai}, {Turner}, {Uchida}, {Usuda}, {Utsumi}, {Yamada}, \&
  {Wang}}]{2015PASJ...67...86T}
{Toba}, Y., {Nagao}, T., {Strauss}, M.~A., {et~al.} 2015, \pasj, 67, 86

\bibitem[{{Toba} {et~al.}(2017){Toba}, {Nagao}, {Kajisawa}, {Oogi}, {Akiyama},
  {Ikeda}, {Coupon}, {Strauss}, {Wang}, {Tanaka}, {Niida}, {Imanishi}, {Lee},
  {Matsuhara}, {Matsuoka}, {Onoue}, {Terashima}, {Ueda}, {Harikane},
  {Komiyama}, {Miyazaki}, {Noboriguchi}, \& {Usuda}}]{2017ApJ...835...36T}
{Toba}, Y., {Nagao}, T., {Kajisawa}, M., {et~al.} 2017, \apj, 835, 36

\bibitem[{{Tonry} {et~al.}(2012){Tonry}, {Stubbs}, {Lykke}, {Doherty},
  {Shivvers}, {Burgett}, {Chambers}, {Hodapp}, {Kaiser}, {Kudritzki},
  {Magnier}, {Morgan}, {Price}, \& {Wainscoat}}]{2012ApJ...750...99T}
{Tonry}, J.~L., {Stubbs}, C.~W., {Lykke}, K.~R., {et~al.} 2012, \apj, 750, 99

\bibitem[{{Wright} {et~al.}(2010){Wright}, {Eisenhardt}, {Mainzer}, {Ressler},
  {Cutri}, {Jarrett}, {Kirkpatrick}, {Padgett}, {McMillan}, {Skrutskie},
  {Stanford}, {Cohen}, {Walker}, {Mather}, {Leisawitz}, {Gautier}, {McLean},
  {Benford}, {Lonsdale}, {Blain}, {Mendez}, {Irace}, {Duval}, {Liu}, {Royer},
  {Heinrichsen}, {Howard}, {Shannon}, {Kendall}, {Walsh}, {Larsen}, {Cardon},
  {Schick}, {Schwalm}, {Abid}, {Fabinsky}, {Naes}, \&
  {Tsai}}]{2010AJ....140.1868W}
{Wright}, E.~L., {Eisenhardt}, P.~R.~M., {Mainzer}, A.~K., {et~al.} 2010, \aj,
  140, 1868

\bibitem[{{Wu} {et~al.}(2012){Wu}, {Tsai}, {Sayers}, {Benford}, {Bridge},
  {Blain}, {Eisenhardt}, {Stern}, {Petty}, {Assef}, {Bussmann}, {Comerford},
  {Cutri}, {Evans}, {Griffith}, {Jarrett}, {Lake}, {Lonsdale}, {Rho},
  {Stanford}, {Weiner}, {Wright}, \& {Yan}}]{2012ApJ...756...96W}
{Wu}, J., {Tsai}, C.-W., {Sayers}, J., {et~al.} 2012, \apj, 756, 96

\bibitem[{{Yamashita} {et~al.}(2018){Yamashita}, {Nagao}, {Akiyama}, {He},
  {Ikeda}, {Tanaka}, {Niida}, {Kajisawa}, {Matsuoka}, {Nobuhara}, {Lee},
  {Morokuma}, {Toba}, {Kawaguchi}, \& {Noboriguchi}}]{2018ApJ...866..140Y}
{Yamashita}, T., {Nagao}, T., {Akiyama}, M., {et~al.} 2018, \apj, 866, 140

\bibitem[{{York} {et~al.}(2000){York}, {Adelman}, {Anderson}, {Anderson},
  {Annis}, {Bahcall}, {Bakken}, {Barkhouser}, {Bastian}, {Berman}, {Boroski},
  {Bracker}, {Briegel}, {Briggs}, {Brinkmann}, {Brunner}, {Burles}, {Carey},
  {Carr}, {Castander}, {Chen}, {Colestock}, {Connolly}, {Crocker}, {Csabai},
  {Czarapata}, {Davis}, {Doi}, {Dombeck}, {Eisenstein}, {Ellman}, {Elms},
  {Evans}, {Fan}, {Federwitz}, {Fiscelli}, {Friedman}, {Frieman}, {Fukugita},
  {Gillespie}, {Gunn}, {Gurbani}, {de Haas}, {Haldeman}, {Harris}, {Hayes},
  {Heckman}, {Hennessy}, {Hindsley}, {Holm}, {Holmgren}, {Huang}, {Hull},
  {Husby}, {Ichikawa}, {Ichikawa}, {Ivezi{\'c}}, {Kent}, {Kim}, {Kinney},
  {Klaene}, {Kleinman}, {Kleinman}, {Knapp}, {Korienek}, {Kron}, {Kunszt},
  {Lamb}, {Lee}, {Leger}, {Limmongkol}, {Lindenmeyer}, {Long}, {Loomis},
  {Loveday}, {Lucinio}, {Lupton}, {MacKinnon}, {Mannery}, {Mantsch}, {Margon},
  {McGehee}, {McKay}, {Meiksin}, {Merelli}, {Monet}, {Munn}, {Narayanan},
  {Nash}, {Neilsen}, {Neswold}, {Newberg}, {Nichol}, {Nicinski}, {Nonino},
  {Okada}, {Okamura}, {Ostriker}, {Owen}, {Pauls}, {Peoples}, {Peterson},
  {Petravick}, {Pier}, {Pope}, {Pordes}, {Prosapio}, {Rechenmacher}, {Quinn},
  {Richards}, {Richmond}, {Rivetta}, {Rockosi}, {Ruthmansdorfer}, {Sandford},
  {Schlegel}, {Schneider}, {Sekiguchi}, {Sergey}, {Shimasaku}, {Siegmund},
  {Smee}, {Smith}, {Snedden}, {Stone}, {Stoughton}, {Strauss}, {Stubbs},
  {SubbaRao}, {Szalay}, {Szapudi}, {Szokoly}, {Thakar}, {Tremonti}, {Tucker},
  {Uomoto}, {Vanden Berk}, {Vogeley}, {Waddell}, {Wang}, {Watanabe},
  {Weinberg}, {Yanny}, {Yasuda}, \& {SDSS Collaboration}}]{2000AJ....120.1579Y}
{York}, D.~G., {Adelman}, J., {Anderson}, Jr., J.~E., {et~al.} 2000, \aj, 120,
  1579

\end{thebibliography}
%%%%%%%%%%%%%%%%%%%%%%%%%%%%
\appendix
{
\section{Catalog of DOGs in this study} \label{sec:app_cd}
In Table \ref{tab:allDOGsTable}, we list photometric data (ID, R.A., Decl., magnitudes and those errors in each band, photo-$z$, and classification) of the DOGs studied in this paper. The sample of DOGs is selected from a combining catalog using the HSC, VIKING, and ALLWISE clean samples (see Section \ref{sec:ss}), and includes 571 DOGs (51 Bump DOGs, 257 PL DOGs, and 263 Unclassified DOGs).
%%%%%%%%%%
\begin{longrotatetable}
%\begin{splitdeluxetable*}{rrrrrrrrrBrrrrrrrrrc}
\begin{splitdeluxetable*}{crrccccccBcccccccccc}
\tablecaption{Catalog of DOGs in this study}
\tablehead{
\colhead{HSC ID} & \colhead{R.A. (J2000)} & \colhead{Decl. (J2000)} & \colhead{$g$} & \colhead{$r$} & \colhead{$i$} & \colhead{$z$} & \colhead{$y$} &
\colhead{$Z$} & \colhead{$Y$} & \colhead{$J$} & \colhead{$H$} & \colhead{$Ks$} &
\colhead{$W1$} & \colhead{$W2$} & \colhead{$W3$} & \colhead{$W4$} & \colhead{Photo-$z$} & \colhead{DOG classification}\\
(1) & (2) & (3) & (4) & (5) & (6) & (7) & (8) & (9) & (10) & (11) & (12) & (13) & (14) & (15) & (16) & (17) & (18)  & (19) 
} 
\startdata
HSC J021647.48$-$041334.6 & 34.19785 & $-$4.22628 & $24.23\pm0.03$ & $23.44\pm0.03$ & $22.59\pm0.01$ & $21.89\pm0.01$ & $21.70\pm0.02$ & $22.16\pm0.16$ & \nodata & \nodata & $21.78\pm0.38$ & $21.07\pm0.21$ & $20.53\pm0.21$ & \nodata & \nodata & $15.48\pm0.35$ & $0.96\pm0.05$ & u\\
HSC J021656.62$-$051005.4 & 34.23591 & $-$5.16816 & $24.05\pm0.03$ & $23.64\pm0.03$ & $23.00\pm0.01$ & $22.43\pm0.01$ & $22.20\pm0.03$ & $22.28\pm0.14$ & $21.60\pm0.16$ & $21.52\pm0.18$ & $20.80\pm0.17$ & $20.81\pm0.16$ & $19.27\pm0.07$ & $18.35\pm0.06$ & $17.05\pm0.23$ & $15.49\pm0.36$ & $1.11\pm0.04$ & p\\
HSC J021718.52$-$034350.1 & 34.32716 & $-$3.73057 & $24.96\pm0.07$ & $24.08\pm0.06$ & $23.78\pm0.04$ & $23.29\pm0.06$ & $23.22\pm0.11$ & \nodata & \nodata & \nodata & \nodata & $20.82\pm0.17$ & $19.27\pm0.07$ & $19.02\pm0.12$ & $17.75\pm0.43$ & $14.94\pm0.23$ & $0.39\pm0.13$ & p\\
HSC J021729.07$-$041937.6 & 34.37111 & $-$4.32712 & $22.40\pm0.01$ & $21.85\pm0.01$ & $21.30\pm0.00$ & $20.91\pm0.00$ & $20.71\pm0.01$ & $20.75\pm0.05$ & \nodata & $20.68\pm0.15$ & $20.48\pm0.12$ & $19.90\pm0.07$ & $18.93\pm0.06$ & $17.94\pm0.05$ & $15.68\pm0.07$ & $14.04\pm0.12$ & $0.95\pm0.05$ & p\\
HSC J021742.81$-$034531.0 & 34.42836 & $-$3.75861 & $24.31\pm0.03$ & $23.74\pm0.04$ & $23.36\pm0.02$ & $23.14\pm0.03$ & $23.28\pm0.07$ & $22.37\pm0.19$ & \nodata & \nodata & \nodata & $21.00\pm0.20$ & $20.22\pm0.16$ & $19.32\pm0.16$ & $16.56\pm0.14$ & $14.88\pm0.24$ & $0.64\pm0.21$ & p\\
HSC J021749.02$-$052306.7 & 34.45423 & $-$5.38521 & $23.47\pm0.02$ & $22.70\pm0.01$ & $22.00\pm0.01$ & $21.56\pm0.01$ & $20.46\pm0.01$ & $21.68\pm0.09$ & $20.65\pm0.07$ & $20.62\pm0.08$ & $20.17\pm0.10$ & $19.24\pm0.04$ & $17.70\pm0.03$ & $16.73\pm0.03$ & $15.21\pm0.05$ & $14.15\pm0.14$ & $1.47\pm0.01$ & p\\
\enddata
\tablecomments{Column (1): Object ID named based on the coordinates of HSC-SSP. Column (2)--(3): Object coordinates in HSC-SSP in units of degree. Column (4)--(17): Photometric magnitudes in HSC-SSP ($grizy$), VIKING ($ZYJHKs$), and ${\it WISE}$ ($W1-4$) in units of AB magnitude. The VIKING and ${\it WISE}$ magnitudes were converted from Vega magnitude using the conversion factors (\citealt{2018MNRAS.474.5459G};  \citealt{2011ApJ...735..112J}). The HSC and VIKING data were corrected for Galactic extinction \citep{1998ApJ...500..525S}. Non-detection and non-observation are represented as ``$\cdots$''. Column (18): Photometric redshift estimated with the HSC MIZUKI code (\citealt{2015ApJ...801...20T}; \citealt{2018PASJ...70S...9T}). Column (19): Classification of DOG types. ``p'', ``b'', and ``u'' mean PL DOGs, Bump DOGs, and unclassified DOGs, respectively (see Section \ref{subsec: ClassD} for details). For BluDOGs, its BluDOG ID marked in Figure \ref{fig:image_BLU} is also provided as a suffix. The selection of BluDOGs is described in Section \ref{subsec:BluDOGs}.\newline\newline (This catalog data is available in its entirety in the online journal and the VizieR catalog service. A portion is shown here for introducing the catalog.)}
\label{tab:allDOGsTable}
\end{splitdeluxetable*}
\end{longrotatetable}
%%%%%%%%%%
}

%%%%%%%%%%%%%%%%%%%%%%%%%%%%

\end{document}